\documentclass[12pt]{article}
\pdfoutput=1
\usepackage[]{hyperref}
\usepackage[]{putex}
\usepackage[utf8]{inputenc}
\usepackage[table]{xcolor}
\usepackage{amsmath,amsthm,amssymb}
\usepackage{tikz}
\usepackage{tikz-cd}
\usetikzlibrary{decorations.markings}
\usetikzlibrary{calc}
\usepackage{subcaption}
\usepackage{enumerate}
\usepackage{enumitem}
\usepackage{multirow}
\usepackage{mathtools}
\usepackage{bm}
\allowdisplaybreaks

\definecolor{lightgray2}{gray}{0.96}

\newcommand{\beq}{\begin{equation}}
\newcommand{\eeq}{\end{equation}}
\newcommand{\bea}{\begin{eqnarray}}
\newcommand{\eea}{\end{eqnarray}}

\definecolor{colour1}{rgb}{0.8, 0.6, 1}
\definecolor{cyan2}{rgb}{0, 1, 1}
\renewcommand\fbox{\fcolorbox{gray}{white}}

\usepackage{setspace}
\allowdisplaybreaks

\begin{document}

\vskip-2cm
\title{RG Flows and Fixed Points of $O(N)^r$ Models}

\authors {Christian Jepsen$^1$ \&  Yaron Oz$^2$}%
\institution{KIAS}{$^1$School of Physics, Korea Institute for Advanced Study, Seoul 02455, Korea.}
\institution{Tel-Aviv}{
$^2$School of Physics and Astronomy, Tel-Aviv University, Tel-Aviv 69978, Israel.
}

\date{\today}

\abstract{

By means of $\epsilon$ and large $N$ expansions, we study generalizations of the $O(N)$ model where the fundamental fields are tensors of rank $r$ rather than vectors, and where the global symmetry (up to additional discrete symmetries and quotients) is $O(N)^r$, focusing on the cases $r\leq 5$. Owing to the distinct ways of performing index contractions, these theories contain multiple quartic operators, which mix under the RG flow. At all large $N$ fixed points, melonic operators are absent and the leading Feynman diagrams are bubble diagrams, so that all perturbative fixed points can be readily matched to full large $N$ solutions obtained from Hubbard-Stratonovich transformations. The family of fixed points we uncover extend to arbitrary higher values of $r$, and as their number grows superexponentially with $r$, these theories offer a vast generalization of the critical $O(N)$ model.

We also study sextic $O(N)^r$ theories, whose large $N$ limits are obscured by the fact that the dominant Feynman diagrams are not restricted to melonic or bubble diagrams. For these theories the large $N$ dynamics differ qualitatively across different values of $r$, and we demonstrate that the RG flows possess a numerous and diverse set of perturbative fixed points beginning at rank four.

}

\maketitle
{\setstretch{1.25}
\tableofcontents
}
\section{Introduction}

In the absence of supersymmetry, and setting aside low-dimensional toy models, the means by which to gain an analytic handle on the observables of interacting quantum field theories are preciously rare. The current methods available to tackle such theories are well-illustrated by the family of conformal field theories known as the critical $O(N)$ model, whose experimental motivation derives from their utility in describing critical phenomena in liquids, gases, and magnets, and which can be identified with the non-trivial RG fixed point of a scalar QFT of vector fields $\phi^a$, $a\in \{1,2,...,N\}$, subject to a quartic interaction, 
\begin{align}
S = \int d^dx \bigg( \frac{1}{2}(\partial_i\phi^a)(\partial_i\phi^a)
+ \frac{1}{4!}g\big(\phi^a\phi^a\big)^2
\bigg)\,.
\label{O(N)action}
\end{align}
Formally, the theories can be defined in a continuous range of spacetime dimensions $2<d<4$ and admit perturbative treatments in the $\epsilon$ expansion in $d=4-\epsilon$ as well as in $d=2+\epsilon$ dimensions, in addition to being analytically tractable order by order in $1/N$ across the range of dimensions.\footnote{ In dimensions above four, the $O(N)$ model possesses a UV stable fixed point at large $N$. It was pointed out by Parisi \cite{parisi1975theory,parisi1976non} that the operator $\phi^i\phi^i$ violates the unitarity bound for $d>4$ but that the UV fixed point may be unitary between four and six dimension. More recently, Ref.~\cite{fei2014critical} matched the large $N$ UV fixed point to a $6-\epsilon$ expansion.} Outside these limits, theoretical knowledge of this family of theories derives from numerical studies using Monte Carlo methods and the conformal bootstrap.\footnote{See Ref.~\cite{henriksson2023critical} for a comprehensive review of the critical $O(N)$ model.} 

Since their pioneering usage in the context of the $O(N)$ model, the small $\epsilon$ and large $N$ expansions have been fruitfully brought to bear on a host of other quantum theories, and in the present paper we shall apply these methods to the generalizations of the $O(N)$ model obtained by advancing from a single-index scalar $\phi^a$ to multi-index scalars $\phi^{ab}$, $\phi^{abc}$, etc., where we take each index to transform under its own $O(N)$ group. These theories can be thought of as tensorial generalizations of $\phi^4$ theory in the same sense as the $O(N)$ model can be viewed as the vectorial generalization. While the $O(N)$ model at fixed $N$ has but a single non-trivial fixed point in the $\epsilon$ expansion, the $O(N)^r$ models generally have multiple fixed points whose number depends on $N$. But as we let $N$ increase, the number of perturbative fixed points at fixed $r$ eventually stabilizes to a number whose value exhibits double exponential growth with the tensor rank $r$. For the theories with $r\leq 5$ we will see explicitly how all of the fixed points, exact in $N$ but perturbative in $\epsilon$, can be matched to large $N$ solutions that are exact in $d$ but perturbative in $1/N$. In the strict large $N$ limit, the fixed points all look similar to the free or the interacting large $N$ fixed point of the $O(N)$ model, but their $1/N$ corrections are vastly different and suggest that the $O(N)^r$ models furnish a large set of distinct three-dimensional CFTs.

For the sake of explicitness, let us take a look at the $O(N)^3$ model. The possible types of index contractions among quartic operators in this theory can be illustrated by the three diagrams shown below.
\begin{align}
\begin{matrix}
\text{
		\includegraphics[scale=1]{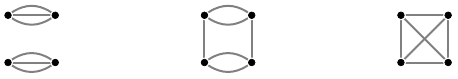}
}
\end{matrix}
\end{align}
We can refer to the three types of interactions as ``double-trace",``pillow", and ``tetrahedron" respectively. If we do not symmetrize with respect to the three $O(N)$ groups of the model, there are actually three different pillow operators, so that the RG flow, after tuning away the quadratic operator, is five-dimensional. When $N\geq 29$, the model has 16 real fixed points in the $\epsilon$ expansion. On scaling the coupling constants with appropriate powers of $N$, one finds that the tetrahedron operator vanishes for all the 16 fixed points as $N$ is taken to infinity. Phrased differently: at large $N$, fixed points with the tetrahedron present require the other coupling constants to be complex. The 16 real fixed points, with only double-trace and pillow operators present, can be matched to 16 large $N$ solutions obtained via Hubbard-Stratonovich transformations. The double trace operator can be integrated through the introduction of the standard $\sigma$ field as in the $O(N)$ model, while the three pillow operators can be integrated out with three fields $\chi^{i_1i_2}$ that each transforms as a traceless symmetric matrix under one of the $O(N)$ groups of the model. The presence or absence of the $\sigma$ field and the three $\chi$ fields results in the $2^4$ total large $N$ solutions. The $O(N)^4$ and $O(N)^5$ models exhibit a similar, simple type of behaviour at large $N$: at any fixed point, only such operators are present as can be integrated out via the Hubbard-Stratonovich transformation, so that leading anomalous dimensions can always be computed exactly in $d$. To determine subleading $1/N$ corrections, however, it is necessary to incorporate all the operators of the models, including, in the rank three case, the tetrahedron. The type of fixed points just described can be shown to exist in increasing numbers for the $O(N)^r$ models with $r>5$, but other types of fixed points may also be present at higher even rank.

In addition to analyzing tensorial $\phi^4$ theory, we also present a study of tensorial $\phi^6$ theory. In the vector case, ie. for sextic $O(N)$ theory, the beta functions become trivial in the strict large $N$ limit. A UV fixed point does appear in three dimensions at sub-leading order in $1/N$ but is situated in a regime where the theory is believed to be unstable at large $N$ (although potentially not at finite $N$), and away from three dimensions this large $N$ fixed point becomes complex. Tensorial $\phi^6$ theory, however, exhibits a substantially different large $N$ limit from the vector model, with non-trivial beta functions and fixed points present already to leading order at large $N$. For tensors at at ranks two and three, a single non-trivial large $N$ fixed point is present in the $3-\epsilon$ dimension, but as we shall see, the number of fixed points increases drastically at rank four.

In the following two subsections, we briefly review aspects of large $N$ and small $\epsilon$ expansions relevant to this paper, before we proceed in Subsection~\ref{subSec:Overview} to provide an overview of the paper and its results.

\subsection{Large \texorpdfstring{$N$}{N} limits}
Within the vast body of literature devoted to studying large $N$ physics, there are three broad classes of known large $N$ limits among the quantum theories with a single interaction term, each of which is dominated by its own type of Feynman diagrams:\footnote{We refer the reader to Ref.~\cite{klebanov2018tasi} for a review and comparison of the different types of large $N$ limits.}
\begin{itemize}
\item The vector model large $N$ limit, dominated by bubble diagrams

This limit, which is realized by the $O(N)$ model and also describes aspects of polymer physics, is the most tractable of the three types of large $N$ limits. At large $N$ the only surviving Feynman diagrams in the $O(N)$ model are bubble diagrams like the one shown below.
\begin{align}
\begin{matrix}
\text{
		\includegraphics[scale=1]{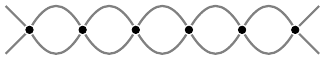}
}
\end{matrix}
\end{align}
The leading large $N$ approximation of the $O(N)$ model was obtain in Ref.~\cite{coleman1974spontaneous} through the use of a Hubbard-Stratonovich transformation that we discuss in more detail later on (in \ref{subSubSec:vector}). Higher-order corrections in $1/N$ were subsequently computed in Refs.~\cite{vasil1981simple,vasil19811,vasil19821}. An important fact about the vectorial large $N$ limit is that on renormalizing the Feynman diagram with a single bubble, the diagrams with multiple bubbles are also rendered finite, in consequence of which the large $N$ beta function is one-loop exact. Ref.~\cite{moshe2003quantum} provides a review of QFTs exhibiting the vector model large $N$ limit.

\item The matrix model large $N$ limit, dominated by planar diagrams. 

This limit is possible in theories of tensors of rank two and higher. A simple example of a theory that realizes the limit is the quartic matrix theory with the following interaction:
\begin{align}
\begin{matrix}
\text{
		\includegraphics[scale=1]{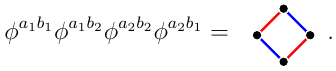}
}
\end{matrix}
\end{align}
The Feynman diagrams that contribute to the four-point function at large $N$ are planar diagrams like the following:
\begin{align}
\nonumber\\[-45pt]
\begin{matrix}
\text{
		\includegraphics[scale=0.8]{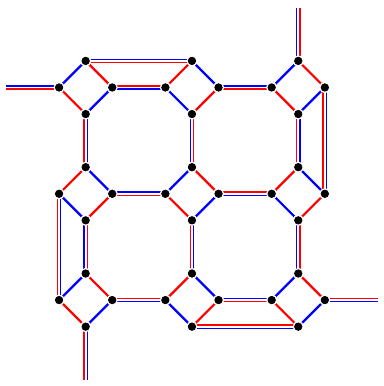}
}
\end{matrix}
\\[-40pt] \nonumber
\end{align}
The set of planar diagrams is generally a large, unwieldy set, unlike the subsets of planar diagrams that dominate in the other two types of large $N$ limits. For this reason, theories in the planar limit are usually much more complicated than theories in the other types of limits, and only for a limited number of low-dimensional examples of theories, such as $2d$ $SU(N)$ theory coupled to quarks in the fundamental representation \cite{hooft2} or general zero-dimensional QFTs \cite{brezin1978planar}, are large $N$ solutions known in the planar limit. Conversely, however, matrix theories offer the benefit of describing an extraordinarily rich range of phenomena, spanning QCD and quantum gravity. Reviews of matrix models at large $N$ and their connections to 2D gravity can be found in Refs.~\cite{klebanov1991string,di19952d}.

\item The melonic large $N$ limit, dominated by so-called melon diagrams. 

This limit was discovered and analyzed by Gurau and collaborators in Refs.~\cite{gurau2011colored,gurau2012colored,gurau20111,gurau2012complete,bonzom2011critical,bonzom2012random} and can appear in theories of tensors of rank three and higher. Foundational work on tensor models \cite{ambjorn1991three,sasakura1991tensor} was motivated by attempts to describe quantum gravity above two dimensions, to which continual effort has been direct in the form of the tensor track \cite{rivasseau2012quantum}, but other physical applications have emerged later on. In particular, the melonic limit provides a way to realize the Schwinger-Dyson equation of the SYK model \cite{sachdev1993gapless,kitaev2015simple,maldacena2016remarks} in genuine quantum mechanical models, rather than in disorder averages of theories, as first shown by Witten \cite{witten2019syk} and also demonstrated for the simple quantum mechanical theory put forward by Klebanov and Tarnopolsky \cite{klebanov2017uncolored}.\footnote{Although they share the same Schwinger-Dyson equation, there are also important differences between the SYK and Klebanov-Tarnopolsky models. The latter contains singlet operators not present in the former \cite{bulycheva2018spectra} as well as additional modes in the fluctuations about the finite temperature saddle point \cite{choudhury2018notes}.}  The simplest version of this theory contains a quartic fermionic interaction: 
\begin{align}
\label{melonicQuartic}
\begin{matrix}
\text{
		\includegraphics[scale=1]{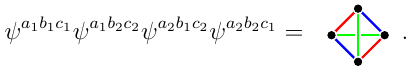}
}
\end{matrix}
\end{align}
To describe the class of diagrams that dominate at large $N$ one introduces the melonic substitution of free propagator given by
\begin{align}
\label{quarticMelonicSub}
\begin{matrix}
\text{
		\includegraphics[scale=1]{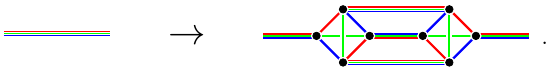}
}
\end{matrix}
\end{align}
In a simplified diagrammatic notation, the melonic substitution can be expressed as
\begin{align}
\begin{matrix}
\text{
		\includegraphics[scale=1]{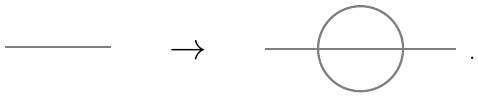}
}
\end{matrix}
\hspace{12mm}
\end{align}
The diagrams that contribute at large $N$, ie. the subset of planar diagrams known as the melonic diagrams, are those generated by iteratively performing the melonic substitution. In Figure~\ref{fig:melons} we display all melonic vacuum diagrams up to eigth order in the coupling constant.
\begin{figure}
    \centering
\begin{align*}
\hspace{-6mm}
\begin{matrix}\text{
\includegraphics[scale=0.9]{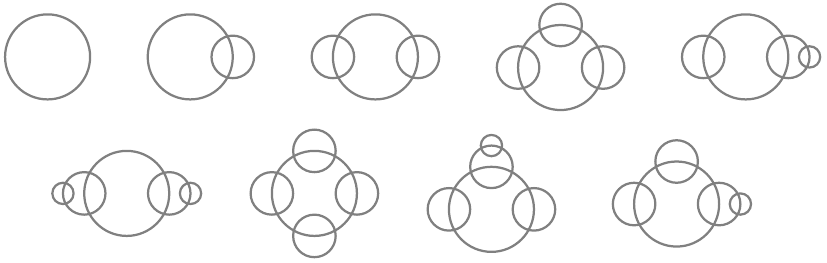}
}
\end{matrix}
\end{align*}
    \caption{Melonic vacuum diagrams to eigth order in the coupling constant} 
    \label{fig:melons}
\end{figure}
The diagrams that contribute to the two-point function admit a re-summation into a simple Schwinger-Dyson equation that can be solved in the IR, and the melonic four-point diagrams posses a recursive structure in term of ladder diagrams from which it is possible to derive the spectrum of bilinear operators 
 \cite{klebanov2017uncolored}, while higher-point functions can be tackled using the machinery developed in Ref.~\cite{gross2017all}. We emphasize that the interaction \eqref{melonicQuartic} is but one example of a much wider class of interactions exhibiting a melonic pattern of diagrams generated by iteratively replacing a free propagator with a propagator containing two interaction vertices, and these interactions need be neither fermionic nor quartic, see for example Refs.~\cite{narayan2017syk,gubser2018higher,ferrari2019new,klebanov2019majorana} for studies of higher-rank interactions. For more details on the melonic limit, we refer the reader to Ref.~\cite{gurau2019notes} for a review of melonic tensor models, Ref.~\cite{benedetti2020melonic} for a review of melonic tensor field theories in $d\geq 2$, and Ref.~\cite{gurau2016special} for a collection of reviews of random tensor models.\footnote{Recent work on melonic theories include studies of spontaneous symmetry breaking in tensor models \cite{diaz2018chiral,diaz2019spontaneous,kim2019symmetry,benedetti2019phase,benedetti2020s}, phase diagrams of melonic quantum mechanics \cite{azeyanagi2018phase,ferrari2019phases}, melonic theories defined over $p$-adic numbers \cite{gubser2018melonic,gubser2018higher}, supersymmetric tensor models \cite{chang2019supersymmetric,popov2020supersymmetric,lettera2022large}, Hagedorn temperature in fermionic tensor quantum mechanics \cite{gaitan2020hagedorn}, and surface defects in tensor models \cite{popov2022non}.
}
\end{itemize}
While the above three limits have each been studied thoroughly, theories with multiple competing interaction terms of different type are generally much less understood than purely vectorial, matricial, or melonic theories. But this is precisely the situation we face when we advance from the $O(N)$ theory of a single-index scalar to theories of multi-index scalars: interactions of every type emerge, arising from the distinct ways of contracting indices amongst the fundamental fields, and renormalized perturbation theory dictates that we omit no relevant operator compatible with the symmetry we choose to consider. While some interactions generate more powers of $N$ than others, a full exploration of the dynamics of a theory with multiple interactions requires that coupling constants be scaled down by suitable powers of $N$ such that all operators contribute on an equal footing. Theories with couplings scaled in this manner are sometimes referred to and studied in the literature under the name of \emph{enhanced tensor theories} \cite{bonzom2013new,bonzom2015enhancing}\footnote{See Ref.~\cite{geloun2023beta} for recent work on quartic theories of this kind.}, with the different scalings of couplings representing the tensorial generalization of the familiar matrix theory fact that coupling constants are scaled according to the number of traces in the associated operators. For rank-three tensor models with $O(N)^3$ symmetry Ref.~\cite{carrozza2016n} worked out the suitable scalings of the coupling constants and established the large $N$ expansion, but in general no algorithm exists for determining large $N$ scalings. Abstractly considered, a theory allows for a set of permitted ranges of scalings that do not lead to a blow up of the theory as $N\rightarrow \infty$, and among these permitted choices Ref.~\cite{ferrari2019new} introduced the term \emph{optimal scaling} to denote the smallest set of scaling exponents compatible with a well-defined $1/N$ expansion. For the specific multi-index generalizations of the $O(N)$ model we study in this paper, it is not difficult to determine the optimal scalings. We are then left with the question, what are the large $N$ RG flows and fixed points under these scalings? By invoking the aid of the $\epsilon$ expansion, we will argue that the large $N$ landscape can be charted out with reasonable confidence.

\subsection{Small \texorpdfstring{$\epsilon$}{epsilon} expansions}
The success of Wilson's and Fisher's outlandish idea \cite{wilson1972critical} to use the number of spacetime dimensions below four as an expansion parameter in $\phi^4$ theory hinges on the fortuitous circumstance that the initial coefficients of the expansions are sufficiently small and decreasing that even an expansion parameter of unit size provides a reasonable approximation to critical exponents in three dimensions. The idea has since proven practicable widely beyond $\phi^4$ theory and spurred a slew of work charting out fixed points in the $\epsilon$ expansion.\footnote{See Refs.~\cite{pelissetto2002critical} and \cite{rychkov2019general} for careful reviews of fixed points in the $4-\epsilon$ expansion, and Ref.~\cite{michel1984renormalization} for early foundational work in this direction.} In the $4-\epsilon$ expansion, the fixed points of four-component \cite{toledano1985renormalization} and five-component \cite{rong2023classifying} scalar theories have now been completely classified, while scalars with six \cite{hatch1985selection,kim1986classification,hatch1986renormalization} and seven components \cite{osborn2021heavy} have been thoroughly if not exhaustively explored, in addition to which several infinite families are known. One of these families consists of the fixed points of the $O(N)$ model itself, while another is the set of bifundamental fixed points of $N_1\times N_2$ matrix fields $\phi^{ab}$ \cite{kawamura1990generalized,vicari2007critical,osborn2018seeking}. The type of fixed points relevant to the present paper are further extensions of these two families into multifundamental fixed points of tensor fields of rank $r$, where we limit the enormous parameter space by taking the dimensions of all the indices to equal the same number: $N_1=N_2=...=N_r\equiv N$.\footnote{There are no fixed points in the bifundamental family with $N_1=N_2$ unless $N_1=N_2=2$, but the same does not apply at higher rank.} These $r$-fundamental fixed points exhibit an $O(N)^r/(\mathbb{Z}_2)^{r-1}$ symmetry, where the $\mathbb{Z}_2$ quotients owe to the fact each $O(N)$ group carries its own $\mathbb{Z}_2$ action, while the theory enjoys but a single overall $\mathbb{Z}_2$ symmetry that sends the fundamental fields to minus themselves. Alternatively, for $N$ odd, the symmetry can be labelled as $\mathbb{Z}_2\times SO(N)^r$. For general $r$, however, and for sufficiently large $N$, there exists not one but numerous distinct multi-fundamental fixed points, and for most of them the symmetry is enhanced to a larger group. Although in some cases the symmetry is enhanced with an extra $S_r$ permutation symmetry, we emphasize that the multi-fundamental fixed points are distinct from the so-called MN fixed points \cite{mukamel1975epsilon,michel1984symmetry,osborn2018seeking}, which possess $O(N)^r \rtimes S_r$ symmetry, but which are fixed points of $(rN)$-component scalar theories, as opposed to the $N^r$-component theories we will study. In uplifting the $O(N)$ model to tensorial generalizations, our approach is ultimately complementary to the $\epsilon$ expansion literature, which generally focuses on theories with small field contents, while we will be dealing with theories with vast numbers of field components. In the case of $r=3$, for example, the full set of tri-fundamental fixed points do not appear until $N\geq 29$, ie. until the scalar fields carry at least $24\,389$ components.

\subsection{Overview of paper}
\label{subSec:Overview}
Using large $N$ and small $\epsilon$ methods, this paper comprehensively elucidates the structure of the RG flows and fixed points of quartic $O(N)^r$ scalar theory with $r\leq 5$ and provides formulas for subleading anomalous dimension at large $N$, which are simply related to known $O(N)$ model formulas. What ultimately renders these $O(N)^r$ models solvable at large $N$ is the fact that the subset of operators that are vectorial, by which we mean operators for which the dominant diagrams at large $N$ are bubble diagrams, close on themselves under the RG flow at large $N$. This means that the RG flow can be consistently truncated by omitting operators that are not vectorial, and in the absence of these operators the large $N$ beta functions terminate at quadratic order and can be explicitly diagonalized by a simple basis transformation. This truncation can also be performed at higher rank $r>5$, and we will write down explicit closed-form expressions that count the number of vectorial fixed points at large yet finite $N$ for arbitrary rank $r$, but it is possible that additional fixed pointsk beyond those in the vectorial class exist at higher even ranks. For the vectorial fixed points, the large $N$ descriptions at any $r$ are obtained by generalizations of the Hubbard-Stratonovich transformation, expressing operators as integrals over intermediary fields. In addition to the standard $\sigma$ field, the additional intermediary fields carry pairs of indices, with each pair transforming as a traceless symmetric or antisymmetric matrix under its own $O(N)$ group. For some fixed points, these intermediate fields give rise to Feynman rules by which $1/N$ corrections to observables are computable to arbitrary order, but for many of the fixed points non-vectorial operators are present at suppressed order, so that no such simple Feynman rules can be formulated. There generally exist numerous relevant deformations of the $O(N)^r$ fixed points, unless particular symmetries are imposed to exclude them. Either scenario renders the challenge of finding concrete real-life applications of the $O(N)^r$ fixed points rather formidable, but it will likely be less challenging to identify statistical mechanical systems with fixed points that fit in this class.

As a contrast and supplement to our analysis of quartic $O(N)^r$ theory, we also present a survey of sextic $O(N)^r$ models for $r\leq 4$ and the challenges they afford. In contradistinction to the simple structure that underlies quartic tensor theory, the space of RG fixed points for tensor theories with sextic interactions is vastly more complicated. This is not apparent from the sextic rank-two and rank-three theories, which have previously been studied in the literature, and for which, in each case, there exists but a single standard large $N$ fixed point in the $3-\epsilon$ expansion, but the complexity becomes discernible at rank four. For the sextic $O(N)^4$ model we find that even the reduced system with additional $S_4$ permutation symmetry possesses 17 Wilson-Fisher fixed points. As the rank $r$ is taken to be even larger, simply determining the number of sextic $O(N)^r$ invariants quickly becomes a difficult task, posing a roadblock to a classification of sextic fixed points. For the new rank-four fixed points we uncover in the $\epsilon$ expansion, the determination of the leading large $N$ behaviour is hindered by the presence of matricial type of operators that contribute to the large $N$ dynamics alongside vectorial operators. Attempts to match known melonic large $N$ solutions of sextic models to $\epsilon$ expansions in renormalized perturbation theory are faced with a separate set of difficulties, as the $d$-dependence of these large $N$ solutions cannot be matched by conventional Wilson-Fisher fixed points but require coupling constants with a leading scaling $g\sim \sqrt{\epsilon}$, meaning that the perturbative fixed points do not appear at quadratic order in beta functions, but rely on a cancellation between linear and cubic terms; and the couplings that scale as $\sqrt{\epsilon}$ feed into other beta functions and induce yet more exotic scalings in other couplings, such as $\epsilon^0$ or $\epsilon^{1/4}$.

The remainder of the paper is organized as follows: Section~\ref{sec:Quartic} analyses tensorial generalizations of the quartic $O(N)$ model, first reviewing and expanding on the literature on the matrix and three-index tensor generalizations, before studying the models with $O(N)^4/(\mathbb{Z}_2)^3$ and $O(N)^5/(\mathbb{Z}_2)^4$ symmetry in the $\epsilon$ expansion and at large $N$, and finally discussing properties of $O(N)^r/(\mathbb{Z}_2)^{r-1}$ theory for general values of $r$. Section~\ref{sec:Sextic} studies $O(N)^r/(\mathbb{Z}_2)^{r-1}$ tensor theories with sextic interactions, reviewing examples in the literature and examining the particular example of the rank-four version of $\phi^6$ theory with global $S_4\ltimes O(N)^4/(\mathbb{Z}_2)^3$ symmetry, illustrating the wealth of fixed points emerging at higher rank but also the obstacles to extracting reliable information from sextic theories. Section~\ref{sec:Outlook} concludes the main portion of the paper by outlining directions of future inquiry. Appendix~\ref{sec:Optimal} contains a discussion of the suitable scalings of coupling constants with $N$ for general tensor operators, while Appendix~\ref{sec:classifying} discusses large $N$ fixed points of quartic $O(N)^r$ theory at general $r$, describing in detail the properties of vectorial fixed points and presenting an argument that melonic operators are always absent. Appendix~\ref{sec:Locations} lists the explicit values of the couplings for the fixed points of the sextic $S_4\ltimes O(N)^4/(\mathbb{Z}_2)^3$ model. 

In submitting this paper to arXiv, we have included a supplementary Mathematica notebook that contains the beta functions and anomalous dimensions to third order in the coupling constants for all the theories studied in this paper.

\section{Tensorial \texorpdfstring{$\phi^4$}{phi-to-the-fourth} Theory}
\label{sec:Quartic}
In this section we study quartic scalar theories of tensors of rank $r$, where each index transforms under a separate $O(N)$ group, endowing the theories with global $O(N)^r/(\mathbb{Z})^{r-1}$ symmetry. The theories all contain a relevant quartic deformation $\phi^2$, which we will tune away. In the special case when $N=1$, every one of the models reduces to $\phi^4$ theory, and in the case when $r=1$, the theories reduce to the $O(N)$ vector model. 

In Subsections~\ref{ON2}, \ref{ON3}, \ref{ON4}, and \ref{ON5}, we analyse in turn the theories with $r\in \{2,3,4,5\}$ and their perturbative fixed points. The theories with $r=2$ and $r=3$ were previously studied in Ref.~\cite{giombi2017bosonic}, which pointed out that they do not posses any melonic large $N$ fixed points with real-valued couplings. We will find that the same applies for $r=4$ and $r=5$, in fact all the real-valued large $N$ fixed points are dominated by operators that realize a vectorial large $N$ limit. For this reason every one of the fixed points can be understood by simple generalizations of vector model techniques. In the last subsection, \ref{generalities}, we explain how these vectorial fixed points extend to higher values of $r$.

Our survey of quartic theories focuses exclusively on dimensions below four: $d=4-\epsilon$ with $\epsilon>0$. In four dimensions, $\phi^4$ theory is well-known to flow to a free theory in the infrared, although it has been recently proposed that the quartic vector model can be made to flow to an interacting infrared theory by starting with a negative bare coupling \cite{romatschke2023if,romatschke2023loophole}, see also \cite{weller2023can} and \cite{romatschke2023negative}.

\subsection{The \texorpdfstring{$O(N)^2$}{O(N)-squared} model}
\label{ON2}
The rank-two theory contains two quartic singlets:
\begin{align}
\label{rank2ops}
\begin{matrix}
\text{
		\includegraphics[scale=1]{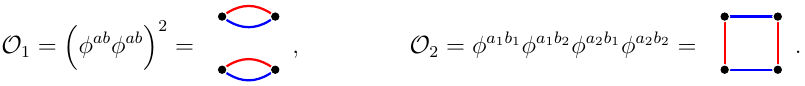}
}
\end{matrix}
\end{align}
At large $N$, the coupling constants $g_1$ and $g_2$ of these operators have the optimal scalings $g_1 \sim N^{-2}$ and $g_2\sim N^{-1}$. In addition to a global $O(N)^2/\mathbb{Z}_2$ symmetry, the model also has an $S_2$ symmetry under interchange of the two matrix indices, owing to the fact that the two operators \eqref{rank2ops} are symmetric with respect to the two colours. At higher ranks $r$, such permutation symmetries are not automatically present for generic values of couplings, but only exist in subsectors of the full $O(N)^r$ theories. 

On deforming the free matrix theory by $\mathcal{O}_1$ and $\mathcal{O}_2$, renormalized perturbation theory in the $\epsilon$ expansion reveals that, unless $N=2$, the two-loop beta functions have but one non-trivial RG fixed-point, and at this fixed point the coupling of the single-trace operator is zero. The fixed point is therefore just the vector model fixed point of an $O(N^2)$ model. For $N=2$ there is one additional fixed point with both couplings non-zero.\footnote{It generally happens in tensor theories that there are linear relations among the operators at low values of $N$. However, for the tensorial $\phi^4$ theories we are studying in this section, it can be checked that all the quartic operators are linearly independent already at $N=2$.}

A straightforward generalization of the $O(N)^2$ model is obtained by allowing the two indices to assume different numbers of values $N_1$ and $N_2$. Without loss of generality, it may be assumed that $N_2 \geq N_1$. It can then be checked that fixed points with non-zero single-trace coupling exist only when $N_1=N_2=2$ or when
\begin{align}
N_2\geq 22\,,
\hspace{15mm}
2 \leq N_1 \leq 5N_2+2-2\sqrt{6(N_2+2)(N_2-1)}\,.
\end{align}
The fixed points with non-zero value of $g_2$ that exist in these cases exhibit $O(N_1)\times O(N_2)\,/\,\mathbb{Z}_2$ symmetry and are known in the literature as the bifundamental fixed points \cite{kawamura1990generalized,vicari2007critical,osborn2018seeking,rychkov2019general}.

\subsection{The \texorpdfstring{$O(N)^3$}{O(N)-cubed} model}
\label{ON3}

Turning to the rank-three theory, there are five quartic singlets: 
\begin{align}
\label{rank3ops}
\begin{matrix}
\text{
		\includegraphics[scale=1]{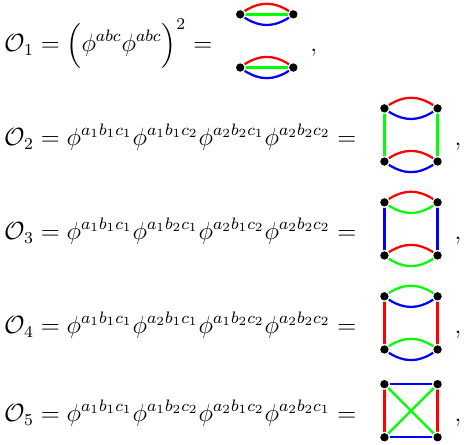}
}
\end{matrix}
\end{align}
and the action of the theory can be written as\footnote{
See Refs.~\cite{benedetti2019line,benedetti2020hints,benedetti2020conformal,benedetti2022f} for studies of variants of this theory with a non-local kinetic term, including indications of large $N$ unitarity and an explicit check of the $F$-theorem. For the generalization of this model to the theory with $O(N_1)\times O(N_2)\times O(N_3)$ symmetry, see Ref.~\cite{benedetti2021trifundamental}, and see Ref.~\cite{carrozza2019syk} for the quartic rank-three model of fermionic tensors with $Sp(N)$ symmetry and Refs.~\cite{benedetti2019phase,pascalie2019large,pascalie2021correlation} for studies of quartic tensors with three or more copies of $U(N)$ symmetry. See Ref.~\cite{bonzom2019diagrammatics} for details on the diagrammatics of $O(N)^3$ theory, and Ref.~\cite{bednyakov2021six} for a study of the beta functions at higher loops.} 
\begin{align}
S = \int d^{4-\epsilon}x
\bigg(
\frac{1}{2}
(\partial_i\phi^{abc})(\partial_i\phi^{abc})
+\frac{1}{4!}\sum_{n=1}^{5}
g_n\,\mathcal{O}_{n}
\bigg)\,.
\end{align}

\subsubsection{Small $\epsilon$ expansion}

Expanding the beta functions of the theory in a perturbative expansion,
\begin{align}
\beta_{g_i}=-\epsilon g_i + \beta^{(2)}_{g_i}+ \beta^{(3)}_{g_i}+\mathcal{O}(g^4)\,,
\end{align}
the quadratic pieces are given by
\begin{align}
\beta_{g_1}^{(2)}=
\frac{1}{48\pi^2}\Big(&(8 + N^3) g_1^2 + 2 (1 + N + N^2) g_1 g_2 + 3 g_2^2 + 
 2 (1 + N + N^2) g_1 g_3 + 2 N g_2 g_3 + 3 g_3^2 
 \nonumber \\&
 + 
 2 (1 + N + N^2) g_1 g_4 + 2 N g_2 g_4 + 2 N g_3 g_4 + 
 3 g_4^2 + 6 N g_1 g_5 + 2 g_2 g_5 + 2 g_3 g_5 
  \nonumber \\&
 + 2 g_4 g_5\Big)\,,
\\
\beta_{g_2}^{(2)}=
\frac{1}{48\pi^2}\Big(&12 g_1 g_2 + (4 + N + N^2) g_2^2 + 2 (1 + N) g_2 g_3 + 
 2 (1 + N) g_2 g_4 + 4 g_3 g_4 + 4 N g_2 g_5 
\nonumber  \\&
 + 4 g_3 g_5 + 
 4 g_4 g_5 + (2 + N) g_5^2
 \Big)\,,
\\
\beta_{g_3}^{(2)}=
\frac{1}{48\pi^2}\Big(&12 g_1 g_3 + 2 (1 + N) g_2 g_3 + (4 + N + N^2) g_3^2 + 
 4 g_2 g_4 + 2 (1 + N) g_3 g_4 + 4 g_2 g_5 
\nonumber  \\&
 + 4 N g_3 g_5 + 
 4 g_4 g_5 + (2 + N) g_5^2
 \Big)\,,
 \\
\beta_{g_4}^{(2)}=
\frac{1}{48\pi^2}\Big(&4 g_2 g_3 + 12 g_1 g_4 + 2 (1 + N) g_2 g_4 + 
 2 (1 + N) g_3 g_4 + (4 + N + N^2) g_4^2 + 4 g_2 g_5 
\nonumber   \\&
 + 
 4 g_3 g_5 + 4 N g_4 g_5 + (2 + N) g_5^2
 \Big)\,,
\\
\beta_{g_5}^{(2)}=
\frac{1}{48\pi^2}\Big(&4 g_2 g_3 + 4 g_2 g_4 + 4 g_3 g_4 + 12 g_1 g_5 + 
 2 (1 + N) g_2 g_5 + 2 (1 + N) g_3 g_5 
 \nonumber   \\&
 + 2 (1 + N) g_4 g_5
 \Big)\,. 
\end{align}
The cubic pieces are listed in the supplementary Mathematica file. Setting to zero the quadratic order beta functions, we can determine the perturbative fixed points to leading order in $\epsilon$, and we can then use the cubic order beta functions to compute the subleading corrections to these. The number of fixed points (including the trivial one) depends on $N$ as follows:\footnote{We have chosen to include the trivial fixed point so that the total number of large $N$ fixed points, 16, equals a power of 2, which will turn out to be a general fact.}
\begin{itemize}
\item For $N$ = 2 there are 16 perturbative fixed points.
\item For $3 \leq N \leq 10$ there are two perturbative fixed points: the trivial fixed point and the $O(N^3)$ model fixed point with only $g_1$ non-zero. These fixed points are present for all $N$.
\item For $11 \leq N \leq 19$ there are eight fixed points.
\item For $20 \leq N \leq 28$ there are 14 fixed points.
\item For $29 \leq N$ there are 16 fixed points.
\end{itemize}
  The above tally of fixed points includes only those present already at quadratic order in the beta functions. At higher order, additional fixed points emerge, with couplings that do not scale with integer powers of $\epsilon$, including a melonic fixed point, which, however, is complex \cite{giombi2017bosonic}. Generally, the additional fixed points are not perturbatively reliable. Indeed, the act of setting beta functions to zero produces a set of polynomial equations in the coupling constants that can be solved to search for RG fixed points; going to higher perturbative order means increasing the order of the polynomials; by mathematical necessity, higher-order polynomial equations have more solutions, but this fact does not convey physical information about CFTs.

\begin{figure}
    \centering
\begin{align*}
\hspace{-6mm}
\begin{matrix}\text{
{
\setlength{\fboxsep}{0pt}
\setlength{\fboxrule}{1pt}
		\fbox{\includegraphics[scale=0.65]{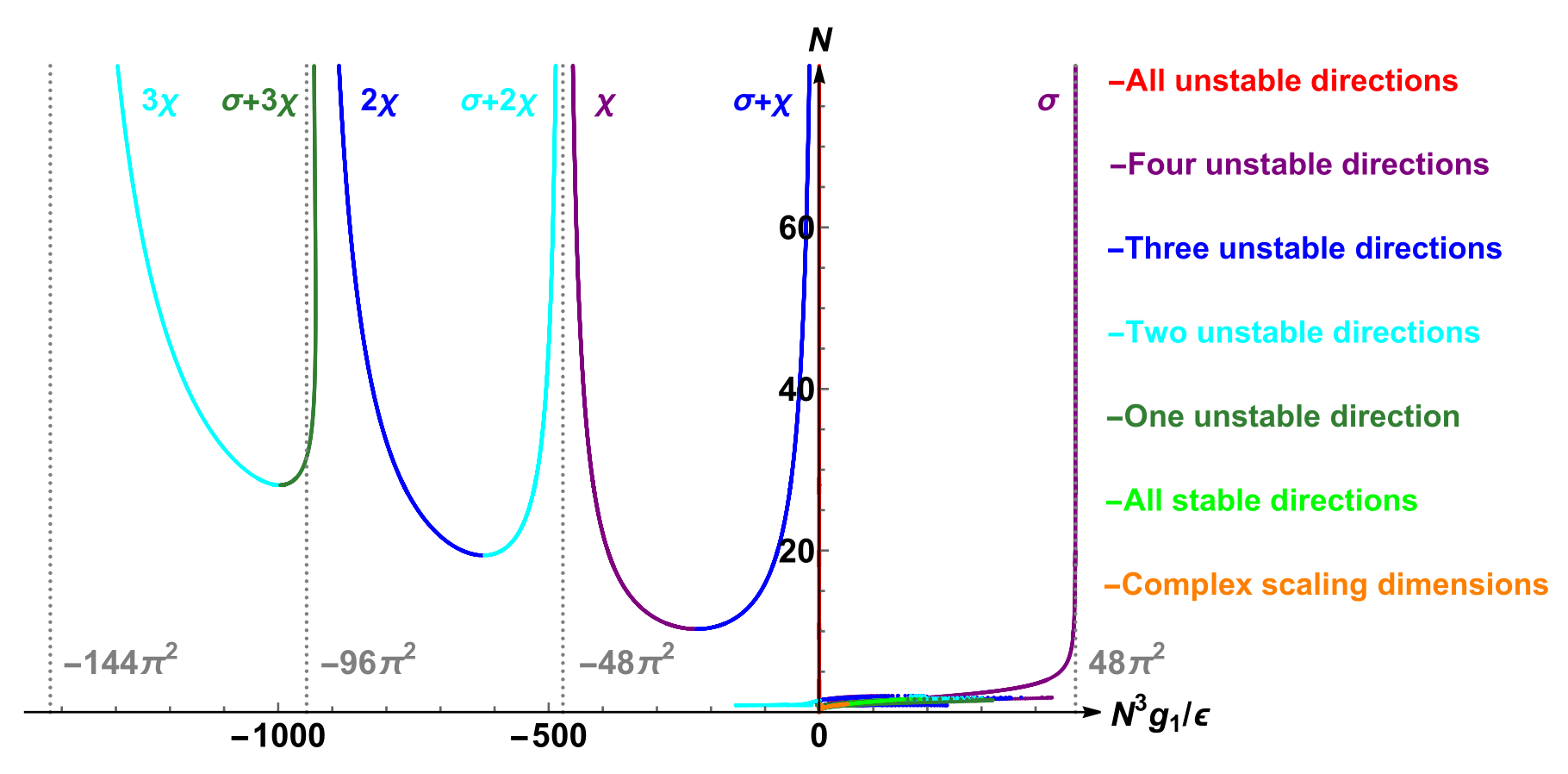}}
}
}\end{matrix}
\end{align*}
    \caption{Perturbative fixed points for quartic $O(N)^3/(\mathbb{Z}_2)^2$ theory in $4-\epsilon$ dimensions. The fixed points are labelled according to their large $N$ field content in terms of Hubbard-Stratonovich fields $\sigma$ and $\chi$. For each point on the curves marked with $\textcolor{violet}{\chi}$, $\textcolor{blue}{2\chi}$, $\textcolor{blue}{\sigma+\chi}$, and $\textcolor{cyan2}{\sigma+2\chi}$ there are actually three separate RG fixed points, related by permutations of the values for the coupling constants $g_2$, $g_3$, and $g_4$. } 
    \label{fig:rank3Overview}
\end{figure}

Figure~\ref{fig:rank3Overview} displays a plot of the values of $g_1$ at the perturbative fixed points as functions $N$. Several fixed points are related by permutations of indices. Fixed points related in this manner assume the same value for $g_1$ as well as for anomalous dimensions of $O(N)^3$ singlets, and for this reason there are not 16 separate fixed points visible for $N\geq 29$ in Figure~\ref{fig:rank3Overview}, but only eight. These eight families of fixed points correspond to orbits under the action of an $S_3$ group that permutes the $O(N)$ groups of the model. We will label these orbits with roman numerals (i) to (viii), where
\\[5pt]
 (i) is the orbit containing only the trivial fixed point with all couplings equal to zero.
 \\[5pt]
(ii) is the orbit containing only the vector model fixed. In the space of couplings, this fixed point is situated at
\begin{align}
\label{vectorModel}
g_1^{\text{(ii)}}= \frac{48\pi^2}{8+N^3}\epsilon + 
\frac{144(14+3N^3)\pi^2}{(8+N^3)^3}\epsilon^2
+\mathcal{O}(\epsilon^3)\,,
\hspace{15mm}
g^{\text{(ii)}}_2=g^{\text{(ii)}}_3=g^{\text{(ii)}}_4=g^{\text{(ii)}}_5=0\,.
\end{align}
The field and mass anomalous dimensions at this fixed point are given by
 \begin{align}
 \label{gamphi(ii)}
 \gamma_\phi^{\text{(ii)}} =\,&
\frac{2+N^3}{4(8+N^3)^2}\epsilon^2
-\frac{(2+N^3)(N^6-56N^3-272)}{16(8+N^3)^4}\epsilon^3
+\mathcal{O}(\epsilon^4)\,,
\\[8pt]
 \label{gamphisq(ii)}
\gamma^{\text{(ii)}}_{\phi^2}=\,& 
\frac{2+N^3}{8+N^3}\epsilon
+\frac{(2+N^3)(44+13N^3)}{2(8+N^3)^3}\epsilon^2
+\mathcal{O}(\epsilon^3)\,.
 \end{align}
At the vector model fixed point, the $O(N)^3$ symmetry of the theory is enhanced to $O(N^3)$ symmetry. The enhanced symmetry is present everywhere along the RG flow whenever $g_1$ is the only non-zero coupling. The operator $\mathcal{O}_1$ alone can never generate $\mathcal{O}_2$ to $\mathcal{O}_5$.
\\[5pt]
(iii) and (iv) are orbits containing three fixed points each. They are characterized by $g_1$ and either $g_2$, $g_3$, or $g_4$ being non-zero, while the other three couplings vanishing, which implies that the fixed points in these orbits are members of the family of bifundamental fixed points. Without loss of generality we can take $g_1$ and $g_2$ to be the non-vanishing couplings, which assume the values
\begin{align}
&g_1^{\text{(iii)}}= \text{\scalebox{0.83}{$\frac{ 80 - 2 N - 3 N^2 - 2 N^3 - 
    N^4 - (4 + N + N^2) \sqrt{
     52 - 4 N - 3 N^2 - 10 N^3 + N^4} }{
 464 - 56 N - 72 N^2 - 16 N^3 - 8 N^4 + 9 N^5 + 2 N^6 + N^7}$}}
 24\pi^2\epsilon+\mathcal{O}(\epsilon^2)\,,
 \nonumber\\
&g_1^{\text{(iv)}}= \text{\scalebox{0.83}{$\frac{ 80 - 2 N - 3 N^2 - 2 N^3 - 
    N^4 + (4 + N + N^2) \sqrt{
     52 - 4 N - 3 N^2 - 10 N^3 + N^4} }{
 464 - 56 N - 72 N^2 - 16 N^3 - 8 N^4 + 9 N^5 + 2 N^6 + N^7}$}}
 24\pi^2\epsilon+\mathcal{O}(\epsilon^2)\,,
 \nonumber\\[-6pt]
 \\[-16pt]
& g_2^{\text{(iii)}}= \text{\scalebox{0.83}{$\frac{-4 - 10 N - 10 N^2 + 4 N^3 + N^4 + N^5 + 
   6 \sqrt{52 - 4 N - 3 N^2 - 10 N^3 + N^4}}{464 - 56 N - 
 72 N^2 - 16 N^3 - 8 N^4 + 9 N^5 + 2 N^6 + N^7}$}}
 48\pi^2\epsilon
 +\mathcal{O}(\epsilon^2)\,,
  \nonumber\\
& g_2^{\text{(iv)}}= \text{\scalebox{0.83}{$\frac{-4 - 10 N - 10 N^2 + 4 N^3 + N^4 + N^5 - 
   6 \sqrt{52 - 4 N - 3 N^2 - 10 N^3 + N^4}}{464 - 56 N - 
 72 N^2 - 16 N^3 - 8 N^4 + 9 N^5 + 2 N^6 + N^7}$}}
 48\pi^2\epsilon
 +\mathcal{O}(\epsilon^2)\,.
  \nonumber
\end{align}
We refrain from writing down the order $\epsilon^2$ terms as they are quite lengthy. Similarly, lengthy closed-form expressions are available for $\gamma_\phi$ to order $\epsilon^3$ and for $\gamma_{\phi^2}$ to order $\epsilon^2$. From the radicands in the above expressions, one sees that the (iii) and (iv) fixed points collide and become complex for $N < N_\ast$ where
\begin{align}
N_\ast = \frac{1}{2}
\Big(
5+4\sqrt{3}+\sqrt{33+24\sqrt{3}}
\Big)\approx 10.2818\,,
\end{align}
up to order $O(\epsilon)$ corrections. We observe that the (iii) and (iv) fixed points exhibit an $O(N^2)\times O(N)/\mathbb{Z}_2$ symmetry that rotates the $a$ and $b$ indices of $\phi^{abc}$ together but the $c$ index separately. This symmetry extends beyond the fixed points and is preserved everywhere within a two-dimensional invariant subspace of the full five-dimensional RG flow. When only $\mathcal{O}_1$ and $\mathcal{O}_2$ are present, $\mathcal{O}_3$, $\mathcal{O}_4$, and $\mathcal{O}_5$ will never be generated.
\\[5pt]
The orbits (v) and (vi) also contain three fixed points each. For each of these fixed points all couplings are non-zero and two of the couplings $g_2$, $g_3$, and $g_4$ assume identical values. Each of the fixed points in these two orbits is situated on a four-dimensional invariant manifold characterized by an $S_2$ permutation symmetry acting on two of the indices of the fundamental fields, meaning that the manifolds and fixed points enjoy an $O(N)^3\rtimes S_2$ symmetry. This symmetry reflects the fact that if, e.g., $g_2$ equals $g_3$ at some scale along an RG flow, then $g_2$ equals $g_3$ everywhere along the flowline. The leading values of the couplings at small $\epsilon$ are given by roots of four coupled quadratic equations, which do not generally admit closed form solutions as functions of $N$, although for a given $N$ the leading and sub-leading values for the couplings and anomalous dimensions can easily be evaluated numerically to high precision. The fixed points in these two orbits merge and become complex for $N< N_{\ast\ast}$ with $N_{\ast\ast}\approx 19.3955$.
\\[5pt]
Orbits (vii) and (viii) contain a single fixed point each, for which $g_2$, $g_3$, and $g_4$ assume the same non-zero value, while $g_1$ and $g_5$ assume two other non-zero values. These fixed points are situated on the three-dimensional invariant manifold of coupling configurations where the theory exhibits an $S_3$ permutation symmetry under interchanges of the three tensor indices: if the theory starts out being symmetric with respect to the three indices, then renormalization will not spoil this property. The leading values of the couplings as functions of $N$ are given in terms of roots of sixth order polynomials. The two fixed points in these orbits merge and become complex for $N< N_{\ast\ast\ast}$ with $N_{\ast\ast\ast}\approx 28.0935$.

\subsubsection{Large $N$ expansion}
The operators \eqref{rank3ops} have have an optimal large $N$ scaling, explicitly worked out by Carrozza and Tanasa \cite{carrozza2016n} and given by
\begin{align}
g_1= \frac{\lambda_1}{N^3}\,, \hspace{12mm}
g_2= \frac{\lambda_2}{N^2}\,, \hspace{12mm}
g_3= \frac{\lambda_3}{N^2}\,, \hspace{12mm}
g_4= \frac{\lambda_4}{N^2}\,, \hspace{12mm}
g_5= \frac{\lambda_5}{N^{3/2}}\,.
\end{align}
The fact that the last operator has the smallest scaling exponent can be interpreted to mean that this operator dominates at large $N$ unless its coupling constant is suppressed. Ref.~\cite{giombi2017bosonic} demonstrated how the large $N$ theory of this operator is melonic and solvable, but also pointed out that this melonic theory requires the four other operators to have complex couplings.\footnote{
Specifically,  $\beta_{\lambda_5}$ has a fixed point at
$\lambda_5 = \pm 48\pi^2\sqrt{\epsilon}$. But at this value of $\lambda_5$, the fixed point values for $\beta_{\lambda_2}$, $\beta_{\lambda_3}$, and $\beta_{\lambda_4}$ are given by $\lambda_2$, $\lambda_3$, and $\lambda_4$ equal to $48\pi^2(\epsilon \pm \sqrt{\epsilon^2-2\epsilon})$, which are complex for small $\epsilon$. Incidentally, the fixed point values become real right in two dimensions:
\begin{flalign}
\label{special2Dfp}
&
\epsilon=2:
\hspace{15mm}
\lambda_1 = \pm 96\sqrt{6}\pi^2\,,
\hspace{15mm}
\lambda_2 = \lambda_3 = \lambda_4 = 96\pi^2\,,
\hspace{15mm}
\lambda_5 = \pm 96\pi^2\,.
&&
\end{flalign}
The signs of $\lambda_1$ and $\lambda_5$ can be chosen independently, so that there are a total of four real fixed points. But because of the finite value of $\epsilon$, higher-order corrections will not be parametrically small, and so we cannot trust these fixed points.
} 
To third order in the couplings, the beta functions in the $N\rightarrow\infty$ limit are given by
\begin{align}
\nonumber &\beta_{\lambda_1}=
-\epsilon \lambda_1+
\frac{1}{48\pi^2}\Big(\lambda_1^2 + 2 \lambda_1 \lambda_2 + 2 \lambda_1 \lambda_3 + 
 2 \lambda_2 \lambda_3 + 2 \lambda_1 \lambda_4 + 
 2 \lambda_2 \lambda_4 + 2 \lambda_3 \lambda_4\Big)
\\\nonumber &\hspace{10mm} -\frac{1}{(48\pi^2)^2}\Big(
\frac{5}{2} \lambda_1 \lambda_5^2 + 2 \lambda_2 \lambda_5^2 + 
 2 \lambda_3 \lambda_5^2 + 2 \lambda_4 \lambda_5^2
 \Big)+\mathcal{O}(\lambda^4)\,,
 \\[5pt]\nonumber
&\beta_{\lambda_2}=-\epsilon \lambda_2
+\frac{\lambda_2^2 + \lambda_5^2}{48\pi^2}
-\frac{\lambda_2 \lambda_5^2}{2(48\pi^2)^2}
+\mathcal{O}(\lambda^4)\,,
 \\[5pt]
&\beta_{\lambda_3}=-\epsilon \lambda_3
+\frac{\lambda_3^2 + \lambda_5^2}{48\pi^2}
-\frac{\lambda_3 \lambda_5^2}{2(48\pi^2)^2}
+\mathcal{O}(\lambda^4)\,,
 \\[5pt]\nonumber
&\beta_{\lambda_4}=-\epsilon \lambda_2
+\frac{\lambda_4^2 + \lambda_5^2}{48\pi^2}
-\frac{\lambda_4 \lambda_5^2}{2(48\pi^2)^2}
+\mathcal{O}(\lambda^4)\,,
 \\[5pt]\nonumber
&\beta_{\lambda_5}=-\epsilon \lambda_5
+\frac{\lambda_5^3}{2(48\pi^2)^2}+\mathcal{O}(\lambda^4)\,.
\end{align}
These beta functions have 16 real fixed points towards which the 16 finite $N$ fixed points that exist for $N\geq 29$ converge as $N$ goes to infinity. For all the large $N$ fixed points, the melonic operator is absent or suppressed below its combinatorial scaling. The explicit locations of these fixed points in the space of $\lambda$-couplings are listed below, where it is to be understood that unspecified couplings are zero:
\begin{enumerate}[label=(\roman*)]
 \item one fixed point: all $\lambda_i$ equal to zero,
 \item one fixed point: $\lambda_1=48\pi^2\epsilon$,
 \item three fixed points: $\lambda_1=-48\pi^2\epsilon$, one of $\lambda_2$, $\lambda_3$, and $\lambda_4$ equal to $48\pi^2\epsilon$,
  \item three fixed points: one of $\lambda_2$, $\lambda_3$, and $\lambda_4$ equal to $48\pi^2\epsilon$,
   \item three fixed points: $\lambda_1=-96\pi^2\epsilon$, two of $\lambda_2$, $\lambda_3$, and $\lambda_4$ equal to $48\pi^2\epsilon$,
  \item three fixed points: $\lambda_1=-48\pi^2\epsilon$, two of $\lambda_2$, $\lambda_3$, and $\lambda_4$ equal to $48\pi^2\epsilon$,
   \item one fixed point: $\lambda_1=-144\pi^2\epsilon$,  $\lambda_2=\lambda_3=\lambda_4=48\pi^2\epsilon$,
  \item one fixed point: $\lambda_1=-96\pi^2\epsilon$,  $\lambda_2=\lambda_3=\lambda_4=48\pi^2\epsilon$.
\end{enumerate}
The leading anomalous dimensions for $\phi$ and $\phi^2$ at the fixed points are listed in Table~\ref{tab:anomDimsQuarticRank3}.
\begin{table}
\centering
\scalebox{1}{
\renewcommand{\arraystretch}{1.5}
\begin{tabular}{|c|c|c|c|c|} 
\hline
& \# of copies
&$\gamma_\phi-\mathcal{O}(\epsilon^4)$
&$\gamma_{\phi^2}-\mathcal{O}(\epsilon^3)$ 
\\
\hline
\cellcolor{lightgray2}(i) &\cellcolor{lightgray2} 1 
&\cellcolor{lightgray2} 0
&\cellcolor{lightgray2} 0
\\
\hline
(ii) & 1 
& $\frac{1}{N^3}(\frac{1}{4}\epsilon^2-\frac{1}{16}\epsilon^3)+\mathcal{O}(\frac{1}{N^6})$
& $\epsilon+\frac{1}{N^3}(-6\epsilon+\frac{13}{2}\epsilon^2)+\mathcal{O}(\frac{1}{N^6})$
\\
 \hline
(iii)\cellcolor{lightgray2} &\cellcolor{lightgray2} 3 
&\cellcolor{lightgray2} $\frac{1}{N}(\frac{1}{8}\epsilon^2-\frac{1}{32}\epsilon^3)+\mathcal{O}(\frac{1}{N^2})$
&\cellcolor{lightgray2} $\frac{1}{N}(3\epsilon-\frac{13}{4}\epsilon^2)+\mathcal{O}(\frac{1}{N^2})$
\\
 \hline
(iv)  & 3 
& $\frac{1}{N}(\frac{1}{8}\epsilon^2-\frac{1}{32}\epsilon^3)+\mathcal{O}(\frac{1}{N^2})$
& $\epsilon+\frac{1}{N}(-3\epsilon+\frac{13}{4}\epsilon^2)+\mathcal{O}(\frac{1}{N^2})$
\\
 \hline
(v)\cellcolor{lightgray2}  &\cellcolor{lightgray2} 3 
&\cellcolor{lightgray2} $\frac{1}{N}(\frac{1}{4}\epsilon^2-\frac{1}{16}\epsilon^3)+\mathcal{O}(\frac{1}{N^2})$
&\cellcolor{lightgray2} $\frac{1}{N}(6\epsilon-\frac{13}{2}\epsilon^2)+\mathcal{O}(\frac{1}{N^2})$
\\
 \hline
(vi)  & 3 
& $\frac{1}{N}(\frac{1}{4}\epsilon^2-\frac{1}{16}\epsilon^3)+\mathcal{O}(\frac{1}{N^2})$
& $\epsilon+\frac{1}{N}(-6\epsilon+\frac{13}{2}\epsilon^2)+\mathcal{O}(\frac{1}{N^2})$
\\
 \hline
(vii)\cellcolor{lightgray2} &\cellcolor{lightgray2} 1 
&\cellcolor{lightgray2} $\frac{1}{N}(\frac{3}{8}\epsilon^2-\frac{3}{32}\epsilon^3)+\mathcal{O}(\frac{1}{N^2})$
&\cellcolor{lightgray2} $\frac{1}{N}(9\epsilon-\frac{39}{4}\epsilon^2)+\mathcal{O}(\frac{1}{N^2})$
\\
 \hline
(viii)  & 1
& $\frac{1}{N}(\frac{3}{8}\epsilon^2-\frac{3}{32}\epsilon^3)+\mathcal{O}(\frac{1}{N^2})$
& $\epsilon+\frac{1}{N}(-9\epsilon+\frac{39}{4}\epsilon^2)+\mathcal{O}(\frac{1}{N^2})$
\\
 \hline
\end{tabular}}
\caption{Leading anomalous dimensions at the large $N$ fixed points of quartic $O(N)^3$ theory}
    \label{tab:anomDimsQuarticRank3}
\end{table}
There is a pattern to these anomalous dimensions, which can be understood by working out the large $N$ limit of the model outside of perturbation theory in $\epsilon$. To describe the pattern, we first review how to solve for the vector model fixed point at large $N$, before turning to the other fixed points.

\subsubsection{Vector model fixed point}
\label{subSubSec:vector}
\noindent The vector model fixed point (ii) is the large $N$ limit of the fixed point in \eqref{vectorModel}. The anomalous dimensions for the large $N$ vector model were computed in Refs.~\cite{vasil1981simple,vasil19811}, and the large $N$ solution was also nicely reviewed more recently in Ref.~\cite{fei2014critical}. For small $\epsilon$, the large $N$ anomalous dimensions can be found by simply expanding \eqref{gamphi(ii)} and \eqref{gamphisq(ii)}:
\begin{align}
 \gamma_\phi^{\text{(ii)}} =\,& \frac{\epsilon^2}{N^3}
\Big(\frac{1}{4}-\frac{7}{2N^3}+\mathcal{O}(\frac{1}{N^6})\Big)+\frac{\epsilon^3}{N^3}\Big(-\frac{1}{16}+\frac{43}{8N^3}+\mathcal{O}(\frac{1}{N^6})\Big)+\mathcal{O}(\epsilon^4)\,,
\\[6pt]
\gamma^{\text{(ii)}}_{\phi^2}=\,& 
\epsilon\Big(1-\frac{6}{N^3}+\frac{48}{N^6}+\mathcal{O}(\frac{1}{N^6})\Big)
+\frac{\epsilon^2}{N^3}
\Big(\frac{13}{2}-\frac{121}{N^3}+\mathcal{O}(\frac{1}{N^6})\Big)+\mathcal{O}(\epsilon^3)\,.
 \end{align}
But by leveraging the power of large $N$, these anomalous dimensions can be determined for any $d$ between 2 and 4. A standard way to obtain the large $N$ solution of an $O(N)$ or $O(N^r)$ model, dating back to the work of Coleman, Jackiw, and Politzer \cite{coleman1974spontaneous}, is through the use of the Hubbard-Stratonovich transformation. Without changing the path integral of the theory, we can introduce an integral over an auxiliary $\sigma$ field and perform the following substitution in the action:
\begin{align}
\frac{g_1}{4!}(\phi^{abc}\phi^{abc})^2
\rightarrow\, \,&
\frac{g_1}{4!}(\phi^{abc}\phi^{abc})^2
-\frac{3}{2g_1N^3}\Big(\sigma - \frac{g_1N^{3/2}}{6}\phi^{abc}\phi^{abc}\Big)^2
\nonumber\\[-11pt]
\\[-11pt]\nonumber
&=
\frac{1}{2N^{3/2}}\sigma\phi^{abc}\phi^{abc}
-\frac{3}{2g_1N^3}\sigma^2\,.
\end{align}
On flowing to a CFT the $\sigma^2$ term goes away, leaving us with an effective action
\begin{align}
\label{S(ii)}
S^{\text{(ii)}}=\int d^d x
\bigg(\frac{1}{2}(\partial_i\phi^{abc})(\partial_i\phi^{abc})
+\frac{1}{2N^{3/2}}\sigma\phi^{abc}\phi^{abc}
\bigg)\,.
\end{align}
The dimension $\Delta_{\phi^2}$ can be determined from the expectation value of $\sigma$:
\begin{align}
\left<\sigma(k)\sigma(p)\right>=\delta^{(d)}(k+p) \frac{C_\sigma}{|p|^{d-2\Delta_{\phi^2}}}\,,
\end{align}
where $C_\sigma$ is $d$-dependent constant. The leading value in $1/N^3$ of this correlator, which gives us the effective non-local propagator for $\sigma$, can be determined by integrating out the fundamental fields $\phi^{abc}$:
\begin{align}
\text{
\begin{tikzpicture}
\draw [white] (-0.1,0)--(0.1,0);
\draw [very thick,dashed] (-1,0.1)--(1,0.1);
\end{tikzpicture}
}
\hspace{3mm}
\left<\sigma(k)\sigma(p)\right>_0=\delta^{(d)}(k+p)\frac{2^{d+1}(4\pi)^{\frac{d-3}{2}}\Gamma(\frac{d-1}{2})\sin(\frac{\pi d}{2})}{|p|^{d-4}}\,.
\end{align}
Comparing the above two equations we see that $\Delta_{\phi^2}=2+\mathcal{O}(1/N^3)$. Subtracting off the engineering dimension $[\phi^2]=d-2$ from this value, we obtain the anomalous dimension:
\begin{align}
\gamma_{\phi^2}^{\text{(ii)}}=\Delta_{\phi^2}-[\phi^2] 
= 4-d + \mathcal{O}(\frac{1}{N^3})\,.
\end{align}
The anomalous dimension for $\phi$ is determined to leading order in $1/N^3$ by the sunset diagram:
\begin{align}
\label{eq:sunset}
\hspace{-5mm}
\begin{matrix}
\text{
		\includegraphics[scale=1]{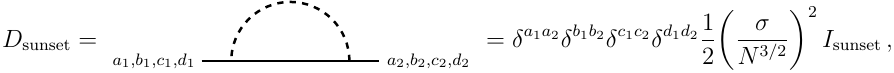}
}
\end{matrix}
\end{align}
where the solid line indicates the propagator for $\phi^{abcd}$, and $I_\text{sunset}$ is the loop integral associated to the diagram. By carrying out the integral, one finds that
\begin{align}
\gamma_\phi^\text{(ii)} =\frac{A_\phi(d)}{N^3}+\mathcal{O}(\frac{1}{N^6})\,,
\hspace{13mm}
\text{where}
\hspace{14mm}
A_\phi(d) \equiv \frac{2\Gamma(d-2)\sin(\frac{\pi d}{2})}{\pi\Gamma(\frac{d}{2}-2)\Gamma(\frac{d}{2}+1)}\,.
\label{eq:gammaPhi}
\end{align}
The $1/N^3$ term in $\gamma_{\phi^2}$ is determined from the following three diagrams:
\begin{align}
\label{diagrams}
\begin{matrix}
\text{
		\includegraphics[scale=1]{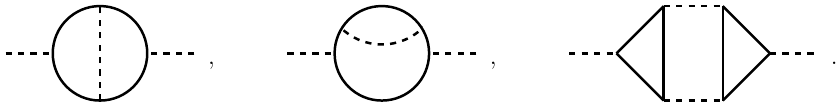}
}
\end{matrix}
\end{align}
Through evaluation of these diagrams, one finds that 
\begin{align}
\gamma_{\phi^2}^\text{(ii)}=4-d+\frac{B_{\phi^2}(d)}{N^3}+\mathcal{O}(\frac{1}{N^6})\,,
\hspace{13mm}
\text{where}
\hspace{14mm}
B_{\phi^2}(d) \equiv 
\frac{4\Gamma(d)\sin(\frac{\pi d}{2})}{\pi\Gamma(\frac{d}{2}-1)\Gamma(\frac{d}{2}+1)}\,.
\label{eq:gammaPhiSquared}
\end{align}
Expanding \eqref{eq:gammaPhi} and \eqref{eq:gammaPhiSquared} in $\epsilon = 4-d$, we recover the values in row (ii) of Table~\ref{tab:anomDimsQuarticRank3} To understand the other rows, we need to introduce additional field content to the large $N$ effective theory. 

\subsubsection{Other fixed points}

We now turn to the three large $N$ fixed points in orbit (iii). Without loss of generality, we can study the fixed point with $\lambda_1=-48\pi^2\epsilon$, $\lambda_2=48\pi^2\epsilon$, and $\lambda_3=\lambda_4=\lambda_5=0$. We note that despite the negative value of $\lambda_1$, the potential is in fact bounded below at this fixed point, as can be seen by applying the Cauchy-Schwarz inequality to $N^2$-component vectors $v$ and $w$ given by $v^{cc'} = \phi^{abc}\phi^{abc'}$ and $w^{cc'}=\delta^{cc'}$.

One could attempt to apply the Hubbard-Stranovich transformation to the operator $\mathcal{O}_2$, but it is not the best way to proceed in the present case. We should pick the large $N$ field content to transform under irreducible representations of the $O(N)$ group.\footnote{Insofar as the Hubbard-Stratonovich transformation is a mathematical re-writing of the path integral, one can perform the transform whichever way one pleases. The issue, we believe, in using an intermediate field, call it $\widetilde{\chi}^{c_1c_2}$, that does not transform as an irrep is that one can no longer simply drop the square term $\widetilde{\chi}^2$ at a CFT since this term will not just be a mass term, but will also contain cross terms of fields that are irreps.} This means we should apply the Hubbard-Stratonovich transformation not to $\mathcal{O}_2$ alone but rather to a specific linear combination of $\mathcal{O}_1$ and $\mathcal{O}_2$: 
\begin{align}
\nonumber
\frac{g_2}{4!}\big(\mathcal{O}_2-\frac{1}{N}\mathcal{O}_1\big)
=\,&\frac{3}{2g_2N^3}
\bigg(\chi_3^{c_1c_2}-\frac{N^{3/2}g_2}{6}\Big(
\phi^{abc_1}\phi^{abc_2}-\frac{1}{N}\delta^{c_1c_2}\phi^{abc}\phi^{abc}
\Big)\bigg)^2
\\[-11pt]
\label{introducingChi}
\\[-11pt]
&-\frac{3}{2g_2N^3}(\chi_3^{c_1c_2})^2
+\frac{1}{2N^{3/2}}\chi_3^{c_1c_2}
\Big(\phi^{abc_1}\phi^{abc_2}
-\frac{\delta^{c_1c_2}}{N}\phi^{abc}\phi^{abc}
\Big)\,.
\nonumber
\end{align}
This particular way of re-expressing $\mathcal{O}_1$ and $\mathcal{O}_2$ has been chosen such that the new auxiliary field $\chi_3^{c_1c_2}$ in the last line multiplies a combination of the fundamental fields that is traceless and symmetric in the $c$ indices.

Taking the $(\chi_3^{c_1c_2})^2$ term to go away at criticality, we are left with the following effective action:
\begin{align}
\label{S(iii)}
S^{\text{(iii)}}=\int d^d x
\bigg(\frac{1}{2}(\partial_i\phi^{abc})(\partial_i\phi^{abc})
+\frac{\chi_3^{c_1c_2}}{2N^{3/2}}
\Big(\phi^{abc_1}\phi^{abc_2}
-\frac{\delta^{c_1c_2}}{N}\phi^{abc}\phi^{abc}
\Big)
\bigg)\,.
\end{align}
The cubic interaction for $\chi_3^{c_1c_2}$ induces the following propagator:
\begin{align}
&\left<\chi_3^{c_1c_2}(k)\chi_3^{c_3c_4}(p)\right>_0
=\delta^{(d)}(k+p) G^{c_1c_2c_3c_4} \frac{C_\sigma}{|p|^{d-4}}\,,
\\[5pt]
&\text{where}\hspace{6mm} G^{c_1c_2c_3c_4} = \frac{\delta^{c_1c_3}\delta^{c_2c_4}+\delta^{c_1c_4}\delta^{c_2c_3}}{2}-\frac{\delta^{c_1c_2}\delta^{c_3c_4}}{N}\,,
\hspace{9mm}
\textcolor{white}{.}
\end{align}
so that $\chi_3^{c_1c_2}$ behaves as a traceless symmetric matrix field. In the presence of such a field, the anomalous dimension for the fundamental field recieves a contribution from the sunset diagram \eqref{eq:sunset} with the $\sigma$ propagator replaced with the $\chi_3^{c_1d_2}$ propagator:
\begin{align}
\label{sunsetSubstitution}
D_\text{sunset}
\rightarrow\,&
\delta^{a_1a_2}
\delta^{b_1b_2}
G^{c_1c_3c_2c_3}
\frac{1}{2}
\bigg(
\frac{\sigma}{N^{3/2}}
\bigg)^2
\,I_\text{sunset}
\\
\nonumber
&=
\delta^{a_1a_2}
\delta^{b_1b_2}
\frac{N^2+N-2}{2N}
\delta^{c_1c_2}
\frac{1}{2}
\bigg(
\frac{\sigma}{N^{3/2}}
\bigg)^2
\,I_\text{sunset}
= \frac{N^2}{2}D_\text{sunset}\Big(1+\mathcal{O}(\frac{1}{N})\Big)\,.
\end{align}
The Hubbard-Stranovich transformation we just applied to $\mathcal{O}_2$ can of course equally well be applied to $\mathcal{O}_3$ and $\mathcal{O}_4$, by respectively introducing traceless symmetric matrices $\chi_2^{b_1b_2}$ and $\chi_1^{a_1a_2}$.  

The above considerations present a clear way of understanding the 16 large $N$ fixed points of $O(N)^3$ theory and the values for the anomalous dimension $\gamma_\phi$ in Table~\ref{tab:anomDimsQuarticRank3}. The possible field content of each large $N$ fixed point is an $O(N)$ scalar $\sigma$ and up to three traceless symmetric matrices $\chi$. Whenever the $\sigma$ field is present, it contributes a term $\frac{A_\phi}{N^3}$ to the anomalous dimension $\gamma_\phi$. The substitution \eqref{sunsetSubstitution} shows that for each traceless symmetric matrix present in the effective large $N$ theory, the anomalous dimension receives a contribution of $\frac{A_\phi}{2N}$. As for the anomalous dimension for $\phi^2$, the $\sigma$ field makes a contribution of $4-d+\frac{B_{\phi^2}}{N^3}$, and by replacing the internal $\sigma$ propagators in the diagrams \eqref{diagrams} with traceless symmetric matrix propagators, it is a simple exercise to show that each traceless symmetric matrix makes a contribution $-\frac{B_{\phi^2}}{2N^2}$ to the anomalous dimension for $\sigma$. For the large $N$ fixed points without $\sigma$ the $\epsilon$ term in $\gamma_{\phi^2}$ should not be present. The relative sign flip in the $1/N$ corrections to $\gamma_{\phi^2}$ may be viewed as ensuing from the fact that, in the absence of $\sigma$, $\Delta_{\phi^2}$ is determined from the correlators of $\phi^{abc}\phi^{abc}$ rather than $\sigma$. Since $\sigma$ couples to $\phi^{abc}\phi^{abc}$, their anomalous dimensions are opposite.

The rows in Table~\ref{tab:anomDimsQuarticRank3} where $\gamma_{\phi^2}$ does not contain an $\mathcal{O}(N^0)$ term $\epsilon$ correspond to fixed points without $\sigma$ present, and these are precisely the fixed points for which the coupling constants satisfy the relation $\lambda_1 = \lambda_2 + \lambda_3 + \lambda_4$, so that in these cases, on eliminating $\mathcal{O}_2$, $\mathcal{O}_3$, $\mathcal{O}_4$ via Hubbard-Stratonovich transformations in $\chi$ fields as in \eqref{introducingChi}, $\mathcal{O}_1$ precisely gets eliminated as well.

The number of fixed points in each of the orbits (i) to (viii) is simply a binomial coefficient associated to choices among the three traceless symmetric matrices. Concretely, the fixed points have the following large $N$ field content, which all interact with the fundamental fields $\phi^{abcd}$ via cubic interactions, along with the following values for $\gamma_\phi$ and $\gamma_{\phi^2}$: 
\begin{enumerate}[label=(\roman*)]
    \item One fixed point: no $\sigma$ or $\chi$ field present, $O(N^3)$ symmetry,
    \begin{flalign}
    \gamma_\phi^\text{(i)} =0\,, 
    \hspace{55mm}
    \gamma_{\phi^2}^\text{(i)}=0\,.
    &&
    \end{flalign}
    \item One fixed point: only $\sigma$ present, $O(N^3)$ symmetry,
    \begin{flalign}
    \gamma_\phi^\text{(ii)} =\frac{A_\phi(d)}{N^3}+\mathcal{O}(\frac{1}{N^6})\,, 
    \hspace{27mm}
    \gamma_{\phi^2}^\text{(ii)}=4-d+\frac{B_{\phi^2}(d)}{N^3}+\mathcal{O}(\frac{1}{N^6})\,.
    &&
    \end{flalign}
     \item  Three fixed points: one $\chi$ present, $\sigma$ absent, $\displaystyle \frac{O(N^2)\times O(N)}{\mathbb{Z}_2}$ symmetry,
    \begin{flalign}
    \gamma_\phi^\text{(iii)} =\frac{A_\phi(d)}{2N}+\mathcal{O}(\frac{1}{N^2})\,, 
    \hspace{25mm}
    \gamma_{\phi^2}^\text{(iii)}=-\frac{B_{\phi^2}(d)}{2N}+\mathcal{O}(\frac{1}{N^2})\,.
    &&
    \end{flalign}
     \item  Three fixed points: one $\chi$ present, $\sigma$ present, $\displaystyle\frac{O(N^2)\times O(N)}{\mathbb{Z}_2}$ symmetry,
     \begin{flalign}
    \gamma_\phi^\text{(iv)} =\frac{A_\phi(d)}{2N}+\mathcal{O}(\frac{1}{N^2})\,, 
    \hspace{25mm}
    \gamma_{\phi^2}^\text{(iv)}=4-d+\frac{B_{\phi^2}(d)}{2N}+\mathcal{O}(\frac{1}{N^2})\,.
    &&
    \end{flalign}
     \item  Three fixed points: two $\chi$ present, $\sigma$ absent, $\displaystyle \frac{O(N)^3\rtimes S_2}{(\mathbb{Z}_2)^2}$ symmetry,
    \begin{flalign}
    \gamma_\phi^\text{(v)} =\frac{A_\phi(d)}{N}+\mathcal{O}(\frac{1}{N^2})\,, 
    \hspace{26mm}
    \gamma_{\phi^2}^\text{(v)}=-\frac{B_{\phi^2}(d)}{N}+\mathcal{O}(\frac{1}{N^2})\,.
    &&
    \end{flalign}
     \item Three fixed points: two $\chi$ present, $\sigma$ present, $\displaystyle \frac{O(N)^3 \rtimes S_2}{(\mathbb{Z}_2)^2}$ symmetry,
         \begin{flalign}
    \gamma_\phi^\text{(vi)} =\frac{A_\phi(d)}{N}+\mathcal{O}(\frac{1}{N^2})\,, 
    \hspace{24mm}
    \gamma_{\phi^2}^\text{(vi)}=4-d+\frac{B_{\phi^2}(d)}{N}+\mathcal{O}(\frac{1}{N^2})\,.
    &&
    \end{flalign}
     \item  One fixed point: all three $\chi$ present, $\sigma$ absent, $\displaystyle \frac{O(N)^3 \rtimes S_3}{(\mathbb{Z}_2)^2}$ symmetry,
    \begin{flalign}
    \gamma_\phi^\text{(vii)} =\frac{3A_\phi(d)}{2N}+\mathcal{O}(\frac{1}{N^2})\,, 
    \hspace{21mm}
    \gamma_{\phi^2}^\text{(vii)}=-\frac{3B_{\phi^2}(d)}{2N}+\mathcal{O}(\frac{1}{N^2})\,.
    &&
    \end{flalign}
     \item  One fixed point: all three $\chi$ and $\sigma$ present, $\displaystyle \frac{O(N)^3 \rtimes S_3}{(\mathbb{Z}_2)^2}$ symmetry,
    \begin{flalign} 
    \gamma_\phi^\text{(viii)} =\frac{3A_\phi(d)}{2N}+\mathcal{O}(\frac{1}{N^2})\,, 
    \hspace{20mm}
    \gamma_{\phi^2}^\text{(viii)}=4-d+\frac{3B_{\phi^2}(d)}{2N}+\mathcal{O}(\frac{1}{N^2})\,.
    &&
    \end{flalign}
\end{enumerate}
For the vector model fixed point, the effective action \eqref{S(ii)} can be used to compute not just the leading $1/N$ corrections we have discussed here, but also higher-order corrections to arbitrary order. We expect the same to apply for the action \eqref{S(iii)}, as well as for the effective actions for the fixed points in orbit (iv), e.g.
\begin{align}
\label{S(iv)}
S^{\text{(iv)}}=\int d^d x
\bigg(\frac{1}{2}(\partial_i\phi^{abc})(\partial_i\phi^{abc})
+\frac{1}{2N^{3/2}}\sigma\phi^{abc}\phi^{abc}
+\frac{\chi_3^{c_1c_2}}{2N^{3/2}}
\Big(\phi^{abc_1}\phi^{abc_2}
-\frac{\delta^{c_1c_2}}{N}\phi^{abc}\phi^{abc}
\Big)
\bigg)\,.
\end{align}
But while the fixed points in orbits (iii) and (iv) exhibit an $O(N^2)$ symmetry that enforces $g_5$ to be zero even at finite $N$, the fixed points in orbits (v) to (viii) enjoy no such symmetry, and for these fixed points $\lambda_5$ is non-vanishing at subleading order in $1/N$. To capture higher-order $1/N$ corrections, it will therefore be necessary to take the melonic operator $\mathcal{O}_5$ into account.

The separate possibilities for the $\sigma$ and $\chi$ fields to be present or absent at the fixed points can be made manifest in the large $N$ beta functions by switching operator basis. First set $\lambda_5=0$ so as to only study the RG flow among the couplings $\lambda_1$ to $\lambda_4$. For this starting configuration, $\lambda_5$ will remain zero to leading order in $1/N$ throughout the RG flow. Next, change operator basis from $\mathcal{O}_n$, $n\in \{1,2,3,4\}$, to a basis given by 
\begin{align}
\mathcal{O}_1\,, \hspace{10mm} \mathcal{O}_n-\mathcal{O}_1 \hspace{3mm} \text{for }n\in\{2,3,4\}\,,
\end{align}
which corresponds to changing coupling constants from $\lambda_n$ to $h_n$ given by
\begin{align}
h_1=\lambda_1+\lambda_2+\lambda_3+\lambda_4\,, \hspace{15mm} h_n = \lambda_n\hspace{4mm}\text{for }n\in\{2,3,4\}\,.
\end{align}
This new basis diagonalizes the large $N$ beta functions:
\begin{align}
\label{betah}
\beta_{h_n} =-\epsilon\,h_n+\frac{h_n^2}{48\pi^2} \hspace{5mm}\text{for }n\in\{1,2,3,4\}\,.
\end{align}
Since the operators $\mathcal{O}_1$ to $\mathcal{O}_4$ are all vectorial, meaning that the Feynman diagrams constructed out of them are dominated by bubble diagrams at large $N$, the beta functions for the $h$ couplings truncate at quadratic order, so that equation \eqref{betah} is valid to all orders in perturbation theory. The 16 fixed points correspond to the $2^4$ possible options of independently choosing $h_n=0$ or $h_n=48\pi^2\epsilon$ for each $n\in\{1,2,3,4\}$. 

The above diagonalization clarifies not only the number of fixed points but also their respective levels of stability under $O(N)^r/(\mathbb{Z}_2)^{r-1}$ symmetric deformations since $h_n=0$ is an unstable fixed point to $\beta_{h_n}$, while $h_n=48\pi^2$ is stable. This means that for each of the Hubbard-Stratonovich fields $\sigma$, $\chi_1$, $\chi_2$, $\chi_3$ present at a fixed point there is a stable direction in the RG flow. The most stable fixed point is therefore fixed point (viii), which is stable along all four $h$-directions. But even this fixed point has one unstable direction as all the fixed points are unstable under deformations wrt. $\mathcal{O}_5$. Therefore, generic configurations of couplings do not flow to a perturbative fixed point, even after tuning away the quadratic operator $\phi^{abc}\phi^{abc}$. At minimum one must also tune $g_5$. Note however that fixed point (ii) and the three fixed points (iv) are all stable under quartic deformations that respect the symmetries present at those fixed points.

\subsection{The \texorpdfstring{$O(N)^4$}{O(N)-to-the-fourth} model}
\label{ON4}
\begin{figure}
    \centering
\begin{align*}
\begin{matrix}\text{
\includegraphics[scale=1]{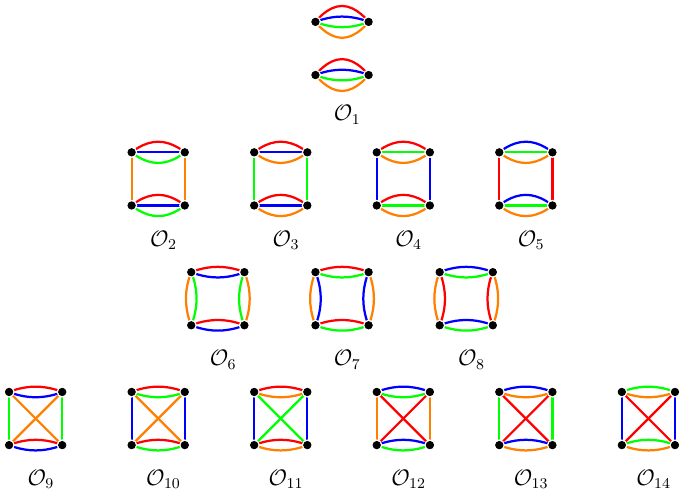}}
\end{matrix}
\end{align*}
    \caption{Quartic singlet operators in $O(N)^4$ theory}
    \label{fig:quarticRank4ops}
\end{figure}

We now turn to the quartic scalar theory of rank-four tensors. We take the action to be
\begin{align}
S = \int d^{4-\epsilon}x
\bigg(
\frac{1}{2}
(\partial_i\phi^{abcd})(\partial_i\phi^{abcd})
+\frac{1}{4!}\sum_{n=1}^{14}
g_n\,\mathcal{O}_{n}
\bigg)\,,
\label{Squartic}
\end{align}
where the 14 operators in the potential are given by
\begin{align}
& \mathcal{O}_1 = \Big(\phi^{abcd}\phi^{abcd}\Big)^2\,,
\\[10pt]
\nonumber& \mathcal{O}_2 = \phi^{a_1b_1c_1d_1}\phi^{a_1b_1c_1d_2}\phi^{a_2b_2c_2d_1}\phi^{a_2b_2c_2d_2}\,,
\\
\nonumber& \mathcal{O}_3 = \phi^{a_1b_1c_1d_1}\phi^{a_1b_1c_2d_1}\phi^{a_2b_2c_1d_2}\phi^{a_2b_2c_2d_2}\,,
\\[-11pt]
\\[-11pt]
\nonumber& \mathcal{O}_4 = \phi^{a_1b_1c_1d_1}\phi^{a_1b_2c_1d_1}\phi^{a_2b_1c_2d_2}\phi^{a_2b_2c_2d_2}\,,
\\
\nonumber& \mathcal{O}_5 = \phi^{a_1b_1c_1d_1}\phi^{a_2b_1c_1d_1}\phi^{a_1b_2c_2d_2}\phi^{a_2b_2c_2d_2}\,,
\\[10pt]
\nonumber & \mathcal{O}_6 = \phi^{a_1b_1c_1d_1}\phi^{a_1b_1c_2d_2}\phi^{a_2b_2c_1d_1}\phi^{a_2b_2c_2d_2}\,,
\\
& \mathcal{O}_7 = \phi^{a_1b_1c_1d_1}\phi^{a_1b_2c_1d_2}\phi^{a_2b_1c_2d_1}\phi^{a_2b_2c_2d_2}\,,
\\
\nonumber & \mathcal{O}_8 = \phi^{a_1b_1c_1d_1}\phi^{a_2b_1c_1d_2}\phi^{a_1b_2c_2d_1}\phi^{a_2b_2c_2d_2}\,,
\\[10pt]
\nonumber & \mathcal{O}_9 = \phi^{a_1b_1c_1d_1}\phi^{a_1b_1c_2d_2}\phi^{a_2b_2c_1d_2}\phi^{a_2b_2c_2d_1}\,,
\\
\nonumber & \mathcal{O}_{10} = \phi^{a_1b_1c_1d_1}\phi^{a_1b_2c_1d_2}\phi^{a_2b_1c_2d_2}\phi^{a_2b_2c_2d_1}\,,
\\
\nonumber & \mathcal{O}_{11} = \phi^{a_1b_1c_1d_1}\phi^{a_1b_2c_2d_1}\phi^{a_2b_1c_2d_2}\phi^{a_2b_2c_1d_2}\,,
\\[-11pt]
\\[-11pt]
\nonumber & \mathcal{O}_{12} = \phi^{a_1b_1c_1d_1}\phi^{a_2b_1c_1d_2}\phi^{a_1b_2c_2d_2}\phi^{a_2b_2c_2d_1}\,,
\\
\nonumber & \mathcal{O}_{13} = \phi^{a_1b_1c_1d_1}\phi^{a_2b_1c_2d_1}\phi^{a_1b_2c_2d_2}\phi^{a_2b_2c_1d_2}\,,
\\
\nonumber & \mathcal{O}_{14} = \phi^{a_1b_1c_1d_1}\phi^{a_2b_2c_1d_1}\phi^{a_1b_2c_2d_2}\phi^{a_2b_1c_2d_2}\,.
\end{align}
Diagrammatically, we can depict these operators as shown in Figure~\ref{fig:quarticRank4ops}.
By standard Feynman diagrammatic methods, one can compute the perturbative beta functions and anomalous dimensions of this model. The expressions one arrives at are rather lengthy owing to the large number of operators that mix in this theory, but we obtain a significant simplification by taking the large $N$ limit of the theory in the next subsection. Another means to simplify the model is to focus on the sector of the theory that is symmetric with respect to the four tensor indices, which we will do in Subsection~\ref{subSec:reduced}.

\subsubsection{Large $N$ limit}
\label{subSec:largeN}

The first step to studying the large $N$ limit of the model is to set down the appropriate scalings of the couplings constants. For this model there exists a unique optimal scaling
\begin{align}
g_i= \frac{\lambda_i}{N^{e_i}}\,,
\end{align}
where the exponents are given by
\begin{align}
e=\{4,\hspace{2mm} 3,3,3,3,\hspace{2mm} 2,2,2,\hspace{2mm} 2,2,2,2,2,2\}\,.
\end{align}
In the supplementary Mathematica notebook we list the large $N$ beta functions and anomalous dimensions obtained from this scaling to third order in the coupling constants. By setting to zero the quadratic beta functions, we determine the Wilson-Fisher fixed points with coupling constants proportional to $\epsilon$. One can subsequently use the cubic terms in the beta functions to compute order $\epsilon^2$ corrections. In total the model has 32 large $N$ fixed points including the trivial one, and they can be grouped into ten orbits according to the action of an $S_4$ group that permutes the four $O(N)$ groups of the model: 
\begin{enumerate}[label=(\roman*)]
    \item one fixed point: The trivial fixed point.
    \item one fixed point: $\lambda_1=48\pi^2\epsilon$. This is the familiar vector model fixed point.
     \item four fixed points: $\lambda_1=-48\pi^2\epsilon$, one of $\lambda_2$ to $\lambda_5$ equal to $48\pi^2\epsilon$,
     \item four fixed points: one of $\lambda_2$ to $\lambda_5$ equal to $48\pi^2\epsilon$,
     \item six fixed points: $\lambda_1=-96\pi^2\epsilon$, $\lambda_2$ to $\lambda_5$ all equal to $48\pi^2\epsilon$ except for any two of them,
     \item six fixed points: $\lambda_1=-48\pi^2\epsilon$, $\lambda_2$ to $\lambda_5$ all equal to $48\pi^2\epsilon$ except for any two of them,
    \item four fixed points: $\lambda_1=-144\pi^2\epsilon$, $\lambda_2$ to $\lambda_5$ all equal to $48\pi^2\epsilon$ except for any one of them,
     \item four fixed points: $\lambda_1=-96\pi^2\epsilon$, $\lambda_2$ to $\lambda_5$ all equal to $48\pi^2\epsilon$ except for any one of them,
     \item one fixed point: $\lambda_1=-192\pi^2\epsilon$, $\lambda_2=\lambda_3=\lambda_4=\lambda_5=48\pi^2\epsilon$,
     \item one fixed point: $\lambda_1=-144\pi^2\epsilon$, $\lambda_2=\lambda_3=\lambda_4=\lambda_5=48\pi^2\epsilon$.
\end{enumerate}
For each fixed point, the unspecified couplings are zero. Note in particular that the couplings $\lambda_6$ to $\lambda_{14}$ are all zero to leading order in $1/N$ at all the fixed points. The leading anomalous dimensions for $\phi$ and $\phi^2$ at each of the fixed points are listed in Table~\ref{tab:anomDimsQuarticRank4}.
\begin{table}
\centering
\scalebox{1}{
\renewcommand{\arraystretch}{1.5}
\begin{tabular}{|c|c|c|} 
\hline
&$\gamma_\phi-\mathcal{O}(\epsilon^4)$
&$\gamma_{\phi^2}-\mathcal{O}(\epsilon^3)$ 
\\
\hline
\cellcolor{lightgray2}(i)
&\cellcolor{lightgray2} 0
&\cellcolor{lightgray2} 0
\\
\hline
(ii)
&
$\frac{1}{N^4}
(
\frac{1}{4}\epsilon^2
-\frac{1}{16}\epsilon^3
)+\mathcal{O}(\frac{1}{N^8})$
&
$\epsilon
+\frac{1}{N^4}
(
-6\epsilon+\frac{13}{2}
\epsilon^2)+\mathcal{O}(\frac{1}{N^8}
)$
\\
 \hline
\cellcolor{lightgray2}  (iii)
&\cellcolor{lightgray2}
$\frac{1}{N^2}(
\frac{1}{8}\epsilon^2
-\frac{1}{32}\epsilon^3
)+\mathcal{O}(\frac{1}{N^3})$
&\cellcolor{lightgray2}
$\frac{1}{N^2}(
3\epsilon-\frac{13}{4}\epsilon^2
)+\mathcal{O}(\frac{1}{N^3})$
\\
 \hline
 (vi)
&
$\frac{1}{N^2}(
\frac{1}{8}\epsilon^2
-\frac{1}{32}\epsilon^3
)+\mathcal{O}(\frac{1}{N^3})$
&
$\epsilon+\frac{1}{N^2}(
-3\epsilon+\frac{13}{4}\epsilon^2
)+\mathcal{O}(\frac{1}{N^3})$
\\
 \hline
\cellcolor{lightgray2} (v)
&\cellcolor{lightgray2}
$\frac{1}{N^2}(
\frac{1}{4}\epsilon^2
-\frac{1}{16}\epsilon^3
)+\mathcal{O}(\frac{1}{N^3})$
&\cellcolor{lightgray2}
$\frac{1}{N^2}(
6\epsilon-\frac{13}{2}\epsilon^2
)+\mathcal{O}(\frac{1}{N^3})$
\\
 \hline
 (vi)
&
$\frac{1}{N^2}(
\frac{1}{4}\epsilon^2
-\frac{1}{16}\epsilon^3
)+\mathcal{O}(\frac{1}{N^3})$
&
$\epsilon+\frac{1}{N^2}(
-6\epsilon+\frac{13}{2}\epsilon^2
)+\mathcal{O}(\frac{1}{N^3})$
\\
 \hline
\cellcolor{lightgray2} (vii)
&\cellcolor{lightgray2}
$\frac{1}{N^2}(
\frac{3}{8}\epsilon^2
-\frac{3}{32}\epsilon^3
)+\mathcal{O}(\frac{1}{N^3})$
&\cellcolor{lightgray2}
$\frac{1}{N^2}(
9\epsilon-\frac{39}{4}\epsilon^2
)+\mathcal{O}(\frac{1}{N^3})$
\\
 \hline
 (viii)
&
$\frac{1}{N^2}(
\frac{3}{8}\epsilon^2
-\frac{3}{32}\epsilon^3
)+\mathcal{O}(\frac{1}{N^3})$
&
$
\epsilon+
\frac{1}{N^2}(
-9\epsilon+\frac{39}{4}\epsilon^2
)+\mathcal{O}(\frac{1}{N^3})$
\\
\hline
\cellcolor{lightgray2} (ix)
&\cellcolor{lightgray2}
$\frac{1}{N^2}(
\frac{1}{2}\epsilon^2
-\frac{1}{8}\epsilon^3
)+\mathcal{O}(\frac{1}{N^3})$
&\cellcolor{lightgray2}
$\frac{1}{N^2}(
12\epsilon-13\epsilon^2
)+\mathcal{O}(\frac{1}{N^3})$
\\
 \hline
 (x)
&
$\frac{1}{N^2}(
\frac{1}{2}\epsilon^2
-\frac{1}{8}\epsilon^3
)+\mathcal{O}(\frac{1}{N^3})$
&
$\epsilon+\frac{1}{N^2}(
-12\epsilon+13\epsilon^2
)+\mathcal{O}(\frac{1}{N^3})$
\\
 \hline
\end{tabular}
}
\caption{Leading anomalous dimensions at the large $N$ fixed points of quartic $O(N)^4$ theory}
    \label{tab:anomDimsQuarticRank4}
\end{table}
Since we know the perturbative fixed points exactly in $N$, it is straightforward to expand the values in the table to higher orders in $1/N$ if desired, as long as we remain at the same order in $\epsilon$. Meanwhile, the exact $d$ dependencies of the leading large $N$ values be determined by a simple adjustment of the analysis we performed on the rank-three model: only the vectorial operators $\mathcal{O}_1$ to $\mathcal{O}_5$ are present at large $N$, and by introducing four traceless symmetric fields $\chi_1^{a_1a_2}$, $\chi_2^{b_1b_2}$, $\chi_3^{c_1c_2}$, and $\chi_4^{d_1d_2}$ we can perform Hubbard-Stratonovich transformations on the linear combinations $\mathcal{O}_i-\frac{1}{N}\mathcal{O}_1$ for $i\in\{2,3,4,5\}$. For the fixed points satisfying the condition $\lambda_1=\lambda_2+\lambda_3+\lambda_4+\lambda_5$, the Hubbard-Stratonovich transformations in $\chi$ eliminate $\mathcal{O}_1$ as well, whereas otherwise we need to also introduce a $\sigma$ field. Each $\chi$ field makes a contribution of $\frac{A_\phi(d)}{2N^2}$ to $\gamma_\phi$, and 
a contribution of $\frac{B_{\phi^2}(d)}{2N^2}$ to $\gamma_{\phi^2}$ if $\sigma$ is present and of $\frac{-B_{\phi^2}(d)}{2N^2}$ if $\sigma$ is absent, where $A_\phi(d)$ and $B_{\phi^2}(d)$ are given in equations \eqref{eq:gammaPhi} and \eqref{eq:gammaPhiSquared} respectively. The possible inclusion or exclusion of each of the five fields $\sigma$, $\chi_1$, $\chi_2$, $\chi_3$, $\chi_4$, yields up the total $2^5$ fixed points. The number of fixed points in each of the orbits (i) to (ix) are the binomial coefficients associated to choices among the four $\chi$ fields. 

Concretely, the fixed points have the following large $N$ field content, which all interact with the fundamental fields $\phi^{abcd}$ via cubic interactions, along with the following anomalous dimensions: 
\begin{enumerate}[label=(\roman*)]
    \item One fixed point: no $\sigma$ or $\chi$ present, $O(N^4)$ symmetry,
    \begin{flalign}
    \gamma_\phi^\text{(i)} =0\,, 
    \hspace{53mm}
    \gamma_{\phi^2}^\text{(i)} =0\,.
    &&
    \end{flalign}
    \item One fixed point: only $\sigma$ present, $O(N^4)$ symmetry,
    \begin{flalign}
    \gamma_\phi^\text{(ii)} =\frac{A_\phi(d)}{N^4}+\mathcal{O}(\frac{1}{N^8})\,, 
    \hspace{26mm}
    \gamma_{\phi^2}^\text{(ii)}=d-4+\frac{B_{\phi^2}(d)}{N^4}+\mathcal{O}(\frac{1}{N^8})\,.
    &&
    \end{flalign}
    \item Four fixed points: one $\chi$ present, $\sigma$ absent, $\displaystyle\frac{O(N^3)\times O(N)}{\mathbb{Z}_2}$ symmetry,
    \begin{flalign}
    \gamma_\phi^\text{(iii)} =\frac{A_\phi(d)}{2N^2}+\mathcal{O}(\frac{1}{N^3})\,, 
    \hspace{25mm}
    \gamma_{\phi^2}^\text{(iii)}=-\frac{B_{\phi^2}(d)}{2N^2}+\mathcal{O}(\frac{1}{N^3})\,.
    &&
    \end{flalign}
    \item Four fixed points: one $\chi$ present, $\sigma$ present, $\displaystyle\frac{O(N^3)\times O(N)}{\mathbb{Z}_2}$ symmetry,
    \begin{flalign}
    \gamma_\phi^\text{(iv)} =\frac{A_\phi(d)}{2N^2}+\mathcal{O}(\frac{1}{N^3})\,, 
    \hspace{24mm}
    \gamma_{\phi^2}^\text{(iv)}=4-d+\frac{B_{\phi^2}(d)}{2N^2}+\mathcal{O}(\frac{1}{N^3})\,.
    &&
    \end{flalign}
    \item Six fixed points: two $\chi$ present, $\sigma$ absent, $\displaystyle\frac{O(N^2)\times O(N)^2}{(\mathbb{Z}_2)^2}\rtimes S_2$ symmetry,
    \begin{flalign}
    \gamma_\phi^\text{(v)} =\frac{A_\phi(d)}{N^2}+\mathcal{O}(\frac{1}{N^3})\,, 
    \hspace{25mm}
    \gamma_{\phi^2}^\text{(v)}=-\frac{B_{\phi^2}(d)}{N^2}+\mathcal{O}(\frac{1}{N^3})\,.
    &&
    \end{flalign}
    \item Six fixed points: two $\chi$ present, $\sigma$ present, $\displaystyle\frac{O(N^2)\times O(N)^2}{(\mathbb{Z}_2)^2}\rtimes S_2$ symmetry,
    \begin{flalign}
    \gamma_\phi^\text{(vi)} =\frac{A_\phi(d)}{N^2}+\mathcal{O}(\frac{1}{N^3})\,, 
    \hspace{23mm}
    \gamma_{\phi^2}^\text{(vi)}=4-d+\frac{B_{\phi^2}(d)}{N^2}+\mathcal{O}(\frac{1}{N^3})\,.
    &&
    \end{flalign}
     \item Four fixed points: three $\chi$ present, $\sigma$ absent, $\displaystyle \frac{O(N)^4\rtimes S_3}{(\mathbb{Z}_2)^3}$ symmetry,
    \begin{flalign}
    \gamma_\phi^\text{(vii)} =\frac{3A_\phi(d)}{2N^2}+\mathcal{O}(\frac{1}{N^3})\,, 
    \hspace{21mm}
    \gamma_{\phi^2}^\text{(vii)}=-\frac{3B_{\phi^2}(d)}{2N^2}+\mathcal{O}(\frac{1}{N^3})\,.
    &&
    \end{flalign}
     \item Four fixed points: three $\chi$ present, $\sigma$ present, $\displaystyle \frac{O(N)^4\rtimes S_3}{(\mathbb{Z}_2)^3}$ symmetry, 
    \begin{flalign}
    \gamma_\phi^\text{(viii)} =\frac{3A_\phi(d)}{2N^2}+\mathcal{O}(\frac{1}{N^3})\,, 
    \hspace{20mm}
    \gamma_{\phi^2}^\text{(viii)}=4-d+\frac{3B_{\phi^2}(d)}{2N^2}+\mathcal{O}(\frac{1}{N^3})\,.
    &&
    \end{flalign}
     \item One fixed point: all four $\chi$ present, $\sigma$ absent, $\displaystyle\frac{O(N)^4\rtimes S_4}{(\mathbb{Z}_2)^3}$ symmetry,
    \begin{flalign}
    \gamma_\phi^\text{(ix)} =\frac{2A_\phi(d)}{N^2}+\mathcal{O}(\frac{1}{N^3})\,, 
    \hspace{21mm}
    \gamma_{\phi^2}^\text{(ix)}=-\frac{2B_{\phi^2}(d)}{N^2}+\mathcal{O}(\frac{1}{N^3})\,.
    &&
    \end{flalign}
     \item One fixed point: all four $\chi$ and $\sigma$ present, $\displaystyle\frac{O(N)^4\rtimes S_4}{(\mathbb{Z}_2)^3}$ symmetry,
              \begin{flalign}
    \gamma_\phi^\text{(x)} =\frac{2A_\phi(d)}{N^2}+\mathcal{O}(\frac{1}{N^3})\,, 
    \hspace{22mm}
    \gamma_{\phi^2}^\text{(x)}=4-d+\frac{2B_{\phi^2}(d)}{N^2}+\mathcal{O}(\frac{1}{N^3})\,. &&
    \end{flalign}
\end{enumerate}
For the fixed points in the first four orbits, the presence of an $O(N^3)$ symmetry bars the operators $\mathcal{O}_6$ to $\mathcal{O}_{14}$ from being generated by renormalization, even at finite $N$. For this reason, Feynman rules derived from the rank-four analogs of the actions \eqref{S(iii)} and \eqref{S(iv)} should correctly capture $1/N$ corrections to arbitrary order. The $O(N^2)$ symmetry of the fixed points in orbits (v) and (vi) prevents the melonic operators $\mathcal{O}_9$ to $\mathcal{O}_{14}$ from being generated, but the matricial operators $\mathcal{O}_6$, $\mathcal{O}_7$, and $\mathcal{O}_{8}$ are present at subleading order in $1/N$. For the fixed points in orbits (vii) to (x), all of the operators $\mathcal{O}_6$ to $\mathcal{O}_{14}$ are present at subleading order. Because of the presence of matricial operators, the task of developing a $1/N$ expansion to arbitrary order for the fixed points in orbits (v) to (x)
is a difficult endeavour, certainly more so than the same task for orbits (v) to (viii) in the rank-three case, where only the contributions of a melonic operator has to be incorporated.

We can clarify the above considerations by performing a change of operator basis. First we set $\lambda_6=\lambda_7=...=\lambda_{14}=0$ to study only the flow among the couplings $\lambda_1$ to $\lambda_5$. It can be checked from the large $N$ beta functions that for this starting configuration, $\lambda_6$ to $\lambda_{14}$ will remain zero throughout the RG flow. Next we change operator basis from $\mathcal{O}_n$, $n\in \{1,2,3,4,5\}$, to a basis given by 
\begin{align}
\mathcal{O}_1\,, \hspace{10mm} \mathcal{O}_n-\mathcal{O}_1 \hspace{3mm} \text{for }n\in\{2,3,4,5\}\,,
\end{align}
which corresponds to the following change of coupling constants:
\begin{align}
h_1=\lambda_1+\lambda_2+\lambda_3+\lambda_4+\lambda_5\,, \hspace{10mm} h_n = \lambda_n\hspace{3mm}\text{for }n\in\{2,3,4,5\}\,.
\end{align}
This new basis diagonalizes the large $N$ beta functions:
\begin{align}
\beta_{h_n} =-\epsilon h_n+\frac{h_n^2}{48\pi^2} \hspace{5mm}\text{for }i\in\{1,2,3,4,5\}\,.
\label{quartich}
\end{align}
The five quadratic equations \eqref{quartich} are true only at leading order in $1/N$ but to all orders in $h_n$.

All of the fixed points are unstable under deformations with respect to any of the nine non-vectorial operators $\mathcal{O}_6$ to $\mathcal{O}_{14}$, so a tremendous amount of tuning is required to reach any of the fixed points if we only impose $O(N)^4/(\mathbb{Z}_2)^3$ symmetry. But the fixed point (ii) and the four fixed points (iv) are stable under quartic deformations that respect their respective symmetries, and the other fixed points too have lessened numbers of relevant deformations under their own symmetries.

\begin{table}
\centering
\scalebox{1}{
\renewcommand{\arraystretch}{1.5}
\begin{tabular}{|c|c|c|c|} 
\hline
$g_a/\pi^2-\mathcal{O}(\epsilon^3)$ 
&$g_b/\pi^2-\mathcal{O}(\epsilon^3)$ 
&$g_c/\pi^2-\mathcal{O}(\epsilon^3)$ 
&$g_d/\pi^2-\mathcal{O}(\epsilon^3)$ 
\\
 \hline
\cellcolor{lightgray2}$2\epsilon+\frac{31}{48}\epsilon^2$ 
&\cellcolor{lightgray2}0
&\cellcolor{lightgray2}0 
&\cellcolor{lightgray2}0
\\
 \hline
$\frac{14}{3}\epsilon+\frac{35}{27}\epsilon^2$ 
&$-16\epsilon-8\epsilon^2$
&$12\epsilon+6\epsilon^2$
&0
\\
 \hline
\cellcolor{lightgray2}$\frac{9}{2}\epsilon+\frac{513}{256}\epsilon^2$ 
&\cellcolor{lightgray2}$-12\epsilon-\frac{171}{32}\epsilon^2$
&\cellcolor{lightgray2}$9\epsilon+\frac{513}{128}\epsilon^2$
&\cellcolor{lightgray2}0
\\
 \hline
$\frac{23}{7}\epsilon-\frac{929}{1029}\epsilon^2$
&$-\frac{72}{7}\epsilon+\frac{984}{343}\epsilon^2$
&$\frac{90}{7}\epsilon+\frac{2046}{343}\epsilon^2$
&$-\frac{36}{7}\epsilon-\frac{2784}{343}\epsilon^2$
\\
 \hline
\cellcolor{lightgray2}$\frac{27}{8}\epsilon-\frac{621}{4096}\epsilon^2$
&\cellcolor{lightgray2}$-9\epsilon+\frac{207}{512}\epsilon^2$
&\cellcolor{lightgray2}$\frac{45}{4}\epsilon+\frac{13\,941}{2048}\epsilon^2$
&\cellcolor{lightgray2}$-\frac{9}{2}\epsilon-\frac{7281}{1024}\epsilon^2$
\\
 \hline
$\frac{9}{5}\epsilon+\frac{27}{25}\epsilon^2$
&$-\frac{24}{5}\epsilon-\frac{72}{25}\epsilon^2$
&$\frac{54}{5}\epsilon+\frac{162}{25}\epsilon^2$
&$-\frac{36}{5}\epsilon-\frac{108}{25}\epsilon^2$
\\
 \hline
\cellcolor{lightgray2}$\frac{13}{6}\epsilon+\frac{1241}{1728}\epsilon^2$
&\cellcolor{lightgray2}$-4\epsilon-\frac{95}{24}\epsilon^2$
&\cellcolor{lightgray2}$9\epsilon+\frac{285}{32}\epsilon^2$
&\cellcolor{lightgray2}$-6\epsilon-\frac{95}{16}\epsilon^2$
 \\
 \hline
\end{tabular}
}
\caption{Locations of peturbative fixed points in quartic $O(2)^4\rtimes S_4$ theory}
    \label{tab:O(2)4points}
\end{table}

\subsubsection{Reduced System with $S_4$ theory}
\label{subSec:reduced}

\begin{figure}
    \centering
\begin{align*}
\begin{matrix}
\scalebox{1.11}{
\text{
{
\setlength{\fboxsep}{0pt}
\setlength{\fboxrule}{1pt}
		\fbox{\includegraphics[scale=0.7]{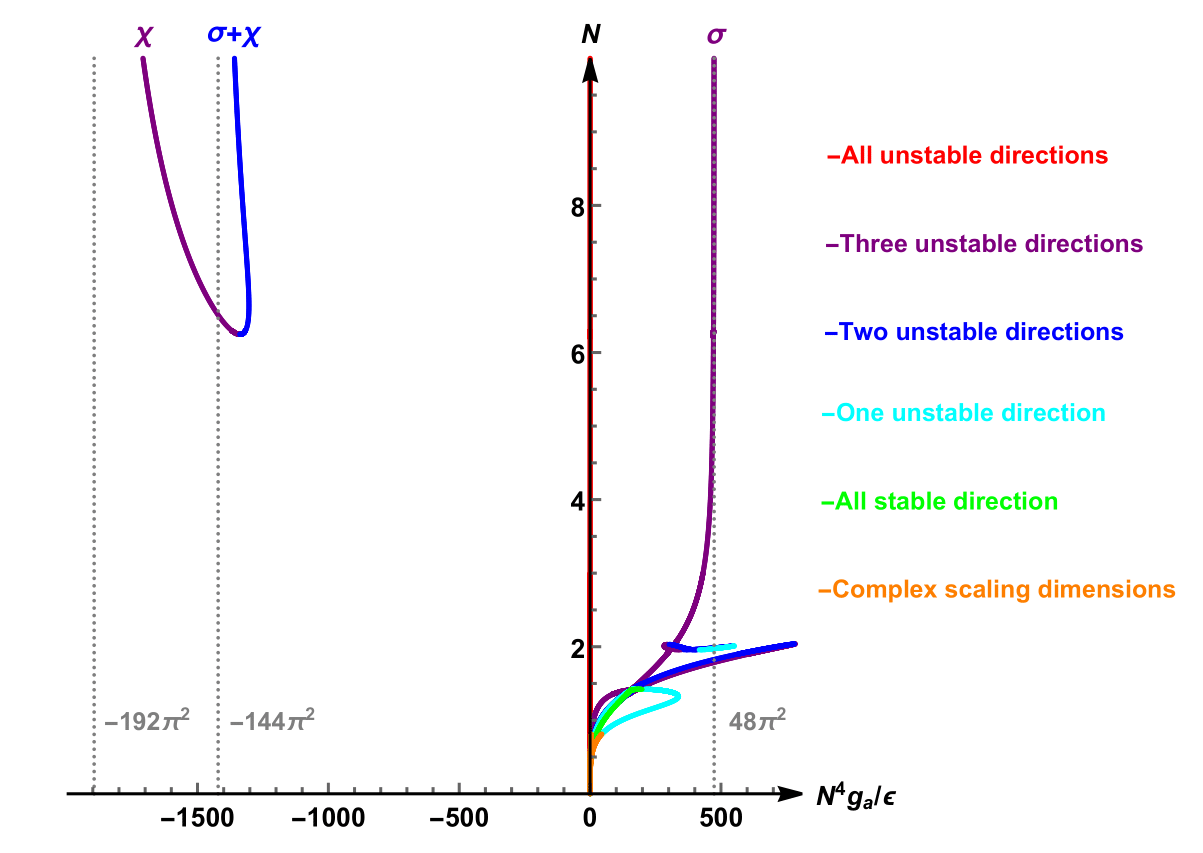}}
}
}}\end{matrix}
\end{align*}
    \caption{Values of the coupling $g_a$ at Wilson-Fisher fixed points in rank-four scalar theory with $S_4\ltimes O(N)^4/(\mathbb{Z}_2)^3$ symmetry in $d=4-\epsilon$ dimensions for different values of $N$. The dotted lines indicate the asymptotic values for $g_a$ at the large $N$ fixed points.}
    \label{fig:phiFourthOverview}
\end{figure}

The 16-dimensional RG flow of the quartic model \eqref{Squartic} contains a four-dimensional invariant subspace, owing to the fact any configuration of coupling constants that is symmetric with respect to a global $S_4$ symmetry that permutes indices will remain symmetric under the RG flow. The flow of this subspace is associated to the following four operators:
\begin{align}
\nonumber
\mathcal{O}_a =\,& \mathcal{O}_1\,,
\\\nonumber
\mathcal{O}_b =\,&\frac{1}{4}\Big(\mathcal{O}_2+\mathcal{O}_3+\mathcal{O}_4+\mathcal{O}_5\Big)\,,
\\[-10.5pt]
\\[-10.5pt] \nonumber
\mathcal{O}_c =\,& \frac{1}{3}\Big(\mathcal{O}_6+\mathcal{O}_7+\mathcal{O}_8\Big)\,,
\\[3pt]
\mathcal{O}_d =\,& \frac{1}{6}\Big(\mathcal{O}_9+\mathcal{O}_{10}+\mathcal{O}_{11}+\mathcal{O}_{12}+\mathcal{O}_{13}+\mathcal{O}_{14}\Big)\,.
\nonumber
\end{align}
The theory living in the invariant subspace is described by the action
\begin{align}
S = \int d^{4-\epsilon}x
\bigg(
\frac{1}{2}
(\partial_i\phi^{abcd})(\partial_i\phi^{abcd})
+\frac{1}{4!}\Big(
g_a\,\mathcal{O}_{a}+
g_b\,\mathcal{O}_{a}+
g_c\,\mathcal{O}_{a}+
g_d\,\mathcal{O}_{a}
\Big)
\bigg)\,.
\end{align}
Rescaling the coupling constants to obtain a smooth large $N$ limit,
\begin{align}
g_a=\frac{\lambda_a}{N^4}\,,
\hspace{15mm}
g_b=\frac{\lambda_b}{N^3}\,,
\hspace{15mm}
g_c=\frac{\lambda_c}{N^2}\,,
\hspace{15mm}
g_d=\frac{\lambda_d}{N^2}\,,
\end{align}
the perturbative fixed points of the reduced system at large $N$ are precisely the four fixed points of the full system that are symmetric under the global $S_4$ symmetry:
\begin{align}
\text{(i)}\hspace{8mm}&\lambda_a=\lambda_b=\lambda_c=\lambda_d=0\,,
   \nonumber\\
\text{(ii)}\hspace{8mm}&\lambda_a=48\pi^2\epsilon\,, \hspace{15.5mm}\lambda_b=\lambda_c=\lambda_d=0\,,
    \nonumber\\[-11pt]
    \\[-11pt]
\text{(ix)}\hspace{8mm}& \lambda_a=-192\pi^2\epsilon\,, \hspace{10mm}\lambda_b=192\pi^2\epsilon\,, 
\hspace{10mm}\lambda_c=\lambda_d=0\,,
     \nonumber\\
\text{(x)}\hspace{8mm}&\lambda_a=-144\pi^2\epsilon\,, \hspace{10mm}\lambda_b=192\pi^2\epsilon\,,
\hspace{10mm}\lambda_c=\lambda_d=0\,.
\nonumber
\end{align}
Because of the relative simplicity of the reduced system compared to the full system, it is not difficult to also chart out the finite $N$ fixed points of the model. Studying the beta functions at quadratic order in the coupling constants but exactly in $N$, one finds that the three non-trivial large $N$ fixed points extend down to finite $N$ values for all $N\geq 7$, but that the fixed points (ix) and (x) merge and disappear in a saddle-node bifurcation at $N\approx 6.247$. For $3\leq N \leq 6$, only the trivial fixed point and the vector model fixed point are present. At $N=2$, seven non-trivial fixed points are present, situated at the values of coupling constants listed in Table~\ref{tab:O(2)4points}. Figure~\ref{fig:phiFourthOverview} displays the perturbative fixed points of the model as a function of $N$.

\subsection{The \texorpdfstring{$O(N)^5$}{O(N)-to-the-fifth} model}
\label{ON5}
\begin{figure}
    \centering
\begin{align*}
\begin{matrix}\text{
		\includegraphics[scale=0.68]{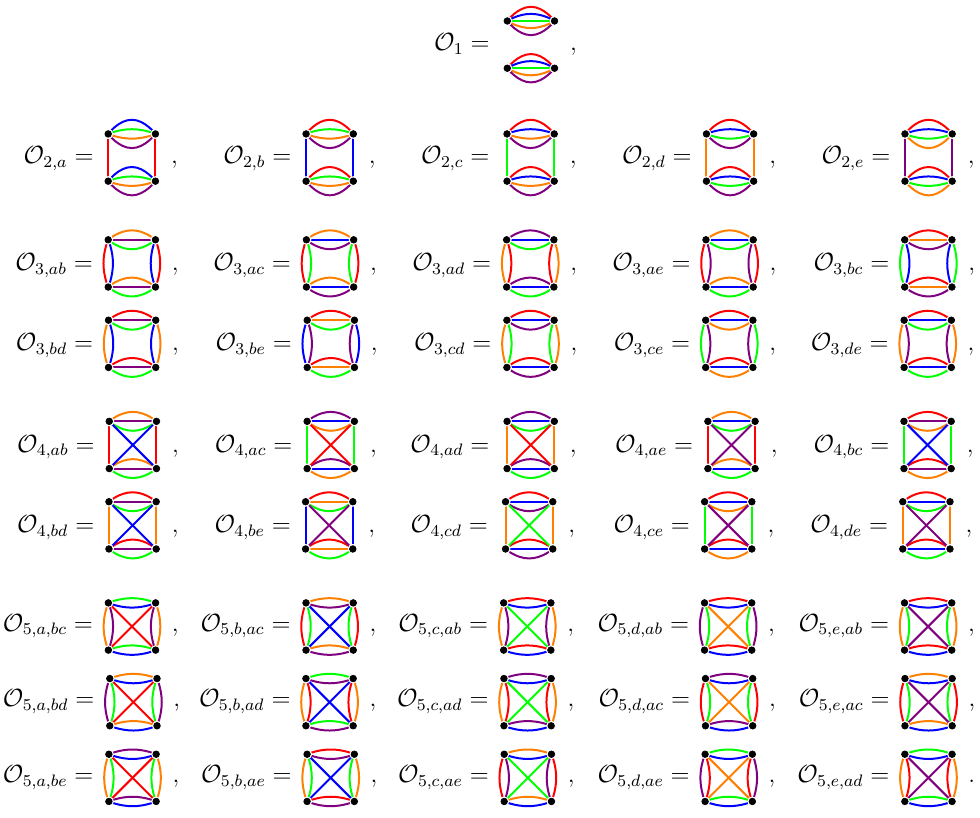}
}\end{matrix}
\end{align*}
    \caption{Quartic singlet operators in $O(N)^5$ theory}
    \label{fig:quarticRank5figs}
\end{figure}

We turn now to the quartic theory of rank-five tensors, with action given by
\begin{align}
S = \int d^{4-\epsilon}x
\bigg(
\frac{1}{2}
(\partial_i\phi^{abcde})(\partial_i\phi^{abcde})
+\frac{1}{4!}\sum_n
g_n\,\mathcal{O}_{n}
\bigg)\,.
\end{align}
The sum over $n$ runs over 41 distinct quartic operators and their associated couplings. If we do not distinguish between indices, there are five possible ways of performing index contractions, and they are depicted diagrammatically below, along with the number of operators each of them gives rise to.
\begin{align*}
\begin{matrix}
\text{
		\includegraphics[scale=1]{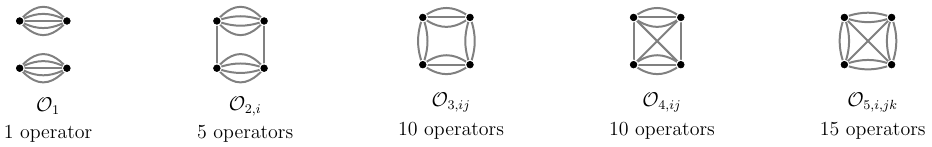}
}
\end{matrix}
\end{align*}
The number of operators associated to a diagram equals the number of distinct ways of colouring the graph with five colours. It is not important for the reader to know the precise manner in which we associate an operator to a colouring, but for the sake of clarity and completeness, we nonetheless provide the identifications in Figure~\ref{fig:quarticRank5figs}.

\subsubsection{Large $N$ limit}
The optimal large $N$ scalings for the operators are given by
\begin{align}
g_1=\frac{\lambda_1}{N^5}\,,
\hspace{8mm}
g_{2,i}=\frac{\lambda_{2,i}}{N^4}\,,
\hspace{8mm}
g_{3,ij}=\frac{\lambda_{3,ij}}{N^3}\,,
\hspace{8mm}
g_{4,ij}=\frac{\lambda_{4,ij}}{N^3}\,,
\hspace{8mm}
g_{5,i,jk}=\frac{\lambda_{5,i,jk}}{N^{5/2}}\,.
\end{align}
The number of large $N$ Wilson-Fisher fixed points increases rather dramatically compared with the rank-four theory. The rank-five model has a total of $10^{26}=67\,108\,864$ perturbative fixed points in the $\epsilon$ expansion including the trivial fixed point. The couplings $\lambda_{5,i,jk}$ vanish on all these fixed points. The set of fixed points and the values of their non-vanishing couplings admit a simple explanation in terms of the following 26 Hubbard-Stratonovich fields:
\begin{itemize}
    \item One $O(N)$ singlet $\sigma$, arising from integrating out $\mathcal{O}_1$.
    \item Five traceless symmetric fields $\chi_i^{i_1i_2}$, arising from integrating out $\mathcal{O}_{2,i}-\frac{1}{N}\mathcal{O}_1$ where $i\in \{a,b,c,d,e\}$.
    \item Ten doubly antisymmetric fields $\tau_{ij}^{i_1i_2\,j_1j_2}$, where $i_1$ and $i_2$ transform antisymmetrically under one $O(N)$ group, while $j_1$ and $j_2$ transform asymmetrically under another $O(N)$ group. The ten versions of these fields correspond to the $\binom{5}{2}$ choices of two out of the total five $O(N)$ groups. The $\tau_{ij}$ field integrates out the linear combination $\mathcal{O}_{3,ij}-\mathcal{O}_{4,ij}$. For example, consider the two operators
    \begin{align}
    \mathcal{O}_{3,de}=\,&
\phi^{a_1b_1c_1d_1e_1}
\phi^{a_1b_1c_1d_2e_2}
\phi^{a_2b_2c_2d_1e_1}
\phi^{a_2b_2c_2d_2e_2}\,,
\nonumber\\[-8pt]
\\[-8pt]\nonumber
    \mathcal{O}_{4,de}=\,&
\phi^{a_1b_1c_1d_1e_1}
\phi^{a_1b_1c_1d_2e_2}
\phi^{a_2b_2c_2d_1e_2}
\phi^{a_2b_2c_2d_2e_1}\,.
    \end{align}
The field $\tau_{de}$ is generated by integrating out the linear combination $\mathcal{O}_{3,de}-\mathcal{O}_{4,de}$ through use of the identity
\begin{align}
&\hspace{18mm}\frac{1}{2}\Big(\tau_{de}^{d_1d_2\,e_1e_2}-\phi^{abcd_1e_1}\phi^{abcd_2e_2}+\phi^{abcd_1e_2}\phi^{abcd_2e_1}\Big)^2=
\nonumber \\[-17pt]
\\[-5pt]
\nonumber
&\frac{1}{2}(\tau_{de}^{d_1d_2\,e_1e_2})^2+\mathcal{O}_{3,de}-\mathcal{O}_{4,de}
+\tau_{de}^{d_1d_2\,e_1e_2}\Big(\phi^{abcd_1e_2}\phi^{abcd_2e_1}-\phi^{abcd_1e_1}\phi^{abcd_2e_2}\Big)\,.
\end{align}
    \item Ten doubly traceless symmetric fields $\rho_{ij}^{i_1i_2\,j_1j_2}$, where $i_1$ and $i_2$ transform as a traceless symmetric matrix under one $O(N)$ group, while $j_1$ and $j_2$ transform as a traceless symmetric matrix under another $O(N)$ group. The field $\rho_{ij}$ integrates out the linear combination $\mathcal{O}_{3,ij}+\mathcal{O}_{4,ij}-\frac{2}{N}(\mathcal{O}_i+\mathcal{O}_j)+\frac{2}{N^2}\mathcal{O}_1$. Take for example the field $\rho_{de}$. Consider the operators
    \begin{align}
    \mathcal{O}_{2,d}=\,&
\phi^{a_1b_1c_1d_1e_1}
\phi^{a_1b_1c_1d_2e_1}
\phi^{a_2b_2c_2d_1e_2}
\phi^{a_2b_2c_2d_2e_2}\,,
\nonumber\\[-8pt]
\\[-8pt]\nonumber
   \mathcal{O}_{2,e}=\,&
\phi^{a_1b_1c_1d_1e_1}
\phi^{a_1b_1c_1d_1e_2}
\phi^{a_2b_2c_2d_2e_1}
\phi^{a_2b_2c_2d_2e_2}\,,
    \end{align}
and consider also the following tensor:
\begin{align}
\Phi^{d_1d_2\,e_1e_2}\equiv \,&
\phi^{abcd_1e_1}\phi^{abcd_2e_2}+\phi^{abcd_1e_2}\phi^{abcd_2e_1}
\\
&
-\frac{2}{N}\delta^{d_1d_2}\phi^{abcde_1}\phi^{abcde_2}
-\frac{2}{N}\delta^{e_1e_2}\phi^{abcd_1e}\phi^{abcd_2e}
+\frac{2}{N^2}\delta^{d_1d_2}\delta^{e_1e_2}\phi^{abcde}\phi^{abcde}\,.
\nonumber
\end{align}
The field $\rho_{de}$ is generated by integrating out the following linear combination of operators:
\begin{align}
\nonumber\\[-40pt]
&\hspace{15mm}\mathcal{O}_{3,de}+\mathcal{O}_{4,de}-\frac{2}{N}(\mathcal{O}_{2,d}+\mathcal{O}_{2,e})+\frac{2}{N^2}\mathcal{O}_1=
\nonumber \\[-17pt]
\\[-5pt]
\nonumber
&\frac{1}{2}(\rho_{de}^{d_1d_2\,e_1e_2}-\Phi^{d_1d_2\,e_1e_2})^2-\frac{1}{2}(\rho_{de}^{d_1d_2\,e_1e_2})^2+2\rho_{de}^{d_1d_2\,e_1e_2}\,\Phi^{d_1d_2\,e_1e_2}\,.
\nonumber
\end{align}
\end{itemize}
The total number of perturbative fixed points equals the $2^{26}$ possible choices of including or not including each of the fields in the above list of possible large $N$ field content. This can be made explicit by switching to the operator basis
\begin{align}
&\mathcal{O}_1\,,
\nonumber\\\nonumber
&\mathcal{O}_{2,i}-\frac{1}{N}\mathcal{O}_1\,,
\\[-11pt]
\\[-11pt]
&\mathcal{O}_{3,ij}-\mathcal{O}_{4,ij}\,,
\nonumber\\\nonumber
&\mathcal{O}_{3,ij}+\mathcal{O}_{4,ij}-\frac{2}{N}(\mathcal{O}_{2,i}+\mathcal{O}_{2,j})+\frac{2}{N^2}\mathcal{O}_1\,,
\end{align}
which amounts to studying the flows of couplings $h_n$ related to the couplings $\lambda_n$ via
\begin{align}
h_1=\,&\lambda_1
+\sum_{i}\lambda_{2,i}
+\sum_{i<j}(\lambda_{3,ij}+\lambda_{4,ij})\,,
\nonumber\\[-2pt]\nonumber
h_{2,i}=\,&
\lambda_{2,i}
+\sum_j(\lambda_{3,ij}+\lambda_{4,ij})\,,
\\[-13pt]
\\[-13pt]
h_{3,ij}=\,&
\frac{1}{2}(\lambda_{3,ij}-\lambda_{4,ij})\,,
\nonumber\\[5pt]\nonumber
h_{4,ij}=\,&
\frac{1}{2}(\lambda_{3,ij}+\lambda_{4,ij})\,.
\end{align}
In the new basis the beta functions are diagonalized as follows:
\begin{align}
\nonumber
\beta_{h_1}\big|_{\lambda_{5,i,jk}=0}\,=&
-\epsilon\beta_{h_1}
+\frac{h_1^2}{48\pi^2}\,,
\\ \nonumber
\beta_{h_{2,i}}\big|_{\lambda_{5,i,jk}=0}\,=&
-\epsilon\beta_{h_{2,i}}
+\frac{h_{2,i}^2}{48\pi^2}\,,
\\[-11pt]
\label{rank5h}
\\[-11pt]
\nonumber
\beta_{h_{3,ij}}\big|_{\lambda_{5,i,jk}=0}\,=&
-\epsilon\beta_{h_{3,ij}}
+\frac{h_{3,ij}^2}{24\pi^2}\,,
\\
\beta_{h_{4,ij}}\big|_{\lambda_{5,i,jk}=0}\,=&
-\epsilon\beta_{h_{4,ij}}
+\frac{h_{4,ij}^2}{24\pi^2}\,.
\nonumber
\end{align}
Of the vast number of large $N$ fixed points of this model, many are related by permutations of tensor indices and so cannot be associated to distinct CFTs. The number of non-isomorphic fixed points is given by the number of orbits of the $2^{26}$ fixed points under the action of $S_5$. By careful counting, we find that this number equals 612\,032.

\begin{table}
\centering
\scalebox{0.91}{
\renewcommand{\arraystretch}{1.5}
\begin{tabular}{|c|c|c|} 
\hline
large $N$ field content &$\gamma_\phi-\mathcal{O}(\epsilon^4)$
&$\gamma_{\phi^2}-\mathcal{O}(\epsilon^3)$ 
\\
\hline
\cellcolor{lightgray2}$\sigma$
&
\cellcolor{lightgray2}$\frac{1}{N^5}
(\frac{1}{4}\epsilon^2-\frac{1}{16}\epsilon^3)+\mathcal{O}(\frac{1}{N^{10}})
$
&\cellcolor{lightgray2}
$
\epsilon
+\frac{1}{N^5}
(-6\epsilon+\frac{13}{2}\epsilon^2)+\mathcal{O}(\frac{1}{N^{10}})
$
\\
 \hline
$\chi$ & 
$\frac{1}{N^3}(
\frac{5}{8}\epsilon^2
-\frac{5}{32}\epsilon^3
)+\mathcal{O}(\frac{1}{N^4})$
& $\frac{1}{N^3}(
15\epsilon
-\frac{65}{4}\epsilon^2
)+\mathcal{O}(\frac{1}{N^4})$
\\
 \hline
\cellcolor{lightgray2}$\sigma$, $\chi$ & \cellcolor{lightgray2}
$\frac{1}{N^3}(\frac{5}{8}\epsilon^2-\frac{5}{32}\epsilon^3)+\mathcal{O}(\frac{1}{N^4})$
&\cellcolor{lightgray2} $\epsilon+\frac{1}{N^3}(
-15\epsilon+\frac{65}{4}\epsilon^2
)+\mathcal{O}(\frac{1}{N^4})$
\\
 \hline
$\tau$ & 
$\frac{1}{N}(\frac{5}{8}\epsilon^2-\frac{5}{32}\epsilon^3)+\mathcal{O}(\frac{1}{N^2})$
&$\frac{1}{N}(15\epsilon-\frac{65}{4}\epsilon^2)+\mathcal{O}(\frac{1}{N^2})$
\\
 \hline
\cellcolor{lightgray2}$\sigma$, $\tau$ & \cellcolor{lightgray2}
$\frac{1}{N}(
\frac{5}{8}\epsilon^2
-\frac{5}{32}\epsilon^3
)+\mathcal{O}(\frac{1}{N^2})$
&\cellcolor{lightgray2} $\epsilon+\frac{1}{N}(-15\epsilon+\frac{65}{4}\epsilon^2)+\mathcal{O}(\frac{1}{N^2})$
\\
 \hline
$\chi$, $\tau$ & 
$\frac{1}{N}(
\frac{5}{8}\epsilon^2-\frac{5}{32}\epsilon^3
)+\mathcal{O}(\frac{1}{N^2})$
&$\frac{1}{N}(15\epsilon-\frac{65}{4}\epsilon^2)+\mathcal{O}(\frac{1}{N^2})$
\\
 \hline
\cellcolor{lightgray2}$\sigma$, $\chi$, $\tau$ &\cellcolor{lightgray2}
$\frac{1}{N}(
\frac{5}{8}\epsilon^2
-\frac{5}{32}\epsilon^3
)+\mathcal{O}(\frac{1}{N^2})$
&\cellcolor{lightgray2} $\epsilon+\frac{1}{N}(
-15\epsilon+\frac{65}{8}\epsilon^2
)+\mathcal{O}(\frac{1}{N^2})$
\\
 \hline
$\rho$ & $\frac{1}{N}(\frac{5}{8}\epsilon^2-\frac{5}{32}\epsilon^3)+\mathcal{O}(\frac{1}{N^2})$
&$\frac{1}{N}(15\epsilon-\frac{65}{4}\epsilon^2)+\mathcal{O}(\frac{1}{N^2})$
\\
 \hline
\cellcolor{lightgray2}$\sigma$, $\rho$ & \cellcolor{lightgray2}
$\frac{1}{N}(\frac{5}{8}\epsilon^2-\frac{5}{32}\epsilon^3)+\mathcal{O}(\frac{1}{N^2})$
&\cellcolor{lightgray2} $\epsilon+\frac{1}{N}(-15\epsilon+\frac{65}{4}\epsilon^2)+\mathcal{O}(\frac{1}{N^2})$
\\
 \hline
$\chi$, $\rho$ & 
$\frac{1}{N}(\frac{5}{8}\epsilon^2-\frac{5}{32}\epsilon^3)+\mathcal{O}(\frac{1}{N^2})$
& $\frac{1}{N}(15\epsilon-\frac{65}{4}\epsilon^2)+\mathcal{O}(\frac{1}{N^2})$
\\
 \hline
\cellcolor{lightgray2}$\sigma$, $\chi$, $\rho$ & \cellcolor{lightgray2}
$\frac{1}{N}(\frac{5}{8}\epsilon^2-\frac{5}{32}\epsilon^3)+\mathcal{O}(\frac{1}{N^2})$
&\cellcolor{lightgray2} $\epsilon+\frac{1}{N}(-15\epsilon+\frac{65}{4}\epsilon^2)+\mathcal{O}(\frac{1}{N^2})$
\\
 \hline
$\tau$, $\rho$ & 
$\frac{1}{N}(\frac{5}{4}\epsilon^2-\frac{5}{16}\epsilon^3)+\mathcal{O}(\frac{1}{N^2})$
& $\frac{1}{N}(30\epsilon-\frac{65}{2}\epsilon^2)+\mathcal{O}(\frac{1}{N^2})$
\\
 \hline
\cellcolor{lightgray2}$\sigma$, $\tau$, $\rho$ &\cellcolor{lightgray2} 
$\frac{1}{N}(\frac{5}{4}\epsilon^2-\frac{5}{16}\epsilon^3)+\mathcal{O}(\frac{1}{N^2})$
&\cellcolor{lightgray2} $\epsilon+\frac{1}{N}(
-30\epsilon+\frac{65}{2}\epsilon^2
)+\mathcal{O}(\frac{1}{N^2})$
\\
 \hline
$\chi$, $\tau$, $\rho$ &
$\frac{1}{N}(\frac{5}{4}\epsilon^2-\frac{5}{16}\epsilon^3)+\mathcal{O}(\frac{1}{N^2})$
&$\frac{1}{N}(30\epsilon-\frac{65}{2}\epsilon^2)+\mathcal{O}(\frac{1}{N^2})$
\\
 \hline
\cellcolor{lightgray2}$\sigma$, $\chi$, $\tau$, $\rho$ & \cellcolor{lightgray2}
$\frac{1}{N}(\frac{5}{4}\epsilon^2-\frac{5}{16}\epsilon^3)+\mathcal{O}(\frac{1}{N^2})$
&\cellcolor{lightgray2} $\epsilon+\frac{1}{N}(-30\epsilon+\frac{65}{2}\epsilon^2)+\mathcal{O}(\frac{1}{N^2})$
\\
 \hline
\end{tabular}}
\caption{Leading anomalous dimensions at the large $N$ fixed points of quartic $O(N)^5\rtimes S_5$ theory}
    \label{tab:anomDimsQuarticRank5}
\end{table}

\subsubsection{Reduced System with $S_5$ symmetry}

\begin{figure}
    \centering
\begin{align*}
\hspace{-11mm}
\begin{matrix}\text{\scalebox{1.1}{
{
\setlength{\fboxsep}{4pt}
\setlength{\fboxrule}{1pt}
		\fbox{\includegraphics[scale=0.51]{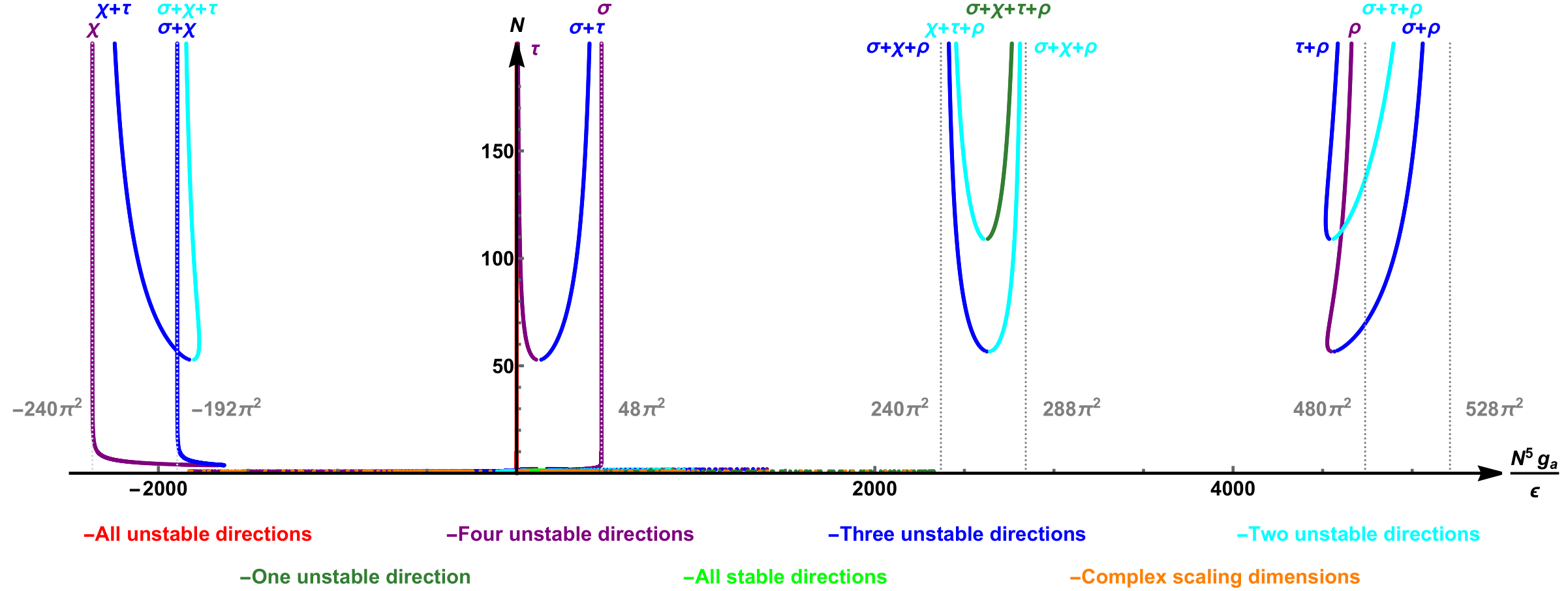}}
}}
}\end{matrix}
\end{align*}
    \caption{Values of the coupling $g_a$ at Wilson-Fisher fixed points in rank-five scalar theory with $S_5\ltimes O(N)^5/(\mathbb{Z}_2)^4$ symmetry in $d=4-\epsilon$ dimensions for different values of $N$. The dotted lines indicate the asymptotic values for $g_a$ at the large $N$ fixed points.}
        \label{fig:phiFifthOverview}
\end{figure}

If we study the subsector of the theory that is symmetric under permutations of the tensor indices and so exhibits a global $\frac{O(N)^5\rtimes S_5}{(\mathbb{Z}_2)^4}$ symmetry, the RG flow reduces to a five-dimensional dynamical system. The fixed points of this flow are the fixed points of the full system that respect the $S_5$ permutation symmetry, and these can be counted by the choices of large $N$ field content--- $\sigma$ or no $\sigma$, all $\chi$ fields or no $\chi$ fields, all $\tau$ fields or no $\tau$ fields, and all $\rho$ fields or no $\rho$ fields--- entailing that there are $2^{4}=16$ fixed points, all of which are non-isomorphic. The leading anomalous dimensions for $\phi$ and $\phi^2$ are listed in Table~\ref{tab:anomDimsQuarticRank5}, while Figure~\ref{fig:phiFifthOverview} shows how the large $N$ fixed points appear at finite values of $N$ and approach their asymptotic values.

It is a simple exercise to work out the general $d$ dependence of the leading $1/N$ corrections listed above by a slight modification of the large $N$ treatment presented in Subsection~\ref{ON3} where the fields $\tau$ and $\rho$ propagate along the dashed lines in diagrams \eqref{eq:sunset} and \eqref{diagrams}. Loop integrals are unchanged, but additional index contraction is involved, which explains why the columnar values are related by simple fractions and powers of $N$, seeing as the full $1/N$ corrections each equal $A_\phi(d)$ or $B_{\phi^2}(d)$ times a rational number and a negative power of $N$.

\subsection{Large \texorpdfstring{$N$}{N} limit at general tensor rank}
\label{generalities}
For quartic field theories of rank-$r$ tensors  with $O(N)^r/(\mathbb{Z}_2)^{r-1}$ or $ S_r \ltimes O(N)^r/(\mathbb{Z}_2)^{r-1}$ symmetry, some of the key characteristics are
\begin{align*}
n_{\text{quartic}}:&\hspace{2mm} \text{the number of quartic $O(N)^r$ singlets,}
\\[2pt]
n_\text{fixed points}:&\hspace{2mm} \text{the number of large $N$ fixed points, and}
\\[2pt]
n_\text{non-isomorphic}:&\hspace{2mm} \text{the number of non-isomorphic large $N$ fixed points, ie. fixed points not} 
\\[-2pt]
&\hspace{2mm}\text{related by a permutation of tensor indices.} 
\end{align*}
For the theories studied in the previous sections, with tensor ranks less than or equal to five, this data can be summarized in Table~\ref{tab:quarticOverview}. We now turn to the question of how the RG flows and fixed points of tensorial $\phi^4$ theory behave for general $r$ and to the question of how the above data extends beyond $r=5$.

\subsubsection{Number of operators}
For models with $O(N)^r/(\mathbb{Z}_2)^{r-1}$ symmetry, the counting of quartic singlets can be performed in generality, and there is a simple argument in \cite{gurau2022duality} establishing that
\begin{flalign}
\label{quarticOperatorCounting}
&\frac{O(N)^r}{(\mathbb{Z}_2)^{r-1}}:\hspace{45mm}  n_\text{quartic}(r) = \frac{3+3^r}{6}\,. &&
\end{flalign}
For models with $S_r\ltimes O(N)^r/(\mathbb{Z}_2)^{r-1}$ symmetry, the data in Table~\ref{tab:quarticOverview} might tempt the reader to think that $n_\text{quartic}$ simply equals $r$, but this is not the case. In the case of $S_6\ltimes O(N)^6/(\mathbb{Z}_2)^5$ symmetry one finds that $n_\text{quartic}=7$:
\begin{align}
\label{sexticOps}
\begin{matrix}
\text{
		\includegraphics[scale=1]{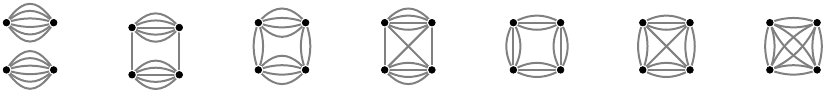}
}
\end{matrix}
\end{align}
\begin{table}
\centering
$\hspace{-4mm}$\scalebox{0.9}{
\renewcommand{\arraystretch}{1.3}
\begin{tabular}{|l|c|c|c|c|c|c|c|c|} 
\hline
global symmetry 
&\cellcolor{lightgray2}$O(N)$ 
&$\frac{O(N)^2\rtimes S_2}{\mathbb{Z}_2}$ 
&\cellcolor{lightgray2}$\frac{O(N)^3}{(\mathbb{Z}_2)^2}$ 
&$\frac{O(N)^3\rtimes S_3}{(\mathbb{Z}_2)^2}$
&\cellcolor{lightgray2}$\frac{O(N)^4}{(\mathbb{Z}_2)^3}$ 
&$\frac{O(N)^4\rtimes S_4}{(\mathbb{Z}_2)^3}$
&\cellcolor{lightgray2}$\frac{O(N)^5}{(\mathbb{Z}_2)^4}$ 
&$\frac{O(N)^5\rtimes S_5}{(\mathbb{Z}_2)^4}$
\\
\hline
$n_{\text{quartic}}$
&\cellcolor{lightgray2} 1
& 2
&\cellcolor{lightgray2} 5
& 3
&\cellcolor{lightgray2} 14
& 4
&\cellcolor{lightgray2} 41
& 5
\\
\hline
$n_\text{fixed points}$
&\cellcolor{lightgray2} 2
& 2
&\cellcolor{lightgray2} 16
& 4
&\cellcolor{lightgray2}
32
& 4
&\cellcolor{lightgray2} 67\,108\,864
& 16
\\
\hline
$n_\text{non-isomorphic}$
&\cellcolor{lightgray2} 2
& 2
&\cellcolor{lightgray2} 8
& 4
&\cellcolor{lightgray2} 10
& 4
&\cellcolor{lightgray2} 612\,032
& 16
\\
\hline
\end{tabular}}
\caption{Numbers of operators and fixed points in quartic $O(N)^r$ theory and $O(N)^r\rtimes S_r$ theory}
    \label{tab:quarticOverview}
\end{table}
The general answer in the $S_r\ltimes O(N)^r/(\mathbb{Z}_2)^{r-1}$ case is that $n_\text{quartic}$ equals $p_3(r)$, ie. the number of integer partitions of $r$ into exactly three parts. This number has the following generating function:
\begin{align}
\frac{1}{(1-x)(1-x^2)(1-x^3)}
=1+x+2x^2+3x^3+4x^4+5x^5+7x^6+8x^7+10x^8+...\,,
\end{align}
and in fact a closed-form expression exists:
\begin{flalign}
\label{quarticOperatorCountingReduced}
&\frac{O(N)^r\rtimes S_r}{(\mathbb{Z}_2)^{r-1}}:\hspace{12mm}  n_\text{quartic}(r) = \frac{1}{72}
\bigg(
47+9(-1)^r+6\,r(6+r)+16\cos\big(\frac{2\pi r}{3}\big)
\bigg)\,.
&&
\end{flalign}

\subsubsection{Number of fixed points}

Through explicit computation we have seen in Sections~\ref{ON2}, \ref{ON3}, \ref{ON4}, and \ref{ON5} that for all the real large $N$ fixed points of quartic $O(N)^r$ theory with $r\leq 5$, the only non-vanishing couplings at the fixed points are those associated to vectorial operators, by which we mean operators, like the first four in \eqref{sexticOps}, for which more than half the indices are contracted between the same pairs of fundamental fields. In other words, operators with melonic or planar large $N$ limits are absent at the perturbative large $N$ fixed points. In Appendix~\ref{sec:classifying} we argue that this same property holds true for all odd values of $r$, while other types of fixed points may be present for even $r$. By diagonalizing the vectorial RG flow at large $N$, we also show that the number of vectorial large fixed points equals two raised to the number of vectorial operators. In consequence, we find that the number of vectorial large $N$ fixed points for general tensor rank $r$ is given by
\begin{flalign}
\label{fpNumber}
&\frac{O(N)^r}{(\mathbb{Z}_r)^{r-1}}:  \hspace{20mm}
\log_2n_{\text{fixed points}}(r)=1+\sum_{m=1}^{\left\lfloor\frac{r-1}{2}\right\rfloor}
\binom{r}{m}2^{m-1}\,.
&&
\end{flalign}
The combinatorial explosion witnessed in Table~\ref{tab:quarticOverview} is made further manifest by this formula. While the fixed points counted by the formula \eqref{fpNumber} include isomorphic copies, the number of non-isomorphic fixed points is bounded below by $n_{\text{vectorial fixed points}}(r)/r!$, and for $r=7$ this lower bound exceeds Avogadro's number, while for $r=9$ it exceeds one googol. The asymptotic behaviour of the number of vectorial fixed points at large rank is given by
\begin{flalign}&
\frac{O(N)^r}{(\mathbb{Z}_r)^{r-1}}: \hspace{20mm}
\log_2n_{\text{fixed points}}(r)
\rightarrow \begin{cases}
\displaystyle
\frac{2^{3r/2}}{\sqrt{\pi r}} \hspace{12mm} \text{for }r\text{ odd}\,,
\\
\\
\displaystyle
\frac{2^{3r/2}}{\sqrt{2\pi r}}\hspace{10mm} \text{for }r\text{ even}\,,
\end{cases}
&&
\end{flalign}
where the limit should be understood to mean that the ratio tends to one (and not that the difference tends to zero) as $r$ goes to infinity. 

The number $\log_2n_{\text{fixed points}}(r)$ equals the number of vectorial operators but also equals the distinct possible Hubbard-Stratonovich fields in $O(N)^r$ theory, as we explain in detail in Appendix~\ref{sec:classifying}. Each of these fields contains up to $\left\lfloor\frac{r-1}{2}\right\rfloor$ index pairs, each of which transforms as a traceless symmetric or as an antisymmetric matric under its own $O(N)$ group, with the number of antisymmetric index pairs being even. For example, the rank seven theory contains $2^{190}$ large $N$ fixed points, and the list of possible Hubbard-Stratonovich fields consists of 
\begin{itemize}
\item one $\sigma$ field, 
\item seven fields $\chi$ with one traceless symmetric index pair, 
\item 21 doubly antisymmetric fields $\tau$, 
\item 21 doubly traceless symmetric fields $\rho$, 
\item 35 triply traceless symmetric fields, and 
\item 105 fields with two antisymmetric index pairs and one traceless symmetric index pair.
\end{itemize}
For all of the vectorial fixed points, the leading large $N$ anomalous dimensions are given by simple modifications of the $O(N)$ model answers, obtained by summing over Feynman diagrams \eqref{eq:sunset} and \eqref{diagrams} for each of the associated Hubbard-Stratonovich fields. Many of the $O(N)^r/(\mathbb{Z}_2)^{r-1}$ fixed points exhibit enhanced symmetry, with each enhanced symmetry being given by the semi-direct product of a group $G$ acting on $\mathbb{Z}_2\times \prod_{i=1}^{i_\text{max}}\frac{O(N^{\ell_i})}{\mathbb{Z}_2}$, where $1\leq \ell_i \leq r$, and $\sum_i\ell_i=r$, and $G$ is a subgroup of $S_r$ that permutes $O(N^{\ell_i})$ groups with the same value for $\ell_i$. For most vectorial fixed points, non-vectorial operators are generated at subleading order in $1/N$, but in cases of an enhancement that includes an $O(N^m)$  subgroup with $m>r/2$, the symmetry prevents any non-vectorial operators from being generated, and so for these theories we expect the $1/N$ expansions developed by the Hubbard-Stratonovich transformation to remain valid to any order. For a given symmetry enhancement of this kind, the fixed points with the most Hubbard-Stratonovich fields associated to them are stable under any quartic deformation that respects the enhanced symmetry. But among the general space of quartic operators with $O(N)^r/(\mathbb{Z}_2)^{r-1}$ symmetry, an inordinate amount of tuning is required to flow to any of the vectorial fixed points since any non-vectorial operator provides a relevant deformation.

The explosion in number of fixed points is significantly slowed down if we restrict ourselves to quartic $O(N)^r$ tensor theories with additional $S_r$ symmetry under permutations of indices. The formula for the number of vectorial large $N$ fixed points is in this case given by
\begin{flalign}
&\frac{O(N)^r\rtimes S_r}{(\mathbb{Z}_2)^{r-1}}: \hspace{3mm} \log_2n_{\text{fixed points}}(r)=
\left\lfloor
\frac{r+3}{4}
\right\rfloor
\,
\left\lfloor
\frac{r+5}{4}
\right\rfloor
=
\begin{cases}
l(l+1) \hspace{3mm}\text{ for }r=4l-1\,,
\\
l(l+1) \hspace{3mm}\text{ for }r=4l\,,
\\
(l+1)^2 \hspace{3mm}\text{ for }r=4l+1\,,
\\
(l+1)^2 \hspace{3mm}\text{ for }r=4l+2\,,
\end{cases}
\hspace{-30mm}
&&
\end{flalign}
all of which fixed points are non-isomorphic.

\section{Tensorial \texorpdfstring{$\phi^6$}{phi-to-the-sixth} Theory}
\label{sec:Sextic}

Following our study of quartic theories in the previous section, we now advance to sextic theories and study tensorial generalizations of $\phi^6$ theory, where each index transforms under its own $O(N)$ group. A high degree of tuning is associated to these theories since not only must the quadratic operator be tuned away, the same applies also to each of the quartic operators, whose number increases exponentially with tensor rank $r$ as shown in equation \eqref{quarticOperatorCounting}, although for the subsector of theories with additional $S_r$ permutation theory the increase with $r$ slows down to a quadratic growth, as attested by equation \eqref{quarticOperatorCountingReduced}.

We will not be able to uncover the general structure of sextic tensor theories at large $N$ for the reason that sextic tensor theories do not exhibit the kind of simple pattern inherent in the large $N$ dynamics of quartic theories, as appears already on comparing the qualitatively different large $N$ limits of sextic vector, matrix, and three-tensor theories. Instead, we will review the theories that have been studied in the literature and then proceed to analyze the next sextic theory theory in order of increasing rank by presenting a case study for rank-four tensor theory at four-loop order, in $d$ just below the critical dimension three. 

For the rank-one theory, ie. the sextic $O(N)$ model, the perturbative beta function has been known since work in the 70ies \cite{stephen1973feynman,lewis1978tricritical}, and a series of subsequent papers demonstrated that at large $N$ all loop corrections vanish at leading order, while the $1/N$ corrections indicate the existence of an interacting UV fixed point in three dimensions \cite{townsend1975spontaneous,townsend1977consistency,appelquist1981three,pisarski1982fixed,appelquist1982vacuum}. More concretely, these papers found that if we write the action as
\begin{align}
S = \int d^{3-\epsilon}x
\bigg(
\frac{1}{2}(\partial_i \phi^{a})(\partial_i \phi^{ab})
+
\frac{1}{6!}\,g\,(\phi^a\phi^a)^3
\bigg)\,,
\end{align}
and we scale the coupling constant according to its optimal scaling, $g = \frac{\lambda}{N^2}$, then the large $N$ beta function to subleading order in $1/N$ terminates at cubic order in the coupling and is given by
\begin{align}
\beta_\lambda =
-2\epsilon \lambda
+\frac{4\lambda^2}{5(8\pi)^2N}\Big(
1-\frac{\lambda}{32\cdot 6!}\Big)+\mathcal{O}(\frac{1}{N^2})\,.
\end{align}
From this equation, we see that at $\lambda = 32\cdot 6!$ the theory has a large $N$ UV fixed point in $3d$.\footnote{Leading large $N$ expressions for several OPE coefficients and anomalous dimensions of operators $(\phi^2)^n$ for $n\geq 3$ at this 3$d$ UV fixed point were computed a few years ago in Ref.~\cite{goykhman2021background}, while Ref.~\cite{jack2020anomalous} determined the scaling dimension of $\phi^n$ to sub-leading order at large $n$.} In principle, the fixed point also exists away from three dimensions, but it becomes complex for $\epsilon>\frac{36}{\pi^2 N}$, so as we take $N$ to infinity, the range of dimensions where the fixed point is real shrinks to just $d=3$. 

The picture of the sextic $O(N)$ model in $3d$ as an asymptotically safe theory suggested by the above considerations, with the RG flow interpolating between a free theory in the IR and an interacting one in the UV, was however cast into doubt by the realization that the effective potential of the theory becomes unstable already for $\lambda > 30(8\pi)^2$ \cite{sarbach1978tricriticality,gudmundsdottir1984more,bardeen1984spontaneous} (note that $30(8\pi)^2<32\cdot 6!$), and a much more recent study of the theory in the presence of a planar boundary also points to an instability \cite{herzog2020n}. But the alternative picture that arose wherein the scale invariance is spontaneously broken before the UV fixed point is reached was itself called into question by the observation that subleading corrections endow the associated dilaton with an order $1/N$ mass that turns out to be tachyonic \cite{omid2016light}.
In brief, the fate of the theory in the UV is not known with certainty, but see Refs.~\cite{fleming2020finite,shrock2023study} for recent discussions. 

The incertitude surrounding the sextic $O(N)$ model exemplifies a broader lesson: once you go beyond the optimal scaling, you open a can of worms, and the large $N$ and small $\epsilon$ expansions generally cease to commute. But fortuitously, multi-index sextic theories differ from the vector model in that they feature loop-corrections not suppressed at large $N$ under the optimal scaling, entailing that these theories enjoy the benefit of possessing Wilson-Fisher that do not become complex at values of $\epsilon$ that scale as $1/N$. In the following subsections, we will inspect the Wilson-Fisher fixed points for the sextic theories of ranks two, three, and four. Whether these fixed points present in the $\epsilon$ expansion actually exist for finite values of $\epsilon$ is a separate issue. In the one case where the answer is known, it is in the affirmative: the non-trivial large $N$ rank-three fixed point can be continued along a finite range of dimensions, although not all the way down to two dimensions \cite{giombi2018prismatic}.

The beta functions and anomalous dimensions of the sextic theories we present have all been computed by modifying the sextic $O(N)$ theory four-loop beta function and six-loop anomalous dimensions calculated in Refs.~\cite{hager1999theta,hager2002six} via suitable index contractions performed on a computer.

\subsection{The \texorpdfstring{$O(N)^2$}{O(N)-squared} model}

The matrix theory with $O(N)^2/\mathbb{Z}_2$ symmetry has three sextic singlets:
\begin{align}
\label{sexticRank2Ops}
\begin{matrix}
\text{
		\includegraphics[scale=1]{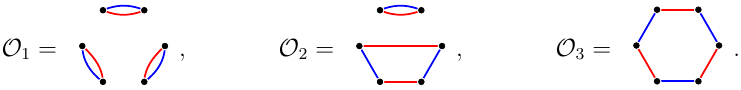}
}
\end{matrix}
\end{align}
In addition to $O(N)^2/\mathbb{Z}_2$ symmetry, the model also enjoys an automatic $S_2$ symmetry under permutations of the two indices---unlike for the higher rank models where permutation theory is not automatic. The action of the theory can be written as
\begin{align}
S = \int d^{3-\epsilon}x
\bigg(
\frac{1}{2}(\partial_i \phi^{ab})(\partial_i \phi^{ab})
+
\frac{1}{6!}\sum_{n=1}^3g_n\,\mathcal{O}_n
\bigg)\,.
\end{align}
The model was previously studied in Ref.~\cite{jepsen2021rg}, which computed the four-loop beta functions and studied the perturbative fixed points as functions of $N$. We will review only the large $N$ fixed points, referring the reader to \cite{jepsen2021rg} for details on the finite $N$ fixed points.\footnote{For the fixed points of the generalization of this model with $O(N_1)\times O(N_2)$ symmetry, see the recent Ref.~\cite{kapoor2023bifundamental}.} For $N> 23.2541-577.350\epsilon +\mathcal{O}(\epsilon^2)$, the beta functions have four fixed points including the trivial one. However, once we impose the optimal large $N$ scaling,
\begin{align}
g_1= \frac{\lambda_1}{N^4}\,, \hspace{20mm}
g_2= \frac{\lambda_2}{N^3}\,, \hspace{20mm}
g_3= \frac{\lambda_3}{N^2}\,,
\end{align}
we are left with but a single non-trivial fixed point in the large $N$ limit:
\begin{align}
&\lambda_1= \phantom{-}(8\pi)^2\cdot 6!\cdot\Big(\frac{295}{108}\epsilon+\frac{4714+6301\pi^2}{1944}\epsilon^2\Big)
+\mathcal{O}(\epsilon^3)\,,
\nonumber \\
&\lambda_2=-(8\pi)^2\cdot 6!\cdot\Big(\frac{\epsilon}{2}+\frac{22+7\pi^2}{36}\epsilon^2\Big)
+\mathcal{O}(\epsilon^3)\,,
\label{sexticRankTwoFP}
\\ \nonumber
&\lambda_3=\phantom{-}(8\pi)^2\cdot 6!\cdot\Big(\frac{1}{36}\epsilon+\frac{17+\pi^2}{324}\epsilon^2\Big)
+\mathcal{O}(\epsilon^3)\,.
\end{align}
The presence of two perturbative fixed points that exist for arbitrarily large $N$ but are suppressed away by the optimal scaling represents a new phenomenon compared with the quartic theories. Unlike the fixed point \eqref{sexticRankTwoFP}, these other two fixed points are not reliable in the limit of large $N$ and small $\epsilon$ unless we also impose that $\epsilon \ll 1/N$.

In the supplementary Mathematica file we display the six-loop anomalous dimensions for $\phi$ and $\phi^2$. Imposing the optimal scaling and taking the large $N$ limit yields the simpler equations
\begin{align}
&\gamma_\phi = 
\frac{48\lambda_3^2}{(6!)^2(8\pi)^4}
-\frac{1152\lambda_3^3}{(6!)^3(8\pi)^6}
+\mathcal{O}(\lambda^4)\,.
\\[6pt]
&\gamma_{\phi^2}=
32\frac{48\lambda_3^2}{(6!)^2(8\pi)^4}
-\frac{4608\big(\pi^2\lambda_2+(50+6\pi^2)\lambda_3\big)\lambda_3^2}{(6!)^3(8\pi)^6}
+\mathcal{O}(\lambda^4)\,.
\end{align}
At the large $N$ fixed point \eqref{sexticRankTwoFP}, these anomalous dimension evaluate to
\begin{align}
&\gamma_\phi \,\,= 
\frac{1}{27}\epsilon^2+\frac{2(14+\pi^2)}{243}\epsilon^3
+\mathcal{O}(\epsilon^4)\hspace{4mm}
\approx\,
0.0370370 \epsilon^2 + 0.196458 \epsilon^3
+\mathcal{O}(\epsilon^4)
\,,
\\[4pt]
&\gamma_{\phi^2}=
\frac{32}{27}\epsilon^2
+\frac{16(22\pi^2-7)}{243}\epsilon^3
+\mathcal{O}(\epsilon^4)
\approx\,
1.18519 \epsilon^2 + 13.8358 \epsilon^3
+\mathcal{O}(\epsilon^4)
\,.
\end{align}
The sizes of these coefficients do not kindle much hope that this $\epsilon$ expansion yields useful information about a putative 2$d$ CFT.

\subsection{The \texorpdfstring{$O(N)^3$}{O(N)-cubed} model}

\begin{figure}
    \centering
\begin{align*}
\begin{matrix}\text{
		\includegraphics[scale=0.95]{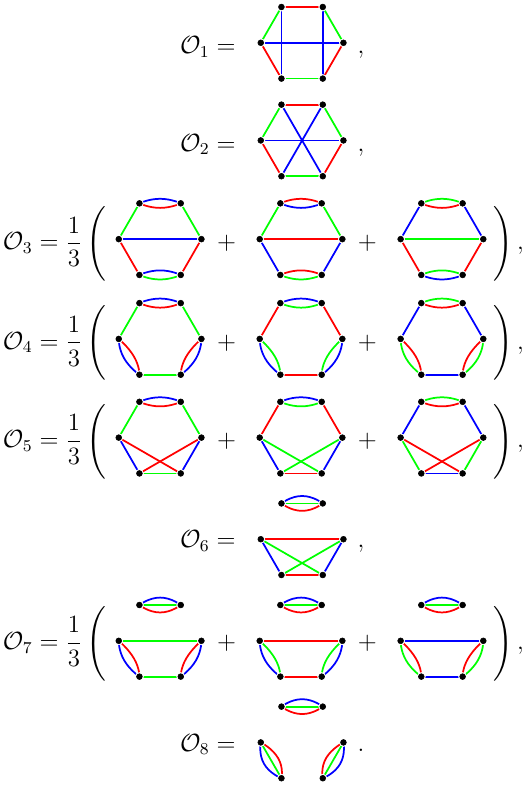}
}\end{matrix}
\end{align*}
    \caption{Sextic singlet operators in $O(N)^3\rtimes S_3$ theory}
    \label{fig:sexticRank3figs}
\end{figure}

The theory with $O(N)^3/(\mathbb{Z}_2)^2$ symmetry contains 16 sextic singlets. It turns out that under the optimal scaling the two-loop large $N$ beta functions of this model contain but a single large $N$ fixed point and that this fixed point exhibits an additional $S_3$ symmetry under permutations of tensor indices. To simplify the presentation, and without loss of large $N$ fixed points, we will therefore restrict our review of rank-three theory to the subsector that possesses this additional $S_3$ symmetry, whereby the 16-dimensional RG flow reduces to an eight-dimensional dynamical system. The action is given by
\begin{align}
\label{sexticRank3action}
S = \int d^{3-\epsilon}x
\bigg(
\frac{1}{2}(\partial_i \phi^{abc})(\partial_i \phi^{abc})
+
\frac{1}{6!}\sum_{n=1}^8g_n\,\mathcal{O}_n
\bigg)\,,
\end{align}
iwth the eight sextic operators in the potential shown in Figure~\ref{fig:sexticRank3figs}.
This precise model was studied in detail in Ref.~\cite{giombi2018prismatic}. The authors of this paper noted the presence of an RG fixed point dominated by the so-called prismatic operator $\mathcal{O}_1$ and demonstrated the separate solvability of this fixed point at large $N$ and at small $\epsilon$, while also observing that the large $N$ fixed point becomes complex for $d$ below approximately 2.81. The model was subsequently studied at four-loop level by Igor Klebanov, Fedor Popov, and one of the present authors in unpublished work, which we review below. The zero-dimensional version of the model was studied from a more mathematical perspective in Ref.~\cite{krajewski2023double}, which analyzed the double scaling limit of the prismatic model and identified the class of sub-leading diagrams at large $N$. 

The optimal scaling of the model \eqref{sexticRank3action} is given as follows:\footnote{For some of the couplings in the theory, Ref.~\cite{giombi2018prismatic} adopted large $N$ scaling exponents bigger than those in equation \eqref{sexticRank3scaling}. Their scaling enjoys the benefit of flushing out the order $\epsilon$ terms in the fixed point \eqref{rank4fixedPT} that are suppressed in $1/N$ but causes a blow-up of the four-loop beta functions at large $N$ unless we take $\epsilon$ to be suppressed in $1/N$.}
\begin{align}
&g_1=\frac{\lambda_1}{N^3}\,,
\hspace{15mm}
g_2=\frac{\lambda_2}{N^3}\,,
\hspace{15mm}
g_3=\frac{\lambda_3}{N^4}\,,
\hspace{15mm}
g_4=\frac{\lambda_4}{N^4}\,,
\nonumber \\[-8pt]
\label{sexticRank3scaling}
\\[-8pt] \nonumber
&g_5=\frac{\lambda_5}{N^{7/2}}\,,
\hspace{12mm}
g_6=\frac{\lambda_6}{N^{9/2}}\,,
\hspace{12mm}
g_7=\frac{\lambda_7}{N^5}\,,
\hspace{15mm}
g_8=\frac{\lambda_8}{N^6}\,.
\end{align}
Adopting this scaling, the two-loop beta functions have a unique large $N$ fixed point. Including four-loop corrections, this fixed point is situated at 
\begin{align}
&\lambda_1= 11\,520\pi^2\epsilon
-34\,560\pi^2\epsilon^2
+\mathcal{O}(\epsilon^3)\,,
\nonumber \\
&\lambda_2=\mathcal{O}(\epsilon^3)\,,
\nonumber \\
&\lambda_3=69\,120\pi^2\epsilon
+138\,240\pi^2\epsilon^2
+\mathcal{O}(\epsilon^3)\,,
\nonumber \\ 
&\lambda_4=-5760\pi^4\epsilon^2
+\mathcal{O}(\epsilon^3)\,,
\nonumber \\[-11pt]
\label{rank4fixedPT}
\\[-11pt]
&\lambda_5=\mathcal{O}(\epsilon^3)\,,
\nonumber \\
&\lambda_6=\mathcal{O}(\epsilon^3)\,,
\nonumber \\
&\lambda_7=34\,560\pi^2\epsilon
+34\,560(83\pi^2-\pi^4)\epsilon^2
+\mathcal{O}(\epsilon^3)\,,
\nonumber \\
&\lambda_8=
11\,520(126\pi^2-\pi^4)\epsilon^2
+\mathcal{O}(\epsilon^3)\,.
\nonumber
\end{align}
Under the optimal scaling, the large $N$ anomalous dimensions of $\phi$ and $\phi^2$ are given by
\begin{align}
&\gamma_\phi=
\frac{16}{(6!)^2(8\pi)^4}(\lambda_1^2+3\lambda_2^2)
-\frac{128}{3(6!)^3(8\pi)^6}\lambda_1^3
+\mathcal{O}(\lambda^4)\,,
\nonumber \\[-11pt]
\label{sexticRank3anomalous}
\\[-11pt] \nonumber 
&\gamma_{\phi^2}=
\frac{512}{(6!)^2(8\pi)^4}(\lambda_1^2+3\lambda_2^2)
-\frac{25\,600}{3(6!)^3(8\pi)^6}\lambda_1^3
+\mathcal{O}(\lambda^4)\,.
\end{align}
At the large $N$ fixed point \eqref{rank4fixedPT}, these anomalous dimensions assume the following values:
\begin{align}
\label{sexticGamma1}
&\gamma_\phi = \epsilon^2-\frac{20}{3}\epsilon^3+\mathcal{O}(\epsilon^4)\,,
\\
&\gamma_{\phi^2} = 32\epsilon^2-\frac{976}{3}\epsilon^3+\mathcal{O}(\epsilon^4)\,.
\label{sexticGamma2}
\end{align}
These values match the large $N$ solution that Ref.~\cite{giombi2018prismatic} obtained by expressing the exponential of the prismatic operator as an integral over an auxiliary field and then writing down a pair of coupled Schwinger-Dyson equations for the fundamental and auxiliary fields. Specifically, the authors of \cite{giombi2018prismatic} found that
\begin{align}
& \Delta_\phi = \frac{1}{2}-\frac{\epsilon}{2}+\epsilon^2-\frac{20\epsilon^3}{3}
+\Big(\frac{472}{9}+\frac{\pi^2}{3}\Big)\epsilon^4
+\Big(
7\zeta(3)-\frac{12\,692}{27}
-\frac{56\pi^2}{9}
\Big)\epsilon^5
+\mathcal{O}(\epsilon^6)\,,
\\
&
\Delta_{\phi^2}
=1-\epsilon+32\epsilon^2
-\frac{976}{3}\epsilon^3
+\Big(\frac{30\,320}{9}+\frac{32\pi^2}{3}\Big)\epsilon^4+\mathcal{O}(\epsilon^5)\,.
\end{align}
The presence of both $\lambda_1$ and $\lambda_2$ in equations \eqref{sexticGamma1} and \eqref{sexticGamma2} suggests that both the prismatic operator $\mathcal{O}_1$ and the wheelic operator $\mathcal{O}_2$ can dominate at large $N$. And indeed Ref.~\cite{prakash2020melonic} established that for precisely these two rank-three operators, the Feynman diagrams that contribute to the free energy at large $N$ are melonic and can be resummed. We may ask then, does the model \eqref{sexticRank3action} posses a whelic fixed point? In this paper we have so far listed only standard type Wilson-Fisher fixed points, which rely on a cancellation between linear and quadratic terms in beta functions, so that the leading coupling constants scale as $\epsilon$, with higher order terms in the beta functions giving rise to $\mathcal{O}(\epsilon^2)$ corrections. But a wheelic fixed point should match the $q=6$ melonic theory described in Ref~\cite{giombi2017bosonic}, whose equations (4.3) and (5.1) tell us that
\begin{align}
& \Delta_\phi = \frac{d}{6} \,,
\\
& \Delta_{\phi^2}
=1+\frac{29}{3}\epsilon
+\frac{400}{9}\epsilon^2
+\frac{160}{27}(237+2\pi^2)\epsilon^3+\mathcal{O}(\epsilon^4)\,,
\end{align}
corresponding to anomalous dimensions given by
\begin{align}
&\gamma_\phi= \Delta_\phi - \frac{d-2}{2}
=\frac{\epsilon}{3}\,,
\nonumber \\[-11pt]
\label{melonicGamma}
\\[-11pt] \nonumber
&\gamma_{\phi^2}= \Delta_{\phi^2} - (d-2)
= \frac{32}{3}\epsilon
+\mathcal{O}(\epsilon^2)\,.
\end{align}
Since the perturbative anomalous dimensions in a sextic theory scale as $\lambda^2$, a wheelic fixed point should manifest itself in pertubation theory as having couplings with leading scaling $\sqrt{\epsilon}$, meaning that such a fixed point arises from a cancellation between linear and cubic terms in the four-loop beta functions. And in fact, the model \eqref{sexticRank3action} does posses a pair of such fixed points:
\begin{align}
&\lambda_1 = 0\,,
\nonumber \\
&\lambda_2 = \pm 3840\pi^2\sqrt{\epsilon}\,,
\nonumber  \\
&\lambda_3 = -17\,280\pi^2(1\pm 2\sqrt{\epsilon})\,,
\nonumber  \\
&\lambda_4 = 0\,,
\label{wheelicPt}
\\
&\lambda_5 = 0\,,
\nonumber \\
&\lambda_6 = 0\,,
\nonumber \\
&\lambda_7 = \frac{34\,560}{7}\pi^2(5\pm 7\sqrt{\epsilon})\,,
\nonumber  \\
&\lambda_8 = -\frac{640}{7}
\pi^2(109\pm 126\sqrt{\epsilon})\,.
\nonumber 
\end{align}
However, since some of the couplings contain terms of order $O(\epsilon^0)$, we cannot trust the perturbative analysis that revealed the above pair of fixed points, irrespectively of how small a value we assign to $\epsilon$. Higher-order terms are liable to change even leading values of the above couplings. But it is compelling to think that the values for $\lambda_1$ and $\lambda_2$ in \eqref{wheelicPt} are robust and that the non-robust couplings do not enter into the large $N$ anomalous dimensions. In any event, plugging the values \eqref{wheelicPt} into \eqref{sexticRank3anomalous}, one does find agreement with \eqref{melonicGamma}.\footnote{Melonic fixed points with order $O(\epsilon^0)$ terms were also observed for the sextic theory with $U(N)^3$ symmetry studied at two loops in Ref.~\cite{benedetti2020sextic} and at four loops in Ref.~\cite{harribey2022sextic}.}

\subsection{The \texorpdfstring{$O(N)^4\rtimes S_4$}{O(N)-to-the-fourth times S4} model}

\begin{figure}
    \centering
\begin{align*}
\begin{matrix}\text{
		\includegraphics[scale=0.95]{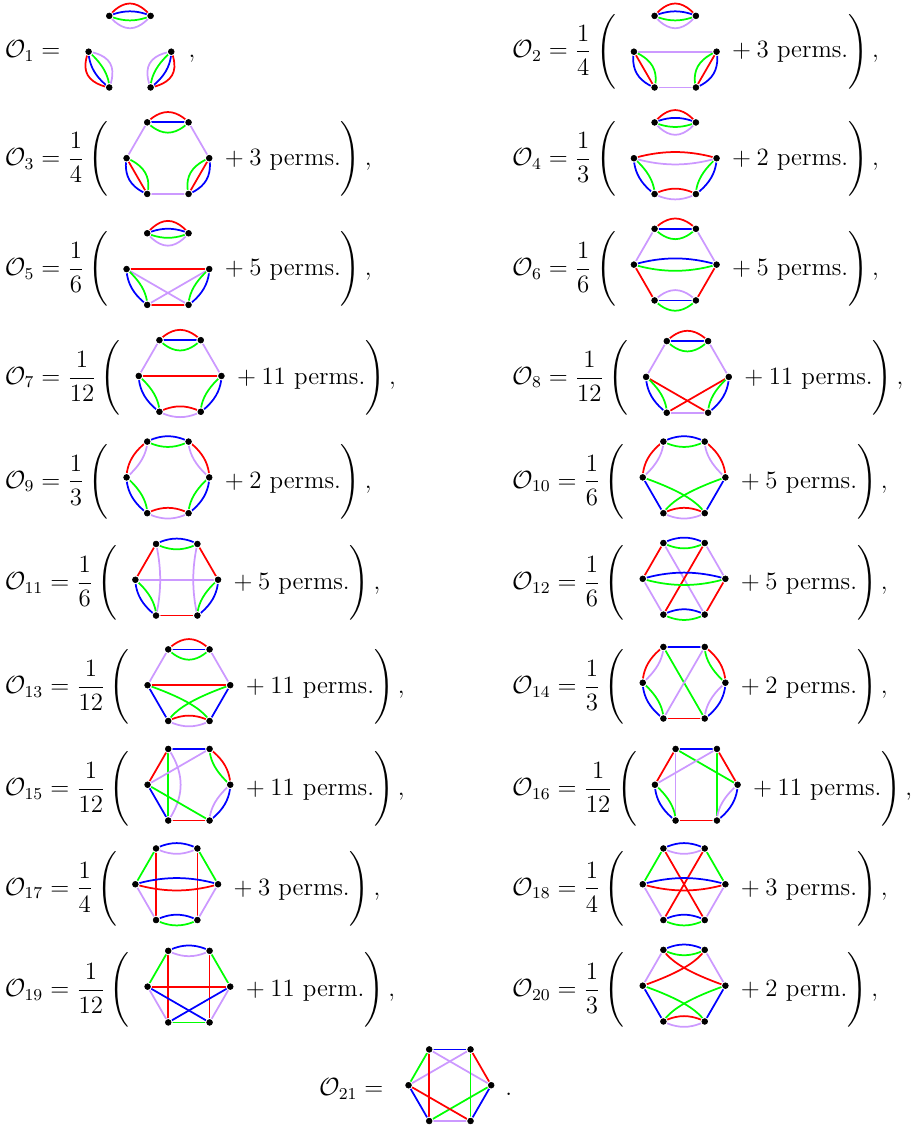}
}\end{matrix}
\end{align*}
    \caption{Sextic singlet operators in $O(N)^4\rtimes S_4$ theory}
    \label{fig:sexticRank4figs}
\end{figure}

We now turn to sextic scalar theory of rank-four tensors. The full theory with $O(N)^4/(\mathbb{Z}_2)^3$ symmetry contains 132 sextic singlet operators. Computing the two-loop beta functions for this system is manageable, but determining the number of fixed points for such a large set of coupled equations is challenging. We therefore impose a global $S_4$ symmetry under permutations of indices, whereby the 132-dimensional RG flow reduces to a more manageable 21-dimensional invariant subspace associated to the operators show in Figure~\ref{fig:sexticRank4figs}.
The reader may observe that the operators $\mathcal{O}_{12}$ and $\mathcal{O}_{18}$ differ only by their colouring. Mathematically speaking, the four-regular graph of order six that is associated to these operators admits two distinct edge colourings not related by permutations of colours or vertices. Physically, this means these operators would not be distinct in a theory of symmetric four-tensors.

The action of the theory we now consider is given explicitly by
\begin{align}
\label{sexticRank4action}
S = \int d^{3-\epsilon}x
\bigg(
\frac{1}{2}
(\partial_i\phi^{abcd})(\partial_i\phi^{abcd})
+\frac{1}{6!}\sum_{n=1}^{21}
g_n\, \mathcal{O}_n
\bigg)\,.
\end{align}
The four-loop beta functions for this theory are listed in the supplementary Mathematica file. They are quite lengthy but simplify somewhat in the large $N$ limit. The optimal scalings associated to the operators are given by
\begin{align}
g_i= \frac{\lambda_i}{N^{e_i}}\,,
\end{align}
where the exponents take on the following values:
\begin{align}
e=\{8, 7, 6, 6, 6, 6, 5, 5, 4, 4, 4, 4, 5, 4, 4, 4, 5, 5, 4, 4, 4\}\,.
\end{align}
Under this scaling, the four-loop anomalous dimensions for $\phi$ and $\phi^2$ are given at large $N$ by
\begin{align}
& \gamma_{\phi}^{(\text{four-loop})}=2\frac{24 \lambda_9^2 + 4 \lambda_{10}^2 + 4 \lambda_{11}^2 + 12 \lambda_{12}^2 + 
 4 \lambda_{14}^2 + \lambda_{15}^2 + \lambda_{16}^2 + 2 \lambda_{19}^2 + 
 8 \lambda_{20}^2 + 12 \lambda_{21}^2}{3(6!)^2(8\pi)^4}
\,,
\nonumber
\\
& \gamma_{\phi^2}^{(\text{four-loop})}=32\gamma_{\phi}^{(\text{four-loop})}\,.
\end{align}
The six-loop anomalous dimensions are listed in the supplementary Mathematica file. Even at large $N$ they are rather lengthy, but we note that $\gamma_{\phi}^{(\text{six-loop})}$ only depends on the same coupling constants $\lambda_i$ as $\gamma_{\phi}^{(\text{four-loop})}$ does, whereas $\gamma_{\phi^2}^{(\text{six-loop})}$ introduces dependencies on $\lambda_4$, $\lambda_5$, $\lambda_7$, $\lambda_8$,  and $\lambda_{13}$.

By setting to zero the large $N$ beta functions to quadratic order, we can determine the perturbative fixed points with couplings proportional to $\epsilon$. This amounts to solving 21 coupled quadratic equations. It can be directly seen from the large $N$ beta functions that $\lambda_{12}$, $\lambda_{19}$, and $\lambda_{21}$ must all be zero to leading order in $\epsilon$ at any fixed point. This leaves 18 coupled equations, which are still difficult to solve analytically. However, numerical evidence strongly suggests that $\lambda_{10}$ must also vanish on the fixed points, and once we enforce the constraint $\lambda_{10}=0$, we are able to solve for the two-loop fixed points analytically. We find a total of 18 perturbative fixed points, which we will label with Roman numerals (i) to (xviii), where (i) is the trivial fixed point. The values of the couplings at each of the fixed points (ii) to (xviii) are listed to leading order in $\epsilon$ in Appendix~\ref{sec:Locations}. The order $\epsilon^2$ corrections can be found in the supplementary Mathematica file. These corrections are required to compute the anomalous dimensions at the fixed points up to order $\epsilon^3$, the results of which computations we display below. For fixed points (ii) to (x), we find that
\begin{align}
&\gamma_{\phi}^{\text{(ii)}}=\gamma_{\phi}^{\text{(iii)}}=\frac{15}{16}\epsilon^2
-\frac{683-3\pi^2}{128}\epsilon^3+\mathcal{O}(\epsilon^4)\,,
\nonumber\\[-11pt]
\\[-11pt] \nonumber
&
\gamma_{\phi^2}^{\text{(ii)}}=\gamma_{\phi^2}^{\text{(iii)}}=30 \epsilon^2 - \frac{1}{4} (1103 + 51 \pi^2) \epsilon^3+\mathcal{O}(\epsilon^4)\,,
\\[8pt]
&
\gamma_{\phi}^{\text{(iv)}}=6\epsilon^2
-(328-6\pi^2)\epsilon^3+\mathcal{O}(\epsilon^4)\,,
\nonumber\\[-11pt]
\\[-11pt] \nonumber
&\gamma_{\phi^2}^{\text{(iv)}}=
192 \epsilon^2 - 32 (349 + 21 \pi^2) \epsilon^3+\mathcal{O}(\epsilon^4)\,,
\\[8pt]
&\gamma_{\phi}^{\text{(v)}}=\frac{1}{9}\epsilon^2
+\frac{2}{81}(12+\pi^2)\epsilon^3+\mathcal{O}(\epsilon^4)\,,
\nonumber\\[-11pt]
\\[-11pt] \nonumber
&\gamma_{\phi^2}^{\text{(v)}}=
\frac{32}{9} \epsilon^2 + \frac{16}{81} (-15 + 22 \pi^2) \epsilon^3+\mathcal{O}(\epsilon^4)\,,
\\[8pt]
&\gamma_{\phi}^{\text{(vi)}}=\gamma_{\phi}^{\text{(vii)}}=\frac{151}{144}\epsilon^2
-\frac{7(10797-133\pi^2)}{10368}\epsilon^3+\mathcal{O}(\epsilon^4)\,,
\nonumber\\[-11pt]
\\[-11pt] \nonumber
&\gamma_{\phi^2}^{\text{(vi)}}=\gamma_{\phi^2}^{\text{(vii)}}=
\frac{302}{9} \epsilon^2 + \frac{49}{324} (-2319 + 469 \pi^2) \epsilon^3+\mathcal{O}(\epsilon^4)\,,
\\[8pt]
&
\gamma_{\phi}^{\text{(viii)}}=\frac{55}{9}\epsilon^2
-\frac{64}{81}(435-11\pi^2)\epsilon^3+\mathcal{O}(\epsilon^4)\,,
\nonumber\\[-11pt]
\\[-11pt] \nonumber
&
\gamma_{\phi^2}^{\text{(viii)}}=
\frac{1760}{9} \epsilon^2 + \frac{80}{81} (-11829 + 3104 \pi^2) \epsilon^3+\mathcal{O}(\epsilon^4)\,,
\\[8pt]
&
\gamma_{\phi}^{\text{(ix)}}=\gamma_{\phi}^{\text{(x)}}=\frac{1395}{1331}\epsilon^2
-\frac{6(261739892 - 2985147\pi^2)}{214358881}\epsilon^3+\mathcal{O}(\epsilon^4)\,,
\nonumber\\[-11pt]
\\[-11pt] \nonumber
&
\gamma_{\phi^2}^{\text{(ix)}}=\gamma_{\phi^2}^{\text{(x)}}=
\frac{44640}{1331} \epsilon^2 + 
 \frac{48 (-3580720525867 + 898956393598 \pi^2)}{488523889799} \epsilon^3+\mathcal{O}(\epsilon^4)\,.
\end{align}
For fixed points (xi) to (xviii) the coefficients of $\epsilon^2$ and $\epsilon^3$ in $\gamma_\phi$ and $\gamma_{\phi^2}$ invole roots of quartic polynomials, so we will merely display their numerical values, which we list Table~\ref{tab:anomDims}. For comparison, we also include the numerical values for the coefficients at the other fixed points.
\begin{table}
\centering
\scalebox{0.67}{
\renewcommand{\arraystretch}{1.5}
\begin{tabular}{|c|c|c|} 
 \hline
 fixed point & $\gamma_\phi-\mathcal{O}(\epsilon^4)$ & $\gamma_{\phi^2}-\mathcal{O}(\epsilon^4)$ 
\\
 \hline
\cellcolor{lightgray2}(i) & \cellcolor{lightgray2}0 & \cellcolor{lightgray2}0
 \\
 \hline
(ii)  
&$0.9375 \epsilon^2 - 5.1046186 \epsilon^3$
&$30 \epsilon^2 - 401.58746 \epsilon^3$ 
 \\
 \hline
\cellcolor{lightgray2}(iii)
 &\cellcolor{lightgray2}$0.9375 \epsilon^2 - 5.1046186 \epsilon^3$
 &\cellcolor{lightgray2}$30 \epsilon^2 - 401.58746 \epsilon^3$ 
 \\
 \hline
(iv)
&$6 \epsilon^2 - 268.78237 \epsilon^3$
&$192 \epsilon^2 - 17800.374 \epsilon^3$
 \\
 \hline
\cellcolor{lightgray2}(v)  
 &\cellcolor{lightgray2}$0.11111111 \epsilon^2 + 0.53999023 \epsilon^3$
 &\cellcolor{lightgray2}$3.5555556 \epsilon^2 + 39.927170 \epsilon^3$ 
 \\
 \hline
(vi)
&$1.0486111 \epsilon^2 - 6.4033949 \epsilon^3$
&$33.555556 \epsilon^2 + 349.32833 \epsilon^3$
 \\
 \hline
\cellcolor{lightgray2}(vii)
 &\cellcolor{lightgray2}$1.0486111 \epsilon^2 - 6.4033949 \epsilon^3$
 &\cellcolor{lightgray2}$33.555556 \epsilon^2 + 349.32833 \epsilon^3$
 \\
 \hline
(viii)
&$6.1111111 \epsilon^2 - 257.92344 \epsilon^3$
&$195.55556 \epsilon^2 + 18574.076 \epsilon^3$
 \\
 \hline
\cellcolor{lightgray2}(ix) 
 &\cellcolor{lightgray2}$1.0480841 \epsilon^2 - 6.5015549 \epsilon^3$
 &\cellcolor{lightgray2}$33.538693 \epsilon^2 + 519.92939 \epsilon^3$
 \\
 \hline
(x) 
&$1.0480841 \epsilon^2 - 6.5015549 \epsilon^3$
&$33.538693 \epsilon^2 + 519.92939 \epsilon^3$ 
 \\
 \hline
\cellcolor{lightgray2}(xi)
 &\cellcolor{lightgray2}$4.8136566  \epsilon^2-177.89072\epsilon^3$
 &\cellcolor{lightgray2}$154.03701 \epsilon^2 - 11073.347 \epsilon^3$
 \\
 \hline
(xii) 
&$0.85551344\epsilon^2-4.4862499 \epsilon^3$
&$27.376430 \epsilon^2 - 342.60322 \epsilon^3$
 \\
 \hline
\cellcolor{lightgray2}(xiii)  
 &\cellcolor{lightgray2}$0.62994776\epsilon^2-2.6804986\epsilon^3$
 &\cellcolor{lightgray2}$20.158328 \epsilon^2 - 205.70175 \epsilon^3$
 \\
 \hline
(xiv)
&$3.1684748\epsilon^2-78.122861\epsilon^3$
&$101.39119 \epsilon^2 - 4647.5684 \epsilon^3$
 \\
 \hline
\cellcolor{lightgray2}(xv) 
 &\cellcolor{lightgray2}$4.9247677\epsilon^2-175.81337\epsilon^3$
 &\cellcolor{lightgray2}$157.59257 \epsilon^2 + 9810.5193 \epsilon^3$
 \\
 \hline
(xvi)
&$0.96662455 \epsilon^2-5.7167280 \epsilon^3$
&$30.931986 \epsilon^2 + 262.06325 \epsilon^3$ 
 \\
 \hline
\cellcolor{lightgray2}(xvii) 
 &\cellcolor{lightgray2}$0.74105887\epsilon^2-3.5240650\epsilon^3$
 &\cellcolor{lightgray2}$23.713884 \epsilon^2 + 131.62751 \epsilon^3$
 \\
 \hline
(xviii)
&$3.2795859\epsilon^2
-80.724969\epsilon^3$
&$104.94675 \epsilon^2 + 3331.7552 \epsilon^3$
\\
\hline
\end{tabular}}
\caption{Table of anomalous dimensions at the large $N$ fixed points of sextic $O(N)^4\rtimes S_4$ theory}
    \label{tab:anomDims}
\end{table}
Based on the sizes of the coefficients in Table~\ref{tab:anomDims}, it is doubtful that the fixed points extend all the way down to $d=2$, nor does it bode well that the $\epsilon^3$ coefficients are everywhere larger than the $\epsilon^2$ coefficients.

\subsubsection{Degeneracy of fixed points}

The fixed point (iv) stands out from the others. It is special in that a one-parameter degeneracy in the values of the couplings emerges at order $\epsilon^2$, as we show explicitly in Appendix~\ref{sec:Locations}. If this one-dimensional continuum of fixed points in the $\epsilon$ expansion were to be meaningfully continued to an integer dimension, the degeneracy would be indicative of the presence of a conformal manifold. The degeneracy may be a perturbative artefact, but either way higher-order corrections in $\epsilon$ or in $1/N$ will not lift the degeneracy for the simple reason that any linearized perturbation at a given order in an expansion always produces 21 equations, $0=\delta \beta_{\lambda_n}$, in 21 unknowns, $\delta \lambda_n$.

\subsubsection{Melonic fixed points?}
As for the rank-three model, we may ask of the sextic rank-four model if it possesses fixed points with couplings that scale as $\sqrt{\epsilon}$. In fact, for the so-called subchromatic sextic theories, ie. the sextic theories of tensors with rank less than five, Ref.~\cite{prakash2020melonic} identified a total of three operators whose contributions to the free energy are melonic and for this reason can be resummed: the prismatic and wheelic operators $\mathcal{O}_1$ and $\mathcal{O}_2$ of the rank-three model and also what they referred to as the octahedral operator, namely $\mathcal{O}_{21}$ of the rank-four model. To search for a fixed point dominated by the octahedron, we set to zero the couplings for the non-octahedral operators that also have a large $N$ scaling exponent of four, ie. we set
\begin{align}
\label{lambdaVan}
\lambda_9=\lambda_{10}=\lambda_{11}=\lambda_{12}=\lambda_{14}=\lambda_{15}=\lambda_{16}=\lambda_{19}=\lambda_{20}=0\,.
\end{align}
Imposing the condition \eqref{lambdaVan}, the octahedral beta function simplifies to
\begin{align}
\label{lambda21}
\beta_{\lambda_{21}}
=-2\epsilon \lambda_{21}
+\frac{48}{(6!)^2(8\pi)^4}\lambda_{21}^3
+\mathcal{O}(\lambda^4)\,.
\end{align}
Plugging either of the fixed point values $\lambda_{21}=\pm 3840\sqrt{6\epsilon}\pi^2$ into the expressions for the six-loop mass and field anomalous dimensions, while still imposing \eqref{lambdaVan}, we recover the melonic values for a sextic theory:
\begin{align}
\gamma_\phi=\frac{\epsilon}{3}+\mathcal{O}(\lambda^4)\,,
\hspace{15mm}
\gamma_{\phi^2}=\frac{32}{3}\epsilon+\mathcal{O}(\lambda^4)\,.
\end{align}
However, in order for the model to posses a genuine fixed point, it is of course necessary that all beta functions vanish, not just $\beta_{\lambda_{21}}$. And it can be checked that on imposing  \eqref{lambdaVan} and setting $\lambda_{21}=\pm 3840\sqrt{6\epsilon}\,\pi^2$, the beta function for the 17th coupling equals 
\begin{align}
\label{lambda17val}
\beta_{\lambda_{17}}
=30\,720\pi^2\epsilon +\mathcal{O}(\lambda^4)\,.
\end{align}
We conclude that the operators whose couplings exhibit a greater large $N$ suppression than that of the octahedron destroy the stability of the melonic fixed point, unless we accept the exotic possibility of fixed points with couplings that scale as $\lambda\sim \epsilon^{1/4}$ or with even smaller fractional powers of epsilon.

\subsection{Overview and discussion}

In Table~\ref{tab:sexticOverview} we present a summary of the number of perturbative fixed points in sextic scalar theories after applying the optimal scalings and taking the large $N$ limit. This tally includes only such fixed points as manifest themselves already at two-loop level, including the trivial fixed point. In contrast to the case of quartic interactions, the sextic vector model does not posses a large $N$ perturbative fixed point under the optimal scaling. At ranks two and three, a single such fixed point appears, while at rank four, even after restricting ourselves to the subsector that is symmetric with respect to permutations of indices, we observe a sudden proliferation, with 18 fixed points present. This number might be significantly higher for the theory without $S_4$ permutation theory, but the computation to check whether this is so or not is hindered by the fact that $n_\text{sextic}=132$ for the theory with only $O(N)^4/(\mathbb{Z}_2)^3$ symmetry. Similarly, in attempting to extend the above table to higher ranks, one encounters an obstruction in the fact that $n_\text{sextic}=52$ for the theory with $S_5 \ltimes O(N)^5/(\mathbb{Z}_2)^4$ symmetry, while $n_\text{sextic}=1439$ for the theory with only $O(N)^5/(\mathbb{Z}_2)^4$ symmetry. Unlike the quartic case, we do not have formulas for the number of operators---let alone the numbers of fixed points---at arbitrary rank, ultimately for the reason that the set of graphs of order six is vastly more complicated than that of order-four graphs. Nor are we able to present general large $N$ solutions as we were for the quartic theories, owing to the fact that while the quartic theories were exclusively dominated by vectorial operators, the dominant operators for the sextic theories also include melonic and matrix-like operators. Purely melonic theories are solvable, and Ref.~\cite{giombi2018prismatic} successfully obtained a large $N$ solution of the non-trivial rank-three fixed point in Table~\ref{tab:sexticOverview}, a fixed point dominated by the the prismatic operator $\mathcal{O}_1$. This is likely to be the only non-trivial fixed point in the table to admit such an elegant large $N$ solution. All the other fixed points in the table contain non-zero matricial couplings among the coupling constants with the smallest suppression in $N$. This entails that the classes of diagrams that dominate at large $N$ include the set of all planar diagrams built out of those matricial operators, and the set of planar diagrams is much more unwieldy than its melonic subset. 
\begin{table}
\centering
$\hspace{-4mm}$\scalebox{1}{
\renewcommand{\arraystretch}{1.3}
\begin{tabular}{|l|c|c|c|c|c|} 
\hline
global symmetry 
&\cellcolor{lightgray2}$O(N)$ 
&$\frac{O(N)^2\rtimes S_2}{\mathbb{Z}_2}$ 
&\cellcolor{lightgray2}$\frac{O(N)^3}{(\mathbb{Z}_2)^2}$ 
&$\frac{O(N)^3\rtimes S_3}{(\mathbb{Z}_2)^2}$
&\cellcolor{lightgray2}$\frac{O(N)^4\rtimes S_4}{(\mathbb{Z}_2)^3}$
\\
\hline
$n_{\text{sextic}}$
&\cellcolor{lightgray2} 1
& 3
&\cellcolor{lightgray2} 16
& 8
&\cellcolor{lightgray2} 21
\\
\hline
$n_\text{fixed points}$
&\cellcolor{lightgray2} 1
& 2
&\cellcolor{lightgray2} 2
& 2
&\cellcolor{lightgray2} 18
\\
\hline
\end{tabular}
}
\caption{Numbers of operators and fixed points in sextic $O(N)^r\rtimes S_r$ theory}
    \label{tab:sexticOverview}
\end{table}

Of all the operators in the sub-chromatic sextic theories we have reviewed, Ref.~\cite{prakash2020melonic} identified three for which the sets of diagrams constructed exclusively out of each of these can be resummed. Besides the prismatic operator, these are the rank-three operator $\mathcal{O}_2$ and the rank-four operator $\mathcal{O}_{21}$. But as we have seen, the melonic fixed points associated to these operators, if they are perturbatively accessible, rely on cancellations between linear and cubic terms in beta functions and require some coupling constants to scale unusually with $\epsilon$, such as $\lambda\sim \epsilon^0$ or $\lambda\sim \epsilon^{1/4}$, and they are not among the Wilson-Fisher fixed points listed in Table~\ref{tab:sexticOverview}.

Moving on from sub-chromatic sextic theories to sextic rank-five tensor theory, one encounters the kind of melonic interaction present in the original Klebanov-Tarnopolsky models proposed in Ref.~\cite{klebanov2017uncolored} and further studied in Ref.~\cite{giombi2017bosonic}, which determined an explicit $d$-dependent melonic large $N$ solution. For their solution, the quadratic operator $\phi^{abcde}\phi^{abcde}$ has a real dimension only for $d_\text{crit}<d \leq 3$ with $d_\text{crit}\approx 2.97$. An analysis of the theory through the lense of renormalized perturbation theory, incorporating the operator mixing of different sextic operators, was carried out in Ref.~\cite{benedetti2020sextic}, which circumvented the issue of the multitudinous sextic rank-five operators by computing the beta functions only for the subset of operators relevant to their analysis and only at large $N$.  Through this computation, the authors arrived at the conclusion that there is no rank-five melonic fixed point in the $\epsilon$ expansion after noticing the same phenomenon as we observed in the rank-four theory with equation \eqref{lambda17val}: the fixed point value for the melonic coupling induces a linear term in another beta function that cannot be cancelled to third order in perturbation theory. Again the presence of couplings with scalings like $\lambda\sim \epsilon^0$ or $\lambda\sim \epsilon^{1/4}$ provide a possible resolution to the issue. If scalings $\lambda\sim \epsilon^0$ are the true reason melonic sextic fixed points elude the $\epsilon$ expansion, then it signifies an intriguing scenario where, as we vary $d$, the melonic theory in $d_\text{crit}<d < 3$ interpolates, not to a trivial theory in 3$d$ in the way that the Ising model interpolates to a trivial theory in $4d$, but to an interacting yet non-melonic three-dimensional theory, which however cannot be accessed through perturbation theory. A less interesting possibility is that a melonic fixed point simply does not exist in this theory.

\section{Outlook}
\label{sec:Outlook}

In this paper we have presented studies of tensorial $\phi^4$ and $\phi^6$ theories with multiple copies of $O(N)$ symmetry, applying small $\epsilon$ and large $N$ methods. As the behaviour of the quartic and sextic models differ drastically, each class of models offers its own set of lessons and further questions. 

The sextic theories are by far the least understood. While the strict large $N$ limit of the sextic $O(N)$ model yields a trivial beta function, the same no longer applies in the multi-index cases. Instead, the large $N$ dynamics exhibit an admixture of vectorial, matricial, and melonic operators. Through our study of sextic $O(N)^4$ theory, we have shown that rank-four tensor theory marks the onset of a sudden increase in the number of perturbative RG fixed points at large $N$ and in the diverseness among them---with couplings given in some cases by simple expressions, in others by perplexingly lengthy ones, and with one perturbative fixed point at subleading order degenerating into a one-parameter family of continuously connected fixed points. Which of these fixed points in the $\epsilon$ expansion admit large $N$ solutions exact in $d$, and for which ranges of dimensions they exist, are questions that remain to be answered. Higher rank theories are bound to produce even more fixed points and yet wider variety among them, but the numbers, types, and ranges of validity of fixed points for general tensor rank elude classification, as the combinatorial explosion of graphs of order six with increasing size quickly stymies searches for underlying patterns.

Contrariwise, for quartic orthogonal tensor theories, owing to the relative simplicity of graphs of order four, we have been able to provide large $N$ solutions of the $O(N)^r$ theories with $r\leq 5$, and to chart out a vast general family of large $N$ fixed points present at any $r$, all of vector-model type, meaning that the dominant Feynman graphs are bubble diagrams. The large $N$ and small $\epsilon$ expansions encouragingly indicate that the fixed points do extend to bona fide $3d$ CFTs, meaning that the critical $O(N)$ model admits a vast generalization to critical $O(N)^2$ models, critical $O(N)^3$ models, etc. Note, however, that our results are all perturbative, either in $1/N$ or in $\epsilon$, and so more work is needed to firmly establish these wider classes of critical models. If this can be achieved, the most intriguing outlook would be a future experimental realization of these prospective CFTs, but the fact that a large number of relevant deformations exist for most of the fixed point poses a serious challenge. On the theoretical side, the $O(N)$ model may just be the single-most studied of all quantum field theories, and if the $O(N)^r$ models indeed, as our study suggests, furnish separate instances of slightly more complicated, but still entirely tractable, relatives of the beloved $O(N)$ model, then all of the QFT phenomena that have been investigated with the $O(N)$ model as a test case can be extended to the larger playfield of the $O(N)^r$ models. Specific avenues for future inquiry include the following:
\begin{itemize}
\item It would be desirable to verify or refute the existence of the vectorial fixed points in three dimensions and at finite $N$ with non-perturbative methods like Monte Carlo simulations or the numerical bootstrap. 
\item 
The critical $O(N)$ model can be studied in $2+\epsilon$ dimensions through a perturbative analysis of the non-linear sigma model \cite{polyakov1975interaction,brezin1976renormalization,bardeen1976phase,brezin1976renormalization,brezin1976spontaneous,hikami1978three}. The status of $O(N)^r$ CFTs would be substantially solidified if $2+\epsilon$ expansions could likewise be developed for these models.
\item  It remains to be investigated whether the large $N$ fixed points we have studied are stable at finite temperature. In the case of the $O(N)$, the free energy at finite temperature was computed to leading order in large $N$ by Ref.~\cite{sachdev1993polylogarithm}, while the subleading correction was computed by Ref.~\cite{chubukov1994theory} and, recently and with increased precision, by Ref.~\cite{diatlyk2023beyond}. As explained in the latter reference, the value of the subleading correction hinges on a delicate renormalization of the fluctuations of the $\sigma$ field, and similar analyses may be possible for the additional Hubbard-Stratonovich fields $\chi$, $\tau$, $\rho$, etc. present in the tensorial theories.
\item In the $4-\epsilon$ expansion, there exist, in addition to the exhaustive list of fixed points for four- and five-component scalar theories, several known infinite families, and it is conceivable that in some cases, for intermediate values of $N$, the tensorial fixed points we studied in Section~\ref{sec:Quartic} can be identified with members of the known families, such as the Paulos fixed points \cite{grinstein1982stable,michel1984symmetry,shpot1988critical,shpot1989critical,osborn2018seeking}, which generalize the MN fixed points. A comprehensive classification of fixed points in the $\epsilon$ expansion requires determining which, if any, of the $O(N)^r$ fixed points that can secretly be identified with known fixed points.
\item Besides the large $N$ fixed points, the perturbative expansions of the quartic $O(N)^r$ theories reveal the presence of numerous fixed points at the specific value of $N=2$. It seems likely that these $O(2)^r$ fixed points can be classified for general values of $r$.
\item The singlet-sector of the critical $O(N)$ model was famously conjectured in Ref.~\cite{klebanov2002ads} to be dual to a higher-spin Vasiliev theory, and an abundance of convincing evidence has since been discovered in favour of the conjecture. How does this duality generalize to tensorial models?\footnote{Initial investigations in this direction can be found in Ref.~\cite{de2020holography}.}
\item It has been known for decades that on sending $N$ to $-N$, the Feynman diagrams in theories with $O(2N)$ symmetry are transformed into Feynman diagrams of theories with $Sp(2N)$ symmetry \cite{mkrtchyan1981equivalence,cvitanovic1982spinors}. This relation extends from vector models to matrix models \cite{mulase2003duality} and to tensor models \cite{gurau2022duality,keppler2023duality}. Consequently, for all the models studied in this paper, the perturbative beta functions and anomalous dimensions for the symplectic versions of the theories are immediately available, so that their fixed points in the $\epsilon$ expansion and their large $N$ limits can be readily analyzed. When the tensor rank is even, the symplectic theory is a theory of commuting scalars, while for odd tensor rank the scalars in the symplectic theory anti-commute, as required for the kinetic term to be non-vanishing. Therefore, the odd rank theories with $Sp(2N)$ symmetry are non-unitary, but it can nonetheless be checked that their perturbative RG flows are gradient and monotonic. And as pointed out in Ref.~\cite{anninos2016higher}, the CFTs relevant to dS/CFT are Euclidean and not continued into Lorentzian signature, meaning that non-unitarity in itself does not despoil symplectic theories from being physically applicable. Higher-dimensional Liouville field theories are examples of non-unitary CFTs that are solvable in Euclidean signature
\cite{levy2018liouville, kislev2022odd}.
\item The large $N$ analysis of the $O(N)$ model indicates the presence of a UV fixed point in dimensions $4<d<6$. Refs.~\cite{fei2014critical,fei2015three} developed a $6-\epsilon$ expansion for this fixed point in terms of a theory with cubic interactions $\sigma \phi^a\phi^a$ and $\sigma^3$. Can theories with cubic terms involving the fields $\chi$, $\tau$, $\rho$, etc. be used to determine similar $6-\epsilon$ expansions for the $O(N)^r$ theories? For theories with symplectic symmetry, $6-\epsilon$ expansions were developed in Refs.~\cite{fei2015critical}, \cite{klebanov2022critical}, and tensorial generalizations likely exist for these expansions as well.
\item The $\epsilon$ expansions of the Gross-Neveu model in $2+\epsilon$ dimensions and of the Gross-Neveu-Yukawa model in $4-\epsilon$ expansion can both be matched against a large $N$ solution using a Hubbard-Stratonovich transformation similar to that of the $O(N)$ model \cite{diab2016and}. Do these fermionic theories also admit similar tensorial generalizations?
\item A straightforward follow-up to the present paper would be to carry out similar analyses for tensors with $U(N)^r$ rather than $O(N)^r$ symmetry. In fact, because symmetric tensors furnish irreps of the $U(N)$ group without the need to subtract off traces, the large $N$ field content of $U(N)^r$ fixed points as well as the associated Hubbard-Stratonovich transformations will almost certainly be simpler than in our case. A study of $U(N)^r$ theories would also offer an avenue to connect with the field of tensor rank decompositions of Hermitian tensors. Other variations of the present work would be to study theories where the fundamental fields transform under different representations than the $r$-fold multi-fundamental representation of the $O(N)$ group, and attempting to classify e.g. the RG fixed points of quartic theories of symmetric traceless tensors, antisymmetric tensors, or mixed representation tensors. Existing results on theories of tensors transforming under non-fundamental representations can be found in Refs.~\cite{klebanov2017large}, \cite{gurau20181}, \cite{carrozza2018large}, \cite{benedetti20191}, \cite{carrozza2022melonic}.
\item Among the topics for which the $O(N)$ model has served as an important testing ground (besides leading to experimental predictions) is the study of line and surface defects, a study which has seen a recent surge of progress. The $O(N)^r$ models offer the opportunity to broadly extend this testing ground.
\item In Appendix~\ref{subsec:melonicabsence}, we present an argument that melonic operators are generally absent at the perturbative large $N$ fixed points of quartic $O(N)^r$ models. For odd $r$, the absence of melonic fixed points implies that all fixed points belong to the vectorial class, whose members we have exhaustively cataloged. But, in the notation of Appendix~\ref{subsec:types}, for even $r$ the possibility remains that matricial and vector-melonic fixed points appear at higher rank. In the interest of completeness, it would be desirable to disprove such fixed points or alternatively flesh out counterexamples at higher rank. A comprehensive understanding of $O(N)^r$ theory at any rank would pave the way for studying the large rank limit of these theories and exploring emergent phenomena in the spirit of Philip Anderson's \emph{More is different} \cite{anderson1972more}. For the tensor theories that are related to the SYK model, the study of the limit of large rank derives motivation from the physical insight that has been gleaned from the double-scaling limit of the SYK model.
\end{itemize}
On a closing note, we remark that while some questions can be answered straightforwardly for quartic tensor models of arbitrarily large rank---such as how many vectorial large $N$ fixed points there are, or what their anomalous dimensions are for $\phi$ and $\phi^2$---there other basic questions that we are unable to fully answer even for rank-three tensors. One such questions is this: what are the single-trace operators of the theory? More precisely, what is the set of operators that do not factor into separate groups of fundamental fields with index contractions only among themselves? In the rank-one case, there is but a single such operator: $\phi^a\phi^a$. For rank-two tensors, there is one operator for each even positive integer $n$, constructed by cyclically contracting $n$ fields together: $\phi^{ab}\phi^{ab}$, $\phi^{a_1b_1}\phi^{a_1b_2}\phi^{a_2b_2}\phi^{a_2b_1}$, etc. In the case of rank-three tensors, the number of such operators proliferates tremendously with increasing number of fields. Determining the single-trace operators for a given number of fundamental fields is tantamount to a graph theoretical problem of a kind so difficult that the mathematical literature does not attempt to provide exhaustive constructions for the full set of graphs or to determine general formulas for their numbers but instead contents itself with searching for efficient algorithms for counting the number of graphs at a given order.\footnote{To get an idea of the complexity involved in enumerating graphs, we refer the reader to the growth of cubic connected graphs in table 2 of Ref.~\cite{robinson1983numbers}. However, for three-tensor invariants, only those cubic graphs are relevant which have three-colourable edges. It turns out that the four colour theorem is equivalent to the statement that the three-colourable cubic graphs are precisely the subset of cubic graphs that bridgeless, or 1PI in physics notation.} Building on and implementing key results of Read \cite{read1959enumeration}, the authors of \cite{avohou2019counting} were able to write down (quite complicated) generating functions for the numbers of $O(N)^r$ tensor invariants and to provide software by which to determine these numbers in the first many cases. Using their results, we can tabulate the number of single trace operators built from $n$ copies of a tensor fields $\phi$ of rank $r$, as shown below.
\begin{align}
\begin{matrix}
\text{
\scalebox{0.9}{
\begin{tabular}{|l|c|c|c|c|c|c|c|} 
\hline
&$n=2$ 
&$n=4$ 
&$n=6$ 
&$n=8$ 
&$n=10$
&$n=12$
&$n=14$
\\
\hline
$r=1$
& 1
& 0
& 0
& 0
& 0
& 0
& 0
\\
\hline
$r=2$
& 1
& 1
& 1
& 1
& 1
& 1
& 1
\\
\hline
$r=3$
& 1
& 4
& 11
& 60
& 318
& 2806
& 29\,359
\\
\hline
$r=4$
& 1
& 13
& 118
& 3931
& 228\,316
& 24\,499\,085
& 3\,816\,396\,556
\\
\hline
\end{tabular}}}
\end{matrix}
\end{align}
As intimated by this table, the mathematical discipline of graph theory, closely tied to the physics of tensorial Feynman diagrams, is one wherein the border region between trivial and impossible inquiries is particularly narrow.

\subsection*{Acknowledgements}

We are thankful to Ofer Aharony, Yasha Neiman, Fedor Popov, Amit Sever, and Grigory Tarnopolskiy for useful discussions and especially to Igor Klebanov for numerous valuable comments offered and insights shared. CBJ expresses his gratitude to the Simons Center for Geometry and Physics (SCGP), Stony Brook University, where he was employed during much of this project. 
YO would like to thank SCGP for a wonderful hospitality during 2021-2022, when this work was initiated.
This work is supported in part by the ISF center of excellence and the U.S.-Israel Binational Science Foundation and by Korea Institute for Advanced Study (KIAS) Grant PG095901.

\appendix

\section{Optimal Scaling}
\label{sec:Optimal}

The subject of this paper is $d$-dimensional scalar theories of rank-$r$ tensor fields $\phi^{a_1a_2...a_r}$, where $a_j \in \{1,..,N\}$, with action given by
\begin{align}
S = \int d^dx\,\bigg(
\frac{1}{2}
(\partial_i\phi^{a_1a_2...a_r})
(\partial_i\phi^{a_1a_2...a_r})
+
\sum_{n=1}^{n_\text{max}} g_n\,\mathcal{O}_n(\phi) \bigg)\ ,
\end{align}
where the interactions $\mathcal{O}_n(\phi)$ are homogeneous polynomials in $\phi$ that are symmetric under separate rotations of each tensor index and thereby exhibit $O(N)^r$ symmetry (quotiented by $(\mathbb{Z}_2)^{r-1}$ to account for overcounting of the $\phi \rightarrow -\phi$ symmetry). For the theories we studied in the paper, the interactions $\mathcal{O}_n$ were all of degree four or all of degree six, but here we will allow for each interaction $\mathcal{O}_n$ to have its own degree $q_n$.

The purpose of this appendix is to describe ways of ascertaining what the appropriate powers of $N$ are by which to scale the coupling constants.  We will denote these powers by $e_n$ and introduce rescaled couplings $\lambda_n$ given, for $n\in\{1,...,n_\text{max}\}$, by
\begin{align}
\lambda_n = \frac{g_n}{N^{e_n}} \,.
\end{align}
We then define the large $N$ limit as
\begin{equation}
g_n \rightarrow 0,~~~~N\rightarrow \infty,~~~~\lambda_n~~~\text{fixed} \ . 
\label{limit}
\end{equation}
In the case of matrix theories, $r=2$, the suitable choice of scaling exponents was determined in the classic work of 't Hooft \cite{hooft}, with the exponent of a given operator being determined by the number of traces of the operator. For general tensorial operators there is no simple known answer for what values scaling exponents should assume. We want all observables of the theory to remain finite as we send $N$ to infinity, which means that we cannot choose the exponents $e_n$ to be too small. We also want all the beta functions for the couplings $\lambda_n$ to remain finite, which means that we cannot choose the exponents to be too large. These conditions generally permit ranges of allowed values for the exponents $e_n$, and among the allowed values it is desirable to choose the exponents to be as small as possible in order not to needlessly suppress any field content and thereby lose generality.

When doing perturbative calculations in an $\epsilon$ expansion, it is straightforward to determine the range of allowed scaling exponents at a given order of perturbation theory. But a scaling allowed at a given order may cause a blow up at higher order. The issue posed by this fact amounts to a question of order of limits: the $\epsilon \rightarrow 0$ limit and the $N \rightarrow \infty$ limits do no commute. Picking a scaling valid only up to a certain order of perturbation theory amounts to studying the regime where $\epsilon$ is much smaller than $1/N$ or than some power thereof, which is usually undesirable since ultimately we would like to set $\epsilon$ equal to one. One could attempt to guess non-perturbative scalings based on perturbative answers, but this is a dangerous path to tread, for scalings of coupling constants are determined by index contractions in Feynman diagrams, and in the field of graph theory it is not uncommon for patterns present in families of graphs at low order to unexpectedly disappear at higher order.\footnote{For example, it was conjecture that the list-chromatic number $\chi_\ell(G)$ of a any plane graph $G$ satisfies the inequality $\chi_\ell(G)\leq \chi(G)+1$, where $\chi(G)$ is the ordinary chromatic number of $G$. The first known counter-example to this conjecture, pointed out by Voigt \cite{voigt1993list}, is a graph on 238 vertices, although Voigt and Wirth \cite{voigt19973} subsequently found a counterexample of order 75, and Mirzakhani \cite{mirzakhani1996small,martin2017early} discovered a counter-example of order 63.} The choice of scaling exponents should therefore be determined non-perturbatively or at least according to arguments that apply at all orders of perturbation theory.

In the context of zero-dimensional quantum field theories with $O(N)^r$ or $U(N)^r$ symmetry, Ref.~\cite{ferrari2019new} introduced the term \emph{optimal scaling} to denote the smallest set of scaling exponents compatible with a well-defined $1/N$ expansion. One way to phrase this definition is by identifying, for a given theory, an optimal scaling as a choice of exponents $\{e_n\}$ such that the free energy of the theory is polynomially bounded in $N$ but on decreasing any one of the exponents $e_n$, the free energy ceases to be polynomially bounded. For tensors of high rank, determining the optimal scaling can be a non-trivial task. In principle, the optimal scaling need not even be unique. In fact, we illustrate in Subsection~\ref{app:example} below that non-unitary theories with $U(N)^r$ symmetry admit multiple inequivalent optimal scalings.

In the following subsection, as a warm-up before we discuss the optimal scaling of tensor operators, we review the arguments by which 't Hooft determined the large $N$ scaling of matrix operators. We then proceed to explain, for general tensor operators, why no simple known prescription exists for determining their large $N$ scalings, but we also provide general upper bounds on scaling exponents.

\subsection{Matricial optimal scaling}

\begin{figure}
    \centering
\hspace*{-0.2cm}\includegraphics[scale=1.1]{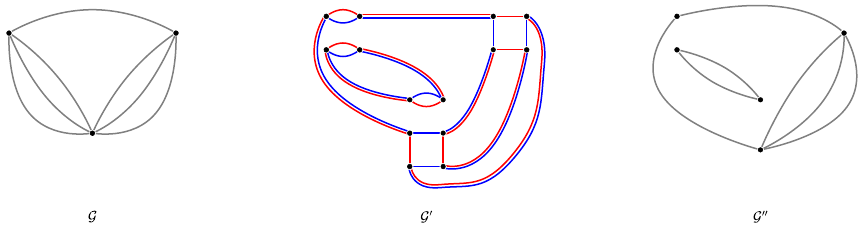}
    \caption{Different graphs corresponding to the same Feynman diagram built from a sextic interaction $\text{Tr}[\phi^2]\text{Tr}[\phi^4]$ ($q_i=6$, $\delta_i=2$, $N_i=1$) and two different quartic interactions $\text{Tr}[\phi^2]^2$ ($q_i=4$, $\delta_i=2$, $N_i=1$) and $\text{Tr}[\phi^4]$ ($q_i=4$, $\delta_i=1$, $N_i=1$) in a matrix theory with $O(N)^2/\mathbb{Z}_2$ symmetry. For this example $C=2$, $F=5$, $E=7$, $V=5$, and $L=6$.}
    \label{fig:graphs}
\end{figure}

Consider a theory of matrix fields $\phi^{ab}$, $a,b\in \{1,...,N\}$, with some set of interaction terms in the action, each of which is dressed with its own coupling constant $g_n$. We wish to determine scaling exponents $e_i$ such that all connected vacuum Feynman diagrams remain finite in the $N \rightarrow \infty$ limit. To this end, we consider an arbitrary such Feynman diagram and apply some elementary graph theory. Since there are multiple distinct ways of representing a Feynman diagram as a graph, we will, for the sake of clarity, distinguish between three graphical representations of the Feynman diagram:
\begin{itemize}
    \item the graph $\mathcal{G}$ which contains one vertex for each interaction term in the Feynman diagram; this diagram contains no information about the index contractions of the Feynman diagram, 
    \item the graph $\mathcal{G}'$ which contains one vertex for each copy of the fundamental matrix field in the Feynman diagram and which explicitly shows the index contractions, and
    \item the graph $\mathcal{G}''$ which contains one vertex for each trace in the interaction terms in the Feynman diagram.
\end{itemize}
For all three types of graphs, the edges are determined by the propagators of the Feynman diagram, but the graph $\mathcal{G}'$ uses a double-edge to indicate the two indices carried by the propagator. See Figure~\ref{fig:graphs} for an example of the three kinds of graphs associated to a Feynman diagram. Note that while we require the graph $\mathcal{G}$ to be connected, the graphs $\mathcal{G}'$ and $\mathcal{G}''$ can be disconnected. 

Consider now some arbitrary Feynman diagram with graphs $\mathcal{G}$, $\mathcal{G}'$ and $\mathcal{G}''$. We will use the symbol $C$ to denote the number of disconnected pieces in $\mathcal{G}'$ and $\mathcal{G}''$. Furthermore, let $F$ denote the number of faces (including the external face), $V$ the number of vertices, and $E$ the number of edges in the graph $\mathcal{G}''$. Normalizing the kinetic term such that the propagator is independent of $N$, the power of $N$ that dresses the Feynman diagram is equal to the number of index loops, which we will label by the letter $L$. When the two types of strands of colours in the graph $\mathcal{G}'$ do not intersect, the number of index loops can be counted from the number of faces plus number of connected pieces minus one. When the strands do intersect, the number of index loops is lessened. Consequently,
\begin{align}
\label{L}
L \leq F + C -1 \,.
\end{align}
Let $\chi$ be the Euler characteristic for the graph $\mathcal{G}''$,
\begin{align}
\label{chi}
\chi = V-E+F\,.
\end{align}
The above formulas tell us that among the set of Feynman diagrams built from a given set of interaction terms and a given number of propagators, the dominant diagrams at large $N$ are those for which the Euler characteristic is the largest, ie. the Feynman diagrams for which $\mathcal{G}''$ is a planar graph, so that $\chi$ equals $C+1$. Combining this maximal value for $\chi$ with \eqref{L} and \eqref{chi}, we obtain the inequality
\begin{align}
\label{ineqL}
L \leq E - V + 2C\,.
\end{align}
Suppose now we use an index $n$ to label the different interaction terms of our theory and that we use the symbol $q_n$ to denote the degrees of each of these interaction terms, and the symbol $\delta_n$ to denote the number of traces in each of them. Letting $N_n$ denote the number of times the interaction term $n$ is present in the Feynman diagram under consideration, we then have the following two formulas 
\begin{align}
\label{V2E}
V = \sum_n N_n\, \delta_n\,,
\hspace{20mm}
2E = \sum_n N_n\, q_n\,.
\end{align}
To determine the general large $N$ scaling for matricial operators, we need one last formula in addition to the above. For any graph, the number of disconnected pieces is less than or equal to the number of vertices: $C \leq V$. But we can obtain a stricter inequality by using the fact that we are dealing with a connected Feynman diagram. This means that the graph $\mathcal{G}$ is connected, and since $\mathcal{G}$ has $\sum_n N_n$ vertices, the implication for the graph $\mathcal{G}''$ is that this latter graph has at minimum $\sum_n N_n-1$ edges connecting together different vertices, so that
\begin{align}
\label{ineqC}
C \leq V - (\sum_n N_n-1) = 1 + \sum_n(\delta_n-1)N_n\,.
\end{align}
Combining \eqref{ineqL}, \eqref{V2E}, and \eqref{ineqC}, we finally arrive at the inequality
\begin{align}
\label{finalIneq}
L \leq 2 + \sum_n N_n \Big(\frac{q_n}{2}+\delta_n-2\Big)\,.
\end{align}
For a given set of values for $N_n$, there always exist Feynman diagrams for which this inequality is saturated. We therefore conclude that the appropriate large $N$ scaling of a coupling constant $g_n$ is given by
\begin{align}
g_n = \frac{\lambda_n}{ N^{\frac{q_n}{2}+\delta_n-2}}\,.
\end{align}
It is sometimes customary to rescale the fields $\phi^{ab} \rightarrow \sqrt{N}\phi^{ab}$, so that each propagator carries a factor of $1/N$ and the scaling exponents become simply $\delta_n-2$, but we will not adopt this convention.

\subsection{Three-tensor optimal scaling}

In a theory where the fundamental fields are tensors $\phi^{abc}$, we can for each Feynman diagram introduce a 3-index generalization of the graph type $\mathcal{G}'$, where now each propagator carries three strands of colours, and the vertices in an interaction term are connected amongst each other with three coloured edges incident to each vertex, as exemplified by the following sextic operator:
\begin{align}
\label{sexticOpExample}
\begin{matrix}
\text{
		\includegraphics[scale=1]{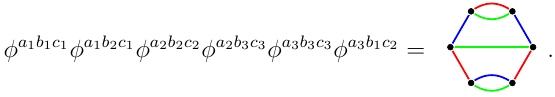}
}
\end{matrix}
\end{align}
For any Feynman diagram we can choose to completely disregard one index, whereby we are left with a matrix-theory Feynman diagram, to which the arguments of the previous subsection and the inequality \eqref{finalIneq} apply. Let $L^{xy}$ denote the number of index loops in indices $x$ and $y$, and let $\delta_n^{xy}$ denote the number of traces in the interaction term $n$ wrt. indices $x$ and $y$, ie. when forgetting about the remaining index and viewing the tensorial interaction as a matrix operator. We then have that
\begin{align}
L = \frac{L^{ab}+L^{ac}+L^{bc}}{2}
\leq 3 + \sum_n N_n \Big(\frac{3q_n}{4}+\frac{\delta^{ab}_n+\delta^{ac}_n+\delta^{bc}_n}{2}-3\Big)
\,.
\end{align}
A Feynman diagram that saturates this inequality possesses the property that the graph $\mathcal{G}'$ remains planar after erasing any one colour. From the above inequality we recover the large $N$ scaling of Ref.~\cite{carrozza2016n}:
\begin{align}
 e_n=\frac{\delta_n^{ab}+\delta_n^{ac}+\delta_n^{bc}}{2}-3+\frac{3}{4}q_n  \,. \end{align}
For example, the operator \eqref{sexticOpExample} has the following three two-colour subdiagrams:
\begin{align}
\begin{matrix}
\text{
		\includegraphics[scale=1]{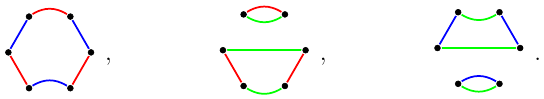}
}
\end{matrix}
\end{align}
Consequently we have that
\begin{align}
\delta_n^{ab}=1\,,
\hspace{15mm}
\delta_n^{ac}=2\,,
\hspace{15mm}
\delta_n^{bc}=2\,,
\end{align}
and so the large $N$ exponent equals
\begin{align}
e_n=\frac{2+1+2}{2}-3+\frac{3}{4}\,6=4\,.
\end{align}

\subsection{Optimal scaling at higher ranks}

At rank four and higher there are multiple ways of adding together index loops from two-colour subdiagrams such as to count the index loops in each colour once. For example, for the number of index loops $L$ of a Feynman diagram in a rank-four theory we have that
\begin{align}
L = L^{ab}+L^{cd} = L^{ac}+L^{bd} = 
\frac{L^{ab}+L^{ac}+L^{ad}+L^{bc}+L^{bd}+L^{cd}}{3}\,.
\end{align}
In general, introducing an ordering $<$ among indices $x$ and $y$ to avoid double counting, we can choose any non-negative coefficients $C^{xy}$ and write
\begin{align}
L=\sum_{x<y} C_{xy}\,L^{xy}
\end{align}
provided that the coefficients satisfy the conditions
\begin{align}
\forall x:\,\,\sum_{y\neq x} C_{xy}=1\,,
\hspace{10mm}
\text{where}
\hspace{10mm}
C_{xy}=C_{yx}\,.
\end{align}
Incidentally, the conditions on the coefficients at rank $r$ are equivalent to the equations satisfied by Mandelstam invariants for $r$ points scattering or by Mellin variables in the context of Mellin amplitudes.\footnote{If we introduce a set of $r$ auxiliary momenta $\vec{k}_x$ satisfying momentum conservation, $\sum_x\vec{k}_x=0$, and carrying unit mass, $\vec{k}_x\cdot \vec{k}_x=-1$, the conditions on the coefficients $C_{xy}$ are automatically satisfied if we set $C_{xy}=\vec{k}_x\cdot \vec{k}_y$.} This entails that the number of degrees of freedom in the choice of coefficients equals $\frac{r(r-3)}{2}$, so that the coefficients are uniquely fixed for $r=3$, while for $r=4$ there are two free parameters in choosing the coefficients, for $r=5$ there are five free parameters, and so on.

For any valid choice of coefficients $C_{xy}$ we can make repeated use of the inequality \eqref{finalIneq} to derive bounds on $L$ and thereby obtain bounds on the scaling exponents $e_n$. Unlike the matrix case, for tensors the bounds we obtain need not ever be saturated. To obtain the sharpest upper bound, we should pick among all valid choices of coefficients the set that minimizes the upper bound on $L$, whereby we obtain the following upper bound on scaling exponents
\begin{align}
\label{exponentBound}
e_n \leq \underset{C_{xy}}{\text{min}}\Big[\sum_{x<y}C_{xy}\,\delta_n^{xy}-r+\frac{r}{4}q_n\Big]\,.
\end{align}
For the specific quartic and sextic theories we study in this paper, the above upper bound is saturated and provides us with the optimal scaling. This can be shown for each operator by constructing explicit classes of Feynman diagrams for which the above scaling is realized. 

In general, however, even the minimal bound \eqref{exponentBound} need not be saturated, and it is not always possible to determine the large $N$ scaling of an operator by separately considering its two-colour subdiagrams. As an example to demonstrate this fact, consider the octic operator in a theory with $O(N)^7$ symmetry shown below.
\begin{align}
\label{octicExample}
\begin{matrix}
\text{
		\includegraphics[scale=1]{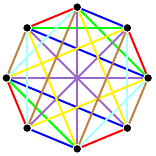}
}
\end{matrix}
\end{align}
All the two-colour subgraphs for this operator consist of two disconnected pieces: $\delta_n^{xy}=2$ for any two colours $x$ and $y$. Consequently, \eqref{exponentBound} tells us that $e_n \leq 14$. To saturate the inequality for any given choice of coefficients $C_{xy}$, one has to construct Feynman diagrams out of the octic operator that are planar with respect to any two colours $x$ and $y$ for which $C_{xy}$ is non-zero. But for the particular operator \eqref{octicExample} it is impossible to impose planarity with respect to all such two colours.

The operator \eqref{octicExample} actually has a large $N$ exponent of $21/2$ and is of melonic type, which means that the leading class of Feynman diagrams at large $N$ are constructed by pairwise grouping together interaction terms and contracting fields that are isomorphic according to the automorphism group of the interaction term under permutations $\sigma$ of fields as shown below.
\begin{align}
\begin{matrix}
\text{
		\includegraphics[scale=1]{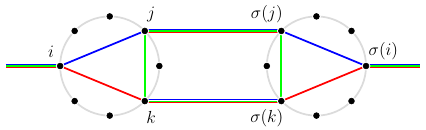}
}
\end{matrix}
\end{align}
This melonic construction of Feynman diagrams is possible for any operator and provides a universal lower bound on scaling exponents:
\begin{align}
e_n \geq \frac{r}{2}\Big(\frac{q}{2}-1\Big)\,.
\end{align}

\subsection{Other scalings}

The optimal scaling we discussed in the previous subsection is rooted in the idea of picking the smallest exponents such that connected Feynman diagrams are bounded by fixed powers of $N$. There are however cases where yet smaller exponents can be considered as well as cases where larger exponents are more appropriate.

Scaling exponents smaller than those determined by the optimal scaling can be appropriate if the connected diagrams that would cause a blow up of the theory for any reason do not contribute to the values of observables. For example, there may be a cancellation of these diagrams, and in some cases the leading diagrams at large $N$ that determine the optimal scaling are tadpole diagrams, which for massless theories in higher dimensions vanish under dimensional regularization---a subtlety that does not apply to the tensor model literature studying QFTs in zero dimensions, where tadpole diagrams contribute on equal footing with any other Feynman diagrams.

Raising scaling exponents to sub-optimal values is relevant for theories where coupling constants vanish to leading but not subleading order at large $N$. In the context of RG flows, the values of coupling constants at fixed points need not---and indeed sometimes do not---scale according to the optimal scaling. In the presence of multiple fixed points, the leading large $N$ scaling of a given coupling constant does not have to be the same across the fixed points. And conversely if a coupling constant grows without bound along all RG flows, then there is no meaningful scaling exponent to associate to that coupling.

\subsection{\texorpdfstring{$U(N)^5$}{U(N)-to-the-fifth} theory with multiple optimal scalings}
\label{app:example}
In a tensor model with multiple operators present in the potential, a kind of synergy may occur wherein combinations of operators together produce Feynman diagrams with a larger scaling with $N$ than the operators do separately. This phenomenon engenders one or more degrees of freedom in the choice of how to scale coupling constants with $N$ while retaining a finite free energy density. For garden variety unitary tensor models with $O(N)$ or $U(N)$ symmetry, this behaviour does not arise, but it may happen for other classes of theories and it certainly occurs in non-unitary tensor models, as the following example illustrates.

Consider a zero-dimensional theory of complex rank-five tensors $\phi^{abcde}$ with $U(N)^5$ symmetry,
\begin{align}
Z(g_1,g_2) =\int_{\mathbb{C}\times\mathbb{C}} d\phi\,d\overline{\phi} \exp\bigg[-\phi^{abcde}\overline{\phi}^{\,abcde}
- g_1\,O_{\text{dec}}(\phi,\overline{\phi}) - g_2\,O^\ast_{\text{dec}}(\phi,\overline{\phi})
\bigg]\,,
\end{align}
where the decic interaction terms are given by
\begin{align}
\begin{matrix}
\text{
		\includegraphics[scale=1]{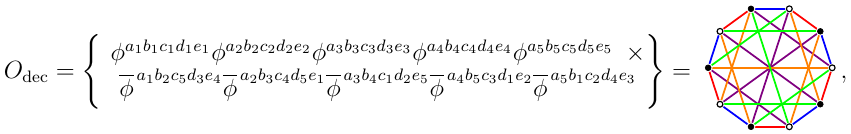}
}
\end{matrix}
\end{align}
and by its complex conjugate. We will not assume that $g_2=g_1^\ast$, and so the action of this theory is generally not real. The free energy density of the theory has the perturbative expansion
\begin{align}
\frac{\log Z}{N^5}
=\,&
\log \pi
-N^5\Big(26+43N+30N^2+17N^3+4N^4\Big)\big(g_1+g_2\big)
\\ \nonumber
&
+
\frac{1}{2} \Big(144 + 2154 N^2 + 32254 N^4 + 50668 N^5 + 195982 N^6 + 
   343550 N^7 + 494568 N^8 
 \\ \nonumber
 &\hspace{10mm}  + 648742 N^9 + 615478 N^{10} + 526730 N^{11} + 
   355811 N^{12} + 194258 N^{13}
   \\ \nonumber
   &\hspace{10mm}
    + 96808 N^{14} +39768 N^{15} + 13601 N^{16} + 
   3484 N^{17} + 400 N^{18}\Big)
   \big(g_1^2+g_2^2\big)
   \\ \nonumber
   &
+\Big(
120 + 2433 N^2 + 32135 N^4 + 50996 N^5 + 194296 N^6 + 342343 N^7 + 
 497778 N^8 
    \\ \nonumber
   &\hspace{8mm}
 + 650311 N^9 + 613461 N^{10} + 525964 N^{11} + 356057 N^{12} + 
 194292 N^{13} + 96912 N^{14} 
    \\ \nonumber
   &\hspace{8mm}
 + 39789 N^{15} + 13607 N^{16} + 3505 N^{17} + 
 400 N^{18} + N^{20}
\Big)g_1g_2
+\mathcal{O}(g^3)\,.
\end{align}
Focusing just on the leading terms in the large $N$ limit and using the elipsis ``..." to signify terms that are suppressed at order $\frac{1}{N}$ or higher, we have that
\begin{align}
\frac{\log Z}{N^5}
=\,&
\log \pi
-4N^9(1+...)\big(g_1+g_2\big)
+
200 N^{18}(1+...)
   \big(g_1^2+g_2^2\big)
+N^{20}(1+...)
g_1g_2
+\mathcal{O}(g^3)\,.
\end{align}
If $g_1$ and $g_2$ scale as $N^{-e_1}$ and $N^{-e_2}$ respectively, then we see that to obtain a finite energy density in the large $N$ limit, we must require that
\begin{align}
e_1\geq 9\,, 
\hspace{15mm}
e_2 \geq 9\,,
\hspace{15mm}
e_1+e_2\geq 20\,.
\end{align}
These inequalities hold true generally at higher orders in perturbation theory, with the leading diagrams built of $g_1$ and $g_2$ being melonic. In order for the scaling exponents to be as small as possible, we can choose $e_1=9+t$ and $e_2=11-t$ with $0\leq t \leq 2$. This produces the following large $N$ limits:
\begin{align}
\lim_{N\rightarrow \infty}\frac{\log \hspace{-1mm}\textcolor{white}{\bigg|}Z\Big(
\displaystyle\frac{\lambda_1}{N^{9+t}},\frac{\lambda_2}{N^{11-t}}\Big)}{N^5}
=\log\pi+\begin{Bmatrix*}[l]
(-4)\lambda_1+200\lambda_1^2 &\hspace{2mm}\text{for }t=0
\\[2pt]
0  &\hspace{2mm}\text{for }0<t<2
\\[2pt]
(-4)\lambda_2+200\lambda_2^2 &\hspace{2mm}\text{for }t=2
\end{Bmatrix*}
+\lambda_1\lambda_2
+\mathcal{O}(\lambda^3)\,.
\label{toyCases}
\end{align}
In this example we see that there is a line segment of optimal scalings parametrized by the parameter $t$. We may interpret the RHS in \eqref{toyCases} as three different theories in the large $N$ limit, of which two are isomorphic. If we impose unitarity, then we single out the theory with $t=1$.

\section{Quartic Large \texorpdfstring{$N$}{N} \texorpdfstring{$O(N)^r$}{O(N)-to-the-rth} Fixed Points at Higher Rank}
\label{sec:classifying}

In this appendix we discuss features of quartic $O(N)^r$ theory at general rank. In Subsection~\ref{subsec:types}, we propose a categorization of quartic singlet operators into vectorial, matricial, melonic, and vector-melonic types and introduce notation to be used later in this appendix. In Subsection~\ref{subsec:melonicabsence}, we argue that melonic operators are always absent at large $N$ fixed points. Subsection \ref{subsec:invariance} explains why the large $N$ RG flow of vectorial operators constitutes an invariant subspace of the full RG flow, and Subsection~\ref{diagonalizing}
shows how to diagonalize the RG flow along this subspace. Lastly, Subsection~\ref{subsec:content} provides an enumeration of vectorial large $N$ fixed points through a description of the large $N$ field content arrived at by Hubbard-Stratonovich transformations.

\subsection{Types of operators}
\label{subsec:types}
A quartic singlet in $O(N)^r$ theory is constructed by contracting indices among four fields $\phi$. Picking any field, the number of times it is respectively contracted with the other three fields is given by three integers $n_1, n_2, n_3 \in \mathbb{N}_0$ with $n_1+n_2+n_3=r$. We can assume without loss of generality that $n_1\geq n_2 \geq n_3$. For example,
\begin{align*}
\begin{matrix}
\text{
		\includegraphics[scale=1]{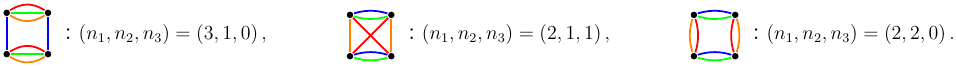}
}
\end{matrix}
\end{align*}
The type of large $N$ limit exhibited by a quartic operator and the optimal large $N$ scaling of its associated coupling constant depend on these three integers. We can group the operators as follows:
\begin{itemize}
    \item vectorial operators: $n_1 > r/2$, large $N$ scaling: $g =\displaystyle \frac{\lambda}{N^{n_1}}\,$,
    \item matricial operators: $n_1 = n_2 = r/2$, large $N$ scaling: $g =\displaystyle \frac{\lambda}{N^{r/2}}\,$,
    \item melonic operators: $n_1< r/2$, large $N$ scaling: $g =\displaystyle \frac{\lambda}{N^{r/2}}\,$,
    \item vector-melonic operators: $n_1 = r/2$, $n_2<n_1$, large $N$ scaling: $g =\displaystyle \frac{\lambda}{N^{r/2}}\,$.
\end{itemize}
The second and fourth classes of operators are only present when $r$ is even. The fourth class of operators is characterized by the fact that the dominant large $N$ Feynman diagrams include both bubble diagrams and melonic diagrams, but not all planar diagrams. In Section~\ref{sec:Quartic} we observed that only vectorial operators are present at the perturbative large $N$ fixed points for $r\leq 5$. Below we present arguments that there are no melonic fixed points for any $r$. This implies that all fixed points are vectorial whenever $r$ is odd. In the case of even $r$, although we observe no large $N$ fixed points with matricial or vector-melonic operators present at ranks two and four, we cannot rule out the possibility that such fixed points appear at higher rank. 

The integers $(n_1,n_2,n_3)$ introduced above do not uniquely identify a quartic singlet since they do not distinguish which of the $r$ indices are contracted where. To fully specify an operator, we can label indices from 1 to $r$ and introduce a triplet $(S_1,S_2,S_3)$ where $S_1$, $S_2$, and $S_3$ are disjoint subsets of the indices whose union equals the full set of indices, ie. $S_1\cup S_2\cup S_3 =\{1,2,...,r\}$. Without loss of generality, we can assume that $|S_1|\geq|S_2|\geq|S_3|$.

\subsection{Absence of melonic fixed points}
\label{subsec:melonicabsence}
For any melonic operator with large $N$ coupling $\lambda_\text{mel}$, the leading beta function only receives a correction from the melonic contribution to the anomalous dimension and so is given by
\begin{align}
\beta_{\lambda_\text{mel}}=-\epsilon\lambda_\text{mel} + \frac{\lambda_\text{mel}}{18(4\pi)^4}\sum_{m}\lambda_{m}^2\,,
\end{align}
where the sum over $m$ runs over all melonic and vector-melonic operators in the theory. In order to have a fixed point with a non-zero value of $\lambda_\text{mel}$, we must therefore have that
\begin{align}
\label{melonicFP}
\sum_{m}\lambda_{m}^2=18(4\pi)^4\epsilon\,,
\end{align}
which means that there must be at least one melonic or vector-melonic coupling $\lambda_{m'}$ that scales as $\sqrt{\epsilon}$. 

Consider first the case when $\lambda_{m'}$ is a vector-melonic coupling. In this case, $\lambda_{m'}$ feeds into a matricial beta function. We can denote the triplet for the corresponding matrix operator by $(S_1,S_2,0)$, where $|S_1|=|S_2|=r/2$. The large $N$ beta function of a matricial operator is given by
\begin{align}
\label{betmat}
\beta_{\lambda_\text{mat}}=\,&
-\epsilon \lambda_\text{mat} +
\frac{2\lambda_\text{mat}^2+\sum_{w}\lambda_w^2}{48\pi^2}
+\mathcal{O}(\lambda^3)\,,
\end{align}
where the index $w$ runs over vector-melonic couplings with triplets $(S_1',S_2',S_3')$ such that $S_1'=S_1$ or $S_1'=S_2$. Now if one of these couplings $\lambda_w$ scales as $\sqrt{\epsilon}$, then the beta function picks up a term given by a positive constant times $\epsilon$. But there is no way to cancel this term at a real perturbative fixed point. We could cancel the term against the linear term $-\epsilon\lambda_\text{mat}$ by letting $\lambda_\text{mat}$ scale as $\epsilon^0$, but then we would not be situated at a perturbative fixed point. And we could cancel the term against the $\lambda_\text{mat}^2$ term or against other other $\lambda_w^2$ terms, but then we would not be situated at a real fixed point.

The other possibility is that $\lambda_{m'}$ is a melonic operator so that it feeds into vectorial beta functions. Let one of the corresponding vectorial operators be associated to the triplet $(\mathcal{S}_1,\mathcal{S}_2,0)$, where $|\mathcal{S}_1|>|\mathcal{S}_2|>0$. The beta function for the vectorial coupling $\lambda_\text{vec}$ is given by
\begin{align}
\beta_{\lambda_\text{vec}}=\,&-\epsilon \beta_{\lambda_\text{vec}} + 
\frac{\lambda_\text{vec}^2+\sum_v\lambda_v^2}{48\pi^2}+\text{terms from other vectorial ops.}
\\&+\text{terms from matrical ops.}+\text{terms from vector-melonic ops.}+\mathcal{O}(\lambda^3)\,,
\nonumber
\end{align}
where the index $v$ runs over melonic or vector-melonic couplings with triplets $(\mathcal{S}'_1,\mathcal{S}'_2,\mathcal{S}'_3)$ such that $\mathcal{S}_1=\mathcal{S}'_1\cup\mathcal{S}'_2$ or $\mathcal{S}_1=\mathcal{S}'_1\cup\mathcal{S}'_3$ or $\mathcal{S}_1=\mathcal{S}'_2\cup\mathcal{S}'_3$. If one of these couplings $\lambda_v$ scales as $\sqrt{\epsilon}$, then the beta function again picks up a term given by a positive constant times $\epsilon$, which cannot be cancelled at a real perturbative fixed point, although it is not quite as easy to show this time. We already ruled out matricial and vector-melonic couplings that scale as $\sqrt{\epsilon}$, so we can disregard those terms. The issue is that the quadratic terms from other vectorial operators contain cross-terms, so that they are not manifestly positive semi-definite. However, on diagonalizing the vectorial beta functions according to the manner we describe in generality below in Subsection~\ref{diagonalizing}, and which we carried out for the theories of ranks three, four, and five in equations \eqref{betah}, \eqref{quartich}, and \eqref{rank5h} respectively, the quadratic terms do become manifestly positive semi-definite, and it can be checked that a term with a positive constant times $\epsilon$ remains present after the diagonalization and so cannot be cancelled to produce a real perturbative fixed point.

\subsection{Invariance of the vectorial RG flow}
\label{subsec:invariance}
For any single vectorial operator, the dominant Feynman diagrams at large $N$ are in one-to-one correspondence with the dominant bubble diagrams of a vector operator at large $N$. For example, there is a correspondence between the large $N$ diagrams constructed from the quartic operators shown below.\footnote{The operator on the left represents a quartic singlet operator in $O(N)^6$ theory, the two colours are not meant to indicate that the contracted indices transform under only two distinct $O(N)$ groups, in this case the colours are merely used for emphasis.}
\begin{align}
\begin{matrix}
\text{
		\includegraphics[scale=1]{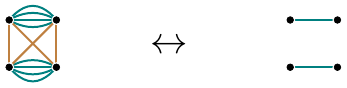}
}
\end{matrix}
\end{align}
The teal indices of the left diagram act in concert like a single dominant vector index determining which diagrams contribute in the large $N$ limit, while the brown indices merely tag along for the ride. When multiple vectorial operators are present, there is instead a many-to-one correspondence. At the level of loop-integrals, the dominant Feynman diagrams remain the same, but for any dominant Feynman diagram in the vector model, there are multiple dominant diagrams in the tensor model, obtained by replacing each vector operator by any vectorial operator. This means that in the tensor case, the dominant operators are built out of multiple distinct vectorial operators. It is, however, possible to diagonalize the theory, such that the dominant diagrams are each built out of only one type of operator, as we will see presently in \ref{diagonalizing}. 

The perturbative beta functions are computed from Feynman diagrams built via index contractions among quartic operators. The first quantum contribution to the beta functions is determined by the operation of taking two contractions across a pair of quartic operators. This operation takes as input two quartic operators and outputs a linear combination of quartic operators. In the case when we have $n$ quartic singlets $\mathcal{O}_i$ in a theory of rank $r$ tensors, we may write the operation as
\begin{align}
\label{operation}
\delta_{AC}\,\delta_{BD}\,
\frac{\partial \mathcal{O}_i}{\partial \phi^A\partial \phi^B}\,
\frac{\partial \mathcal{O}_j}{\partial \phi^C\partial \phi^D}
=\sum_{\ell=1}^n C_{ij}^\ell(N)\,\mathcal{O}_\ell\,,
\hspace{15mm}
\begin{matrix}
\text{{\footnotesize $A=a_1a_2...a_r$}}\,,
\\
\text{{\footnotesize $B=b_1b_2...b_r$}}\,,
\\
\text{{\footnotesize $C=c_1c_2...c_r$}}\,,
\\
\text{{\footnotesize $D=d_1d_2...d_r$}}\,.
\end{matrix}
\end{align}
Below we display an example of one term that contributes to this operation.
\begin{align}
\label{contraction}
\begin{matrix}
\text{
		\includegraphics[scale=1]{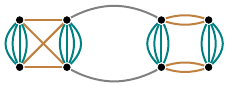}
}
\end{matrix}
\end{align}
In the above example, pairs of vertices connected by teal edges are contracted together, and only this kind of contraction contributes to the large $N$ beta functions. Furthermore, performing this kind of contraction among two vectorial operators necessarily generates a vectorial operator. If the two operators are associated to triplets $(S_1,S_2,S_3)$ and $(S'_1,S'_2,S'_3)$, then they generate a triplet $(S''_1,S''_2,S''_3)$ with $S_1''=S_1\cup S_1'$. On adopting the optimal scaling, it is easy to verify that the powers of $N$ work out correctly. For the original operators carry couplings $\frac{\lambda}{N^{|S_1|}}$ and $\frac{\lambda'}{N^{|S'_1|}}$, while the contraction \eqref{contraction} induces $N^{|S_1\cap S_2|}$ index loops, and the generated operator carries an optimal scaling 
\begin{align}
\frac{\lambda''}{N^{|S_1''|}}=\frac{\lambda''}{N^{|S_1\cup S'_1|}}=\frac{\lambda''}{N^{|S_1|+|S'_1|-|S_1\cap S_1'|}}\,.
\end{align}
In the language of Refs.~\cite{michel1971properties,michel1984symmetry,flodgren2023one}, the operation \eqref{operation} furnishes an algebra on the set of operators $\mathcal{O}_i$. What we are observing now is that at large $N$, the set of vectorial operators close on themselves. Physically, this implies that the RG flow among the couplings of vectorial operators constitutes an invariant subspace of the large $N$ RG flow of the full quartic tensor theory. 

At higher perturbative order, new kinds of index contractions contribute to the beta functions. But the only Feynman diagrams built of vectorial operators that survive in the large $N$ limit are tadpoles and bubble diagrams whose index contractions can be evaluated by iterating the decomposition \eqref{operation}. As in the vector model, once the first bubble is renormalized, all the higher bubbles are rendered finite too, and so the large $N$ RG flow of the vectorial operators terminates at quadratic order.

\subsection{Diagonalizing the vectorial RG flow}
\label{diagonalizing}
The operator basis that diagonalizes the vectorial RG flow at large $N$ is the set of operators of the form $\mathcal{O}=\big(\Phi^{i_1j_1\,i_2j_2\,...\,i_mj_m}\big)^2$, with $m<r/2$, where $\Phi^{i_1j_1\,i_2j_2\,...\,i_mj_m}$ is a homogeneous polynomial in the fundamental fields of degree two that is traceless symmetric or antisymmetric in each index pair $i_p\,j_p$, with an even number of index pairs being antisymmetric. To be more explicit, the $\Phi$ tensors is  constructed by taking the quadratic operator $\phi^{k_1\,...\,k_r}\phi^{l_1\,...\,l_r}$ and multiplying it an even number of times $N_A$ (possibly zero) for different $p\in\{1,...,r\}$ by an antisymmetric tensor 
\begin{align}
A^{i_p\,j_p}_{k_p\,l_p} = \delta^{i_p}_{k_p}\delta^{j_p}_{l_p}-\delta^{i_p}_{l_p}\delta^{j_p}_{k_p}\,,
\end{align}
and multiplying it $N_S=m-N_A$ times for yet other different $p$ by a symmetric traceless tensor 
\begin{align}
S^{i_p\,j_p}_{k_p\,l_p} = \delta^{i_p}_{k_p}\delta^{j_p}_{l_p}+\delta^{i_p}_{l_p}\delta^{j_p}_{k_p}-\frac{2}{N}\delta^{i_p\,j_p}\delta^{k_p\,l_p}\,,
\end{align}
and finally multiplying it with $\delta_{k_p\,l_p}$ a total of $r-m$ times to contract the remaining $\phi$ indices. Schematically,
\begin{align}
\Phi = (A)^{N_A}\,(S)^{m-N_A}\,(\delta)^{r-m}\,\phi\,\phi\,.
\end{align}
The condition that $N_A$ be even is needed in order for $\Phi$ to be non-vanishing. That operators  built according to the above construction diagonalize the operation \eqref{operation} can be seen by
applying the operation to operators given, schematically, by $\mathcal{O}=(\Phi)^2$ and $\mathcal{O}'=(\Phi')^2$, and first noting that, at large $N$, the leading contributions to \eqref{operation} come from the instances where the two $\frac{\partial}{\partial\phi}$ derivatives act on two fields $\phi$ in the same copy of $\Phi$. Roughly speaking, we have that
\begin{align}
\label{schematicProd}
\frac{\partial \mathcal{O}}{\partial \phi\,\partial \phi}\cdot\frac{\partial \mathcal{O}'}{\partial \phi\,\partial \phi} \sim 4\, \Phi\,(A)^{N_A}\,(S)^{m-N_A}\,(\delta)^{r-m}\,\times\, \Phi'\,(A)^{N_A'}\,(S)^{m'-N_A'}\,(\delta)^{r-m'}\,.
\end{align}
 Now unless $\mathcal{O}=\mathcal{O}'$ so that $\Phi=\Phi'$, one of the following will necessarily happen:
 \begin{itemize}
     \item an $A$ tensor from $\Phi$ is contracted with an $S$ tensor or with a $\delta$ function from $\Phi'$,
     \item an $S$ tensor from $\Phi$ is contracted with an $S$ tensor or with a $\delta$ function from $\Phi'$,
     \item a $\delta$ function from $\Phi$ is contracted with an $A$ tensor or an $S$ tensor from $\Phi'$.
 \end{itemize}
In any of the above cases the result will be zero on account of the antisymmery and tracelessness of the $A$ and $S$ tensors. The product \eqref{schematicProd} is non-zero only when $\mathcal{O}=\mathcal{O}'$, in which case the product is proportional to $\mathcal{O}$. In other words, if $\mathcal{O}_i$ and $\mathcal{O}_j$ in equation \eqref{operation} are vectorial operators, then all  other coefficients $C_{ij}^\ell(N)$ are subdominant at large $N$ compared to $C_{jj}^j(N)$ and $C_{ii}^i(N)$.

We have seen now that if we set to zero the couplings for the non-vectorial operator at any scale, they will remain zero along the large $N$ RG flow, while the beta functions for the vectorial operators on this subspace can be explicitly decoupled from each other. For positive $\epsilon$, these decoupled beta functions all have a negative linear coefficient and a positive quadratic coefficient, which entails that each of the decoupled beta functions has its own Wilson-Fisher fixed point, which in turn implies that the number of vectorial fixed points is two raised to the number of vectorial operators. 

\subsection{Large \texorpdfstring{$N$}{N} field content}
\label{subsec:content}
The $O(N)$ model admits a familiar large $N$ description via a Hubbard-Stratonovich transformation in terms of an $O(N)$ scalar field $\sigma$. We have seen in Subsections~\ref{ON3}, \ref{ON4}, and \ref{ON5} that in quartic $O(N)^r$ theory with $r\geq 3$, the large $N$ field content additionally includes $r$ traceless symmetric fields $\chi_p^{i_p\,j_p}$. And when $r\geq 5$, we must also introduce $\binom{r}{2}$ doubly antisymmetric fields $\tau_{ij}^{i_1j_2\,j_1j_2}$ as well as the same number of doubly traceless symmetric fields $\rho_{ij}^{i_1j_2\,j_1j_2}$.
The diagonalization we performed above in \ref{diagonalizing} in terms of $\Phi$ tensors immediately answers the question of which further fields appear at arbitrary higher tensor rank. For the Hubbard-Stratonovich fields by which we integrate out vectorial operators precisely couple to the $\Phi$ tensors and therefore possess the same index structure. We conclude that the Hubbard-Stratonovich fields present at a given rank $r$ are tensor fields with anywhere from 0 to $\left\lfloor\frac{r-1}{2}\right\rfloor$ pairs of incides, whereof an even number transform as antisymmetric matrices under separate $O(N)$ groups, while the remaining index pairs transform as traceless symmetric matrices under yet other separate $O(N)$ groups. Another way to say this, using multinomial coefficients $\binom{n_1+n_2+n_3}{n_1,\,n_2,\,n_3}$ to count fields, is by stating that on increasing the rank $r$ of the theory, new types of fields with an extra pair of indices appear at each odd rank, and that for $r= 4l+1$ the fields that appear, and which carry $2l$ pairs of indices, are
\begin{itemize}
\item $\binom{r}{2l,\,0,\,r-2l} $ fields with $2l$ pairs of traceless symmetric indices,
\item $\binom{r}{2l-2,\,2,\,r-2l}$ fields with $2l-2$ pairs of traceless symmetric indices and two antisymmetric index pairs,
\item $\binom{r}{2l-4,\,4,\,r-2l}$ fields with $2l-4$ pairs of traceless symmetric indices and four antisymmetric index pairs,
\begin{align*}
\\[-45pt]
\vdots
\\[-40pt]
\end{align*}
\item $\binom{r}{0,\,2l,\,r-2l}$ fields with $2l$ antisymmetric index pairs,
\end{itemize}
while for $r= 4l+3$ the new fields appearing, which carry $2l+1$ pairs of indices, are
\begin{itemize}
\item $\binom{r}{2l+1,\,0,\,r-2l-1} $ fields with $2l+1$ pairs of traceless symmetric indices,
\item $\binom{r}{2l-1,\,2,\,r-2l-1}$ fields with $2l-1$ pairs of traceless symmetric indices and two antisymmetric index pairs,
\item $\binom{r}{2l-3,\,4,\,r-2l-1}$ fields with $2l-3$ pairs of traceless symmetric indices and four antisymmetric index pairs,
\begin{align*}
\\[-45pt]
\vdots
\\[-40pt]
\end{align*}
\item $\binom{r}{1,\,2l,\,r-2l-1}$ fields with one pair of traceless symmetric indices and $2l$ antisymmetric index pairs.
\end{itemize}

\section{Locations of Fixed Points in Sextic \texorpdfstring{$O(N)^4\rtimes S_4$}{O(N)-to-the-fourth times S4} theory}
\label{sec:Locations}

The sextic rank-four tensor theory with action \eqref{sexticRank4action} has 18 perturbative fixed points, which we enumerate with Roman numerals (i) to (xviii). The first is simply the trivial fixed point with all couplings zero. To leading order in $\epsilon$, fixed points (ii) to (viii) are situated at the following loci in the space of coupling constants :
\begin{align}
\vec{\lambda}^{\,\text{(ii)}}=
(8\pi)^2\epsilon\,\Big\{ &
\frac{135}{4} (47 + 4 \sqrt{3})\,,\, 
\frac{4995}{4} - 540 \sqrt{3}\,,\, 
\frac{2295}{4}\,,\, 
270\,,\, 
 270 (2 +\sqrt{3})\,,\, 
 \frac{8235}{4} + 540 \sqrt{3}\,,
 \nonumber \\
 \, &
 270\,,\, 
 -1080\,,\, 
 0\,,\, 
 0\,,\, 
 270\,,\, 
 0\,,\, 
 -540 \sqrt{3}\,,\,
 0\,,\, 
 0\,,\, 
 0\,,\, 
 0\,,\, 
 0\,,\, 
 0\,,\, 
 135 \sqrt{3}\,,\,
 0\Big\}\,,
 \\[6pt]
\vec{\lambda}^{\,\text{(iii)}}=
(8\pi)^2\epsilon\,\Big\{ &
\frac{6345}{4} - 135 \sqrt{3}\,,\, 
 \frac{4995}{4} + 540 \sqrt{3}\,,\, 
 \frac{2295}{4}\,,\, 
 270\,,\, 
 270 (2 - \sqrt{3})\,,
 \frac{8235}{4} - 540 \sqrt{3}\,,
 \nonumber \\ &
 270\,,\, 
 -1080\,,\, 
 0\,,\, 
 0\,,\, 
 270\,,\, 
 0\,,\, 
 540 \sqrt{3}\,,\, 
 0\,,\, 0\,,\, 0\,,\, 0\,,\, 0\,,\, 0\,,\, -135 \sqrt{3}, 0\Big\}\,,
\\[6pt]
\vec{\lambda}^{\,\text{(iv)}}=
(8\pi)^2\epsilon\,\Big\{ &69120\,,\, 157680\,,\, 17280\,,\, 1080\,,\, 2160\,,\, 66960\,,\, 4320\,,\, -4320\,,\, 0\,,\, 0\,,\, 1080\,,
\nonumber \\ &
0\,,\, \
0\,,\, 0\,,\, 0\,,\, 0\,,\, 0\,,\, 0\,,\, 0\,,\, 0\,,\, 0\Big\}\,,
 \\[6pt]
\vec{\lambda}^{\,\text{(v)}}=
(8\pi)^2\epsilon\,\Big\{ &5900\,,\, 0\,,\, 0\,,\, -1080\,,\, 0\,,\, 0\,,\, 0\,,\, 0\,,\, 60\,,\, 0\,,\, 0\,,\, 0\,,\, 0\,,\, 0\,,\, 0\,,\, 0\,,\, 0\,,\, 0\,,\, 0\,,\, 0\,,\,
  0\Big\}\,,
 \\[6pt]
\vec{\lambda}^{\,\text{(vi)}}=
(8\pi)^2\epsilon\,\Big\{ &
\frac{601625}{4} + 135 \sqrt{3}\,,\,
-\frac{45}{4} (7225 + 48 \sqrt{3})\,,\, 
\frac{12015}{4}\,,\, 
-5670\,,\, 
 270 (2 + \sqrt{3})\,,\, 
 \nonumber \\ &
 \frac{45}{4} (935 + 48 \sqrt{3})\,,\, 
 1890\,,\, -1080\,,\, 60\,,\, 0\,,\, 270\,,\, 0\,,\, -540 \sqrt{3}\,,\, 0\,,\, 0\,,\, 0\,,\, 0\,,\, 0\,,\, 0\,,\,
 \nonumber \\ &
 135 \sqrt{3}\,,\, 0\Big\}\,,
  \\[6pt]
\vec{\lambda}^{\,\text{(vii)}}=
(8\pi)^2\epsilon\,\Big\{ &
\frac{601625}{4} - 135 \sqrt{3}\,,\, 
 \frac{45}{4} (-7225 + 48 \sqrt{3})\,,\, 
 \frac{12015}{4}\,,\, 
 -5670\,,\, 
 270 (2 - \sqrt{3})\,,\, 
  \nonumber \\ &
 \frac{42075}{4} - 540 \sqrt{3}\,,\, 
 1890\,,\, -1080\,,\, 60\,,\, 0\,,\, 270\,,\, 0\,,\, 
 540 \sqrt{3}\,,\, 0\,,\, 0\,,\, 0\,,\, 0\,,\, 0\,,\, 0\,,\,
  \nonumber \\ &
 -135 \sqrt{3}\,,\, 0\Big\}\,,
\\[6pt]
\vec{\lambda}^{\,\text{(viii)}}=
(8\pi)^2\epsilon\,\Big\{ &6543500\,,\, -4212720\,,\, 172800\,,\, -38880\,,\, 2160\,,\, 560880\,,\, 17280\,,\, -4320\,,\, 60\,,\, 
  \nonumber \\ &
0\,,\, 1080\,,\, 0\,,\, 0\,,\, 0\,,\, 0\,,\, 0\,,\, 0\,,\, 0\,,\, 0\,,\, 0\,,\, 0\Big\}\,.
\end{align}
Like the above fixed points, the leading values of the couplings at the next two fixed points, (ix) and (x), are given simply in terms of ratio of integers and square roots of integers, up to an overall factor of $\pi^2$. But in cases (ix) and (x) the integers involved are particularly lengthy. Although their precise form is probably of little interest to the reader, we nonetheless display these solutions explicitly below, simply to give an idea of the complexity involved in any tentative large $N$ solution that matches these fixed points in the $\epsilon$ expansion:
\begin{align}
\text{\scalebox{0.65}{$\frac{\vec{\lambda}^{\,\text{(ix)}}}{(8\pi)^2\epsilon}$=\bigg\{}}
&
\text{\scalebox{0.65}{$\frac{1620}{156420504648617} (19607370583144755 - 1815344122912 \sqrt{23699}) \,,\,
 \frac{8640}{2951330276389} (-28078989400213 + 17602913268 \sqrt{23699})\,,\,$}} 
   \nonumber \\ &
\text{\scalebox{0.65}{$  \frac{4320}{1295011091} (793909403 - 2211396 \sqrt{23699})\,,\, 
  -  \frac{1080}{567236263} (3067843601 + 436084 \sqrt{23699})\,,\,
  \frac{164160}{567236263} (3379651 - 70585 \sqrt{23699})\,,\, 
   $}} 
    \nonumber \\ &
\text{\scalebox{0.65}{$  \frac{2160}{55685476913} (253780228967 - 486150604 \sqrt{23699})\,,\, 
 \frac{4320}{10702571} (3019649 + 4114 \sqrt{23699})\,,\, 
 \frac{17280}{10702571} (-650012 + 4569 \sqrt{23699})\,,\, $}} 
    \nonumber \\ &
\text{\scalebox{0.65}{$     \frac{7020}{121}\,,\, 0\,,\, \frac{32400}{121}\,,\, 0\,,\, 
 \frac{51840}{10702571} (-187187 + 1523 \sqrt{23699})\,,\, 
 \frac{28080}{5203}\,,\, 
 -\frac{4320}{5203} \sqrt{23699}\,,\, 
 -\frac{56160}{5203}\,,\, 
  \frac{112320}{27071209} (3126 - \sqrt{23699})\,,\, $}} 
      \nonumber \\ &
\text{\scalebox{0.65}{$   \frac{4320}{27071209} (48243 - 5134 \sqrt{23699})\,,\,
  0\,,\, 
  \frac{28080}{121}\,,\, 0\bigg\}\,,$}} 
\\[4pt]
\text{\scalebox{0.65}{$  \frac{\vec{\lambda}^{\,\text{(x)}}}{(8\pi)^2\epsilon}=\Big\{$}}
 &
 \text{\scalebox{0.65}{$ 
\frac{1620}{156420504648617} 
(19607370583144755 + 1815344122912 \sqrt{23699})\,,\, 
    -\frac{8640}{2951330276389} (28078989400213 + 17602913268 \sqrt{23699})\,,\,     $}}
            \nonumber \\ &
  \text{\scalebox{0.65}{$ 
    \frac{4320}{1295011091} (793909403 + 2211396 \sqrt{23699})\,,\,  
   \frac{1080}{567236263} (-3067843601 + 436084 \sqrt{23699})\,,\, 
    \frac{164160}{567236263} (3379651 + 70585 \sqrt{23699})\,,\, 
    $}}
            \nonumber \\ &
          \text{\scalebox{0.65}{$  
    \frac{2160}{55685476913} (253780228967 + 486150604 \sqrt{23699})\,,\, 
    \frac{4320}{10702571} (3019649 - 4114 \sqrt{23699})\,,\, 
    -\frac{17280}{10702571} (650012 + 4569 \sqrt{23699})\,,\, 
         $}} 
     \nonumber \\ &
          \text{\scalebox{0.65}{$ 
    \frac{7020}{121}\,,\, 
    0\,,\, 
    \frac{32400}{121}\,,\, 
    0\,,\, 
    -\frac{51840}{10702571} (187187 + 1523 \sqrt{23699})\,,\, 
    \frac{28080}{5203}\,,\, 
    \frac{4320}{5203}\sqrt{23699}\,,\, 
    -\frac{56160}{5203}\,,\, 
    \frac{112320}{27071209} (3126 + \sqrt{23699})\,,\, 
         $}} 
              \nonumber \\ &
          \text{\scalebox{0.65}{$ 
    \frac{4320}{27071209} (48243 + 5134 \sqrt{23699})\,,\, 
    0\,,\, 
    \frac{28080}{121}\,, 
    0\Big\}\,.
     $}} 
\end{align}
The locations of the last eight fixed points are even more complicated. These have couplings given in terms of roots of quartic equations, meaning that closed-form expressions exist but are very lengthy. For this reason, we will merely display their numerical values:
\begin{align}
\vec{\lambda}^{\,\text{(xi)}}=
(8\pi)^2\epsilon\,\Big\{
&
39771.238\,,\, 89713.610\,,\, 8849.7258\,,\, 874.73176\,,\, 7.3847425\,,\, 38900.990\,,\, 
\nonumber \\ &
3133.3241\,,\, -1890.5304\,,\, 0\,,\, 0\,,\, 874.73176\,,\, 0\,,\, 1875.7610\,,\, 0\,,\, 0\,,\,
-804.19829\,,\, 
\nonumber \\ &
-1003.2853\,,\, 149.70715\,,\, 0\,,\, -66.841092\,,\, 0
\Big\}\,,
\\
\vec{\lambda}^{\,\text{(xii)}}=
(8\pi)^2\epsilon\,\Big\{
&
1140.6966\,,\, 1406.3408\,,\, 73.935497\,,\, 228.84964\,,\, 132.13957\,,\, 1615.4471\,,\,
\nonumber \\ &
255.25587\,,\, -187.72991\,,\, 0\,,\, 0\,,\, 228.84964\,,\, 0\,,\, -76.549228\,,\, 0\,,\, 0\,,\,
\nonumber \\ &
-363.83433\,,\, -92.906483\,,\, 30.642458\,,\, 0\,,\, 201.05447\,,\, 0
\Big\}\,,
\\
\vec{\lambda}^{\,\text{(xiii)}}=
(8\pi)^2\epsilon\,\Big\{
&
947.47330\,,\, -1228.6423\,,\, 716.93412\,,\, 146.05887\,,\, 1584.1817\,,\, 2129.2941\,,\, 
\nonumber \\ &
200.95056\,,\, -1610.6616\,,\, 0\,,\, 0\,,\, 146.05887\,,\, 0\,,\, -1557.7019\,,\, 0\,,\, 0\,,\, 
513.21307\,,\, 
\nonumber \\ &
260.75229\,,\, 60.969364\,,\, 0\,,\, 132.81893\,,\, 0
\Big\}\,,
\\
\vec{\lambda}^{\,\text{(xiv)}}=
(8\pi)^2\epsilon\,\Big\{
&
14455.314\,,\, 33686.191\,,\, 4452.0897\,,\, 610.35973\,,\, 2716.2939\,,\, 12960.658\,,\, 
\nonumber \\ &
1750.4695\,,\, -4231.0780\,,\, 0\,,\, 0\,,\, 610.35973\,,\, 0\,,\, -1201.5099\,,\, 0\,,\, 0\,,\, 
894.81955\,,\, 
\nonumber \\ &
1382.1062\,,\, 185.34769\,,\, 0\,,\, -147.03231\,,\, 0
\Big\}\,,
\\
\vec{\lambda}^{\,\text{(xv)}}=
(8\pi)^2\epsilon\,\Big\{
&
3789647.1\,,\, -2416308.7\,,\, 97152.004\,,\, -29501.994\,,\, 7.3847425\,,\, 
320793.13\,,\, 
\nonumber \\ &
12898.899\,,\, -1890.5304\,,\, 60\,,\, 0\,,\, 874.73176\,,\, 0\,,\, 
1875.7610\,,\, 0\,,\, 0\,,\, 
\nonumber \\ &
-804.19829\,,\, -1003.2853\,,\, 149.70715\,,\, 0\,,\, -66.841092\,,\, 0
\Big\}\,,
\\
\vec{\lambda}^{\,\text{(xvi)}}=
(8\pi)^2\epsilon\,\Big\{
&
122723.36\,,\, -64076.212\,,\, 1956.5468\,,\, -5128.8813\,,\, 132.13957\,,\, 8172.7084\,,\, 
\nonumber \\ &
1681.1662\,,\, -187.72991\,,\, 60\,,\, 0\,,\, 228.84964\,,\, 0\,,\, -76.549228\,,\, 0\,,\, 0\,,\, 
\nonumber \\ &
-363.83433\,,\, -92.906483\,,\, 30.642458\,,\, 0\,,\, 201.05447\,,\, 0
\Big\}\,,
\\
\vec{\lambda}^{\,\text{(xvii)}}=
(8\pi)^2\epsilon\,\Big\{
&
71023.715\,,\, -35338.456\,,\, 1617.3650\,,\, -3892.3509\,,\, 1584.1817\,,\, 5292.0617\,,\, 
\nonumber \\ &
1187.0872\,,\, -1610.6616\,,\, 60\,,\, 0\,,\, 146.05887\,,\, 0\,,\, -1557.7019\,,\, 0\,,\, 0\,,\, 
\nonumber \\ &
513.21307\,,\, 260.75229\,,\, 60.969364\,,\, 0\,,\, 132.81893\,,\, 0
\Big\}\,,
\\
\vec{\lambda}^{\,\text{(xviii)}}=
(8\pi)^2\epsilon\,\Big\{
&
1475365.0\,,\, -917652.49\,,\, 37148.251\,,\, -18296.774\,,\, 2716.2939\,,\,
\nonumber \\ &
118579.60\,,\, 7692.8473\,,\, -4231.0780\,,\, 60\,,\, 0\,,\, 610.35973\,,\, 0\,,\, 
-1201.5099\,,\, 0\,,
\nonumber \\ &
\, 0\,,\, 894.81955\,,\, 1382.1062\,,\, 185.34769\,,\, 0\,,\, -147.03231\,,\, 0
\Big\}\,.
\end{align}
For the fixed point (iv), there is a one-parameter degeneracy in the order $\epsilon^2$ corrections to the values of the couplings. Parameterizing the degeneracy with the number $u$, the explicit values of the couplings are given up to $\mathcal{O}(\epsilon^3)$ contributions by the following:
\begin{align}
&\frac{\lambda^{\text{(iv)}}_{1}}{6!(8\pi)^2}=96\epsilon+u\epsilon^2\,,
&&\frac{\lambda^{\text{(iv)}}_{2}}{6!(8\pi)^2}=219\epsilon+(-14454+286\pi^2+u)\epsilon^2\,,
\nonumber \\
&\frac{\lambda^{\text{(iv)}}_{3}}{6!(8\pi)^2}=24\epsilon+\frac{1}{2}(381420-19123\pi^2-4u)\epsilon^2\,,
&&\frac{\lambda^{\text{(iv)}}_{4}}{6!(8\pi)^2}=\frac{3}{2}\epsilon+\frac{1}{4}(97281-4954\pi^2-u)\epsilon^2\,,
\nonumber \\
&\frac{\lambda^{\text{(iv)}}_{5}}{6!(8\pi)^2}=3\epsilon-\frac{3}{2}(-94713+4713\pi^2+u)\epsilon^2\,,
&&\frac{\lambda^{\text{(iv)}}_{6}}{6!(8\pi)^2}=93\epsilon-\frac{3}{2}(-96781+5002\pi^2+u)\epsilon^2\,,
\nonumber \\
&\frac{\lambda^{\text{(iv)}}_{7}}{6!(8\pi)^2}=6\epsilon+\frac{1}{2}(-94815+4669\pi^2+u)\epsilon^2\,,
&&\frac{\lambda^{\text{(iv)}}_{8}}{6!(8\pi)^2}=-6\epsilon+(-189372+9425\pi^2+2u)\epsilon^2\,,
\nonumber \\
&\frac{\lambda^{\text{(iv)}}_{9}}{6!(8\pi)^2}=-\frac{1}{4}\pi^2\epsilon^2\,,
&&\frac{\lambda^{\text{(iv)}}_{10}}{6!(8\pi)^2}=0\,,
\nonumber \\
&\frac{\lambda^{\text{(iv)}}_{11}}{6!(8\pi)^2}=\frac{3}{2}\epsilon+\frac{3}{4}(-54+\pi^2)\epsilon^2\,,
&&\frac{\lambda^{\text{(iv)}}_{12}}{6!(8\pi)^2}=0\,,
 \\
&\frac{\lambda^{\text{(iv)}}_{13}}{6!(8\pi)^2}=2(-94767+4714\pi^2+u)\epsilon^2\,,
&&\frac{\lambda^{\text{(iv)}}_{14}}{6!(8\pi)^2}=0\,,
\nonumber \\
&\frac{\lambda^{\text{(iv)}}_{15}}{6!(8\pi)^2}=\frac{1}{4}(94767-4714\pi^2-u)\epsilon^2\,,
&&\frac{\lambda^{\text{(iv)}}_{16}}{6!(8\pi)^2}=0\,,
\nonumber \\
&\frac{\lambda^{\text{(iv)}}_{17}}{6!(8\pi)^2}=18\epsilon^2\,,
&&\frac{\lambda^{\text{(iv)}}_{18}}{6!(8\pi)^2}=(94767-4714\pi^2-u)\epsilon^2\,,
\nonumber \\
&\frac{\lambda^{\text{(iv)}}_{19}}{6!(8\pi)^2}=0\,,
&&\frac{\lambda^{\text{(iv)}}_{20}}{6!(8\pi)^2}=0\,,
\nonumber \\
&\frac{\lambda^{\text{(iv)}}_{21}}{6!(8\pi)^2}=0\,.
\nonumber 
\end{align}

\section{Gradient Flow and Linear Relations}
\label{sec:Gradient}

The RG flows of standard perturbative QFTs are gradient flows.\footnote{See Refs.~\cite{wallace1974gradient,wallace1975gradient} for foundational work pertaining to this fact
and Refs.~\cite{klebanov2011f,giombi2015interpolating,fei2015generalized,jack2015gradient,jack2017function} for modern discussions and more recent results.} What this means concretely for a given QFT with $n$ coupling constants $g_i$ is that there exits a function $F$ and an $n\times n$ matrix $G_{ij}$, both of which depend on the coupling constants, such that
\begin{align}
\label{gradient}
\frac{\partial F}{\partial g_i}= G_{ij}\,\beta_{g_j}\,.
\end{align}
If we define $t \equiv \log \mu$ where $\mu$ is the RG scale, such that $\beta_{g_i}=\frac{d}{dt}g_i$, then the gradient condition \eqref{gradient} implies that
\begin{align}
\frac{dF}{dt} = \frac{\partial F}{\partial g_i} \beta_{g_i} = \beta_{g_i} G_{ij}\, \beta_{g_j}\,,
\end{align}
from which we see that the RG flow is monotonic if the matrix $G_{ij}$ is positive definite or negative definite. Since the partial derivatives of $F$ with respect to couplings $g_i$ and $g_k$ commute, \eqref{gradient} tells us that
\begin{flalign}
\label{curl}
&\forall i,k:\hspace{50mm}\frac{\partial G_{ij} \beta_{g_j}}{\partial g_k}-\frac{\partial G_{kj} \beta_{g_j}}{\partial g_i}=0\,.
&&
\end{flalign}
The above equation provides a powerful way of checking that perturbative beta functions have been computed correctly. To carry out the check one must first determine $G_{ij}$, which can be computed by taking second derivatives of the free energy of the given theory. To the perturbative order we are working at, the free energy, which equals the sum of connected vacuum Feynman diagrams, also equals the anomalous dimension for $\phi$, up to rescalings of the coupling constants that, like the overall normalization of $G_{ij}$, carry no significance in this context. For the theories studied in this paper, we computed $\gamma_\phi$ to third order in the couplings, which means we can determine $G_{ij}$ to linear order:
\begin{align}
G_{ij}^{(1)}=\frac{\partial^2\gamma_\phi^{(3)}}{\partial g_i\partial g_j}\,.
\end{align}
Consequently, we are able to verify the curl condition \eqref{curl} to second order:
\begin{flalign}
&\forall i,j:\hspace{50mm}\frac{\partial G^{(1)}_{ij} \beta^{(3)}_{g_j}}{\partial g_k}-\frac{\partial G^{(1)}_{kj} \beta^{(3)}_{g_j}}{\partial g_i}= \mathcal{O}(g^3)\,.
&&
\end{flalign}
Other checks one can perform that beta functions are computed correctly include comparing the double trace beta function $\beta_{g_1}$ against the well-known beta function of the $O(N)$ model and checking that the beta functions respect the linear relations that exists among operators at finite values of $N$. The linear relations can be determined by direct computation but can also be extracted from $G_{ij}$, which depends not only on the coupling constants but also on $N$. In fact, the linear relations can be conveniently read off already from the piece $G_{ij}^{(0)}$ that is constant with respect to the coupling constants. When evaluated at values of $N$ for which linear relations exist among the operator, $G_{ij}^{(0)}$ will contain zero eigenvalues, and the corresponding eigenvectors reveal which linear combinations of operators vanish. Below we list all the linear relations that exist in the theories studied in this paper, except the obvious relation that when $N=1$, all quartic operators equal one another and ditto for the sextic operators. In the quartic theories, the operatos are all linearly independent for all $N\geq 2$, but linear relationships exist at $N=-2$, ie. in the corresponding $Sp(2)^r$ theories. In the sextic theories linear relations exist also at $N=2$ and $N=-4$. 
\begin{itemize}
\item quartic $O(N)^2$ theory:
\begin{flalign}
\nonumber& N = -2:
\hspace{10mm}
\mathcal{O}_1=-2\,\mathcal{O}_2\,,
&&
\end{flalign}
\item quartic $O(N)^3$ theory:
\begin{flalign}
\nonumber& N = -2:
\hspace{10mm}
\mathcal{O}_1
=-2\,\mathcal{O}_2
=-2\,\mathcal{O}_3
=-2\,\mathcal{O}_4
=\mathcal{O}_5
\,,
&&
\end{flalign}
\item quartic $O(N)^4$ theory:
\begin{flalign}
\nonumber& N = -2:
\hspace{10mm}
\mathcal{O}_1 = -2\mathcal{O}_2 = -2\mathcal{O}_3 = -2\mathcal{O}_4 = -2\mathcal{O}_5 =  2\,\mathcal{O}_8-2\,\mathcal{O}_{13}-2\,\mathcal{O}_{14}
\,,
&&
\\
\nonumber& \hspace{28mm}
\mathcal{O}_6 =  \mathcal{O}_8-\mathcal{O}_{13}-2\,\mathcal{O}_{14}
\,,
&&
\\
\nonumber& \hspace{28mm}
\mathcal{O}_7 =  \mathcal{O}_8-2\,\mathcal{O}_{13}-\mathcal{O}_{14}
\,,
&&
\\
\nonumber& \hspace{28mm}
\mathcal{O}_9 = \mathcal{O}_{14}
\,,
&&
\\
\nonumber& \hspace{28mm}
\mathcal{O}_{10} = \mathcal{O}_{13}
\,,
&&
\\
\nonumber& \hspace{28mm}
\mathcal{O}_{11} =  \mathcal{O}_{12} = -\mathcal{O}_{13}-\mathcal{O}_{14}\,,
&&
\end{flalign}
\item quartic $O(N)^5$ theory:
\begin{flalign}
\nonumber& N = -2:
\hspace{10mm}
\mathcal{O}_1 
= -2\,\mathcal{O}_{2,a} 
= -2\,\mathcal{O}_{2,b} 
= -2\,\mathcal{O}_{2,c} 
= -2\,\mathcal{O}_{2,d}
= -2\,\mathcal{O}_{2,e}
&&
\\
\nonumber& \hspace{33.5mm}
= 2\,\mathcal{O}_{3,ab}+2\,\mathcal{O}_{4,ab}
= 2\,\mathcal{O}_{3,ac}+2\,\mathcal{O}_{4,ac}
= 2\,\mathcal{O}_{3,ad}+2\,\mathcal{O}_{4,ad}
&&
\\
\nonumber& \hspace{33.5mm}
= 2\,\mathcal{O}_{3,ae}+2\,\mathcal{O}_{4,ae}
= 2\,\mathcal{O}_{3,bc}+2\,\mathcal{O}_{4,bc}
= 2\,\mathcal{O}_{3,bd}+2\,\mathcal{O}_{4,bd}
&&
\\
\nonumber& \hspace{33.5mm}
= 2\,\mathcal{O}_{3,be}+2\,\mathcal{O}_{4,be}
= 2\,\mathcal{O}_{3,cd}+2\,\mathcal{O}_{4,cd}
= 2\,\mathcal{O}_{3,ce}+2\,\mathcal{O}_{4,ce}
&&
\\
\nonumber& \hspace{33.5mm}
= 2\,\mathcal{O}_{3,de}+2\,\mathcal{O}_{4,de}
=-\mathcal{O}_{5,a,bc}-\mathcal{O}_{5,b,ac}-\mathcal{O}_{5,c,ab}-\mathcal{O}_{5,d,ab}-\mathcal{O}_{5,e,ab}
\,,
&&
\\
\nonumber& \hspace{28mm}
\mathcal{O}_{5,a,bc} = \mathcal{O}_{5,a,bd}
=\mathcal{O}_{5,a,be}
\,,
&&
\\
\nonumber& \hspace{28mm}
\mathcal{O}_{5,b,ac} = \mathcal{O}_{5,b,ad}
=\mathcal{O}_{5,b,ae}
\,,
&&
\\
\nonumber& \hspace{28mm}
\mathcal{O}_{5,c,ab} = \mathcal{O}_{5,c,ad}
=\mathcal{O}_{5,c,ae}
\,,
&&
\\
\nonumber& \hspace{28mm}
\mathcal{O}_{5,d,ab} = \mathcal{O}_{5,d,ac}
=\mathcal{O}_{5,d,ae}
\,,
&&
\\
\nonumber& \hspace{28mm}
\mathcal{O}_{5,e,ab} = \mathcal{O}_{5,e,ac}
=\mathcal{O}_{5,e,ad}
\,,
&&
\\
\nonumber& \hspace{28mm}
\mathcal{O}_{4,ab} 
= -\mathcal{O}_{5,a,bc}-\mathcal{O}_{5,b,ac}
\,,
&&
\\
\nonumber& \hspace{28mm}
\mathcal{O}_{4,ac} 
= -\mathcal{O}_{5,a,bc}-\mathcal{O}_{5,c,ab}
\,,
&&
\\
\nonumber& \hspace{28mm}
\mathcal{O}_{4,ad} 
= -\mathcal{O}_{5,a,bc}-\mathcal{O}_{5,d,ab}
\,,
&&
\\
\nonumber& \hspace{28mm}
\mathcal{O}_{4,ae} 
= -\mathcal{O}_{5,a,bc}-\mathcal{O}_{5,e,ab}
\,,
&&
\\
\nonumber& \hspace{28mm}
\mathcal{O}_{4,bc} 
= -\mathcal{O}_{5,b,ac}-\mathcal{O}_{5,c,ab}
\,,
&&
\\
\nonumber& \hspace{28mm}
\mathcal{O}_{4,bd} 
= -\mathcal{O}_{5,b,ac}-\mathcal{O}_{5,d,ab}
\,,
&&
\\
\nonumber& \hspace{28mm}
\mathcal{O}_{4,be} 
= -\mathcal{O}_{5,b,ac}-\mathcal{O}_{5,e,ab}
\,,
&&
\\
\nonumber& \hspace{28mm}
\mathcal{O}_{4,cd} 
= -\mathcal{O}_{5,c,ab}-\mathcal{O}_{5,d,ab}
\,,
&&
\\
\nonumber& \hspace{28mm}
\mathcal{O}_{4,ce} 
= -\mathcal{O}_{5,c,ab}-\mathcal{O}_{5,e,ab}
\,,
&&
\\
\nonumber& \hspace{28mm}
\mathcal{O}_{4,de} 
= -\mathcal{O}_{5,d,ab}-\mathcal{O}_{5,e,ab}
\,,
&&
\end{flalign}
\item sextic $O(N)^2$ theory:
\begin{flalign}
\nonumber& N = 2:
\hspace{13mm}
\mathcal{O}_1=3\,\mathcal{O}_2-2\,\mathcal{O}_3\,,
&&
\\
\nonumber& N = -2:
\hspace{10mm}
\mathcal{O}_1=-2\,\mathcal{O}_2=4\,\mathcal{O}_3\,,
&&
\\
\nonumber& N = -4:
\hspace{10mm}
\mathcal{O}_1=-6\,\mathcal{O}_2-8\,\mathcal{O}_3\,,
&&
\end{flalign}
\item sextic $O(N)^3\rtimes S_3$ theory:
\begin{flalign}
\nonumber& N = 2:
\hspace{13mm}
\mathcal{O}_1=2\,\mathcal{O}_5-\frac{1}{2}\,\mathcal{O}_6-\mathcal{O}_7+\frac{1}{2}\,\mathcal{O}_8\,,
&&
\\
\nonumber& 
\hspace{28mm}
\mathcal{O}_2=3\,\mathcal{O}_5-\frac{3}{2}\,\mathcal{O}_6-\frac{3}{2}\,\mathcal{O}_7+\mathcal{O}_8\,,
&&
\\
\nonumber& 
\hspace{28mm}
\mathcal{O}_3=\mathcal{O}_5-\frac{1}{2}\,\mathcal{O}_6+\frac{1}{2}\,\mathcal{O}_7\,,
&&
\\
\nonumber& 
\hspace{28mm}
\mathcal{O}_4=\frac{3}{2}\,\mathcal{O}_7-\frac{1}{2}\,\mathcal{O}_8\,,
&&
\\
\nonumber& N = -2:
\hspace{10mm}
\mathcal{O}_1=-\frac{4}{5}\,\mathcal{O}_2
=4\,\mathcal{O}_3
=4\,\mathcal{O}_4
=-2\,\mathcal{O}_5
=\mathcal{O}_6
=-2\,\mathcal{O}_7
=\mathcal{O}_8
\,,
&&
\\
\nonumber& N = -4:
\hspace{10mm}
\mathcal{O}_1=-\frac{1}{3}\,\mathcal{O}_2
-\frac{3}{2}\,\mathcal{O}_5
+\frac{1}{8}\,\mathcal{O}_6
+\frac{3}{4}\,\mathcal{O}_7
+\frac{1}{12}\,\mathcal{O}_8\,,
&&
\\
\nonumber& 
\hspace{28mm}
\mathcal{O}_3=-\frac{1}{2}\,\mathcal{O}_5-\frac{1}{8}\,\mathcal{O}_6-\frac{1}{4}\,\mathcal{O}_7\,,
&&
\\
\nonumber& 
\hspace{28mm}
\mathcal{O}_4=-\frac{3}{4}\,\mathcal{O}_7-\frac{1}{8}\,\mathcal{O}_8\,,
&&
\end{flalign}
\item sextic $O(N)^4\rtimes S_4$ theory:
\begin{flalign}
\nonumber& N = 2:
\hspace{13mm}
\mathcal{O}_1=3\,\mathcal{O}_5
-30\,\mathcal{O}_{16}
+4\,\mathcal{O}_{18}
+18\,\mathcal{O}_{19}
+12\,\mathcal{O}_{20}
-6\,\mathcal{O}_{21}\,,
&&
\\
\nonumber& 
\hspace{28mm}
\mathcal{O}_2=\mathcal{O}_5
+2\,\mathcal{O}_{13}
-8\,\mathcal{O}_{16}
+2\,\mathcal{O}_{18}
+2\,\mathcal{O}_{19}
+2\,\mathcal{O}_{20}\,,
&&
\\
\nonumber& 
\hspace{28mm}
\mathcal{O}_3=
3\,\mathcal{O}_{13}
+3\,\mathcal{O}_{16}
+\,\mathcal{O}_{18}
-6\,\mathcal{O}_{19}
-3\,\mathcal{O}_{20}
+3\,\mathcal{O}_{21}\,,
&&
\\
\nonumber& 
\hspace{28mm}
\mathcal{O}_4=
\mathcal{O}_{5}
+4\,\mathcal{O}_{15}
-8\,\mathcal{O}_{16}
+2\,\mathcal{O}_{18}
-4\,\mathcal{O}_{19}
+2\,\mathcal{O}_{20}
+4\,\mathcal{O}_{21}
\,,
&&
\\
\nonumber& 
\hspace{28mm}
\mathcal{O}_6=
2\,\mathcal{O}_{13}
-2\,\mathcal{O}_{16}
+\mathcal{O}_{18}
\,,
&&
\\
\nonumber& 
\hspace{28mm}
\mathcal{O}_7=
\mathcal{O}_{13}
+2\,\mathcal{O}_{15}
-2\,\mathcal{O}_{16}
+\mathcal{O}_{18}
-3\,\mathcal{O}_{19}
+2\,\mathcal{O}_{21}
\,,
&&
\\
\nonumber& 
\hspace{28mm}
\mathcal{O}_8=
\mathcal{O}_{13}
+2\,\mathcal{O}_{16}
-\,\mathcal{O}_{19}
-\,\mathcal{O}_{20}
\,,
&&
\\
\nonumber& 
\hspace{28mm}
\mathcal{O}_9=
6\,\mathcal{O}_{15}
-9\,\mathcal{O}_{16}
+\mathcal{O}_{18}
-3\,\mathcal{O}_{19}
+3\,\mathcal{O}_{20}
+3\,\mathcal{O}_{21}
\,,
&&
\\
\nonumber& 
\hspace{28mm}
\mathcal{O}_{10}=
2\,\mathcal{O}_{15}
-2\,\mathcal{O}_{16}
+\,\mathcal{O}_{20}
\,,
&&
\\
\nonumber& 
\hspace{28mm}
\mathcal{O}_{11}=
\mathcal{O}_{16}
+\mathcal{O}_{19}
-\mathcal{O}_{21}
\,,
&&
\\
\nonumber& 
\hspace{28mm}
\mathcal{O}_{12}=
\mathcal{O}_{18}
\,,
&&
\\
\nonumber& 
\hspace{28mm}
\mathcal{O}_{14}=
2\,\mathcal{O}_{15}
-\mathcal{O}_{16}
+\mathcal{O}_{18}
-3\,\mathcal{O}_{19}
+2\,\mathcal{O}_{21}
\,,
&&
\\
\nonumber& 
\hspace{28mm}
\mathcal{O}_{17}=
3\,\mathcal{O}_{19}
-2\,\mathcal{O}_{21}
\,,
&&
\\
\nonumber& N = -2:
\hspace{10mm}
\mathcal{O}_1
=-2\,\mathcal{O}_2
=4\,\mathcal{O}_3
=2\,\mathcal{O}_4
=4\,\mathcal{O}_6
=-4\,\mathcal{O}_7
=-4\,\mathcal{O}_{18}+8\,\mathcal{O}_{21}
\,,
&&
\\
\nonumber& 
\hspace{28mm}
\mathcal{O}_5=\mathcal{O}_8=\mathcal{O}_{13}=\mathcal{O}_{19}=0
\,,
&&
\\
\nonumber& 
\hspace{28mm}
\mathcal{O}_9=
-\mathcal{O}_{18}
+4\,\mathcal{O}_{21}
\,,
&&
\\
\nonumber& 
\hspace{28mm}
\mathcal{O}_{10}=
-\mathcal{O}_{11}
=-2\,\mathcal{O}_{21}
\,,
&&
\\
\nonumber& 
\hspace{28mm}
\mathcal{O}_{12}=
-\mathcal{O}_{18}
\,,
&&
\\
\nonumber& 
\hspace{28mm}
\mathcal{O}_{14}=
-\mathcal{O}_{18}
+\mathcal{O}_{21}
\,,
&&
\\
\nonumber& 
\hspace{28mm}
\mathcal{O}_{15}=
-\mathcal{O}_{16}=
-\frac{1}{2}\,\mathcal{O}_{17}=
\frac{1}{2}\,\mathcal{O}_{20}=
\mathcal{O}_{21}
\,,
&&
\\
\nonumber& N = -4:
\hspace{10mm}
\mathcal{O}_3=-\frac{1}{8}\,\mathcal{O}_1
-\frac{3}{4}\,\mathcal{O}_2\,,
&&
\\
\nonumber& 
\hspace{28mm}
\mathcal{O}_8=
-\frac{1}{4}\,\mathcal{O}_2
-\frac{1}{4}\,\mathcal{O}_4
-\frac{1}{4}\,\mathcal{O}_5
-\mathcal{O}_6
-\mathcal{O}_7
\,,
&&
\\
\nonumber& 
\hspace{28mm}
\mathcal{O}_{12}=
\frac{1}{8}\,\mathcal{O}_1
+\frac{3}{2}\,\mathcal{O}_2
+\frac{3}{4}\,\mathcal{O}_4
+\frac{3}{4}\,\mathcal{O}_5
+3\,\mathcal{O}_6
-\mathcal{O}_9
-3\,\mathcal{O}_{10}
-3\,\mathcal{O}_{11}
\,,
&&
\\
\nonumber& 
\hspace{28mm}
\mathcal{O}_{13}=-\frac{1}{4}\,\mathcal{O}_5\,,
&&
\\
\nonumber& 
\hspace{28mm}
\mathcal{O}_{18}=
\frac{1}{2}\,\mathcal{O}_5
-\mathcal{O}_6
-2\,\mathcal{O}_7
-2\,\mathcal{O}_{14}
-4\,\mathcal{O}_{15}
-2\,\mathcal{O}_{16}
-\mathcal{O}_{17}
\,,
&&
\\
\nonumber & 
\hspace{28mm}
\mathcal{O}_{19}=
-\frac{1}{24}\,\mathcal{O}_1
-\frac{1}{2}\,\mathcal{O}_2
-\frac{1}{4}\,\mathcal{O}_4
-\frac{5}{12}\,\mathcal{O}_5
-\frac{2}{3}\,\mathcal{O}_6
+\frac{2}{3}\,\mathcal{O}_7
+\frac{1}{3}\,\mathcal{O}_9
+\mathcal{O}_{10}
&&
\\
\nonumber & \hspace{39mm}
+\mathcal{O}_{11}
+\frac{2}{3}\,\mathcal{O}_{14}
+\frac{4}{3}\,\mathcal{O}_{15}
+\frac{2}{3}\,\mathcal{O}_{16}
+\frac{1}{3}\,\mathcal{O}_{17}
\,,
&&
\\
\nonumber & 
\hspace{28mm}
\mathcal{O}_{20}=
\frac{1}{12}\,\mathcal{O}_1
+\frac{3}{2}\,\mathcal{O}_2
+\mathcal{O}_4
+\frac{19}{12}\,\mathcal{O}_5
+\frac{10}{3}\,\mathcal{O}_6
+\frac{2}{3}\,\mathcal{O}_7
-\frac{2}{3}\,\mathcal{O}_9
-3\,\mathcal{O}_{10}
&&
\\
\nonumber & \hspace{39mm}
-2\,\mathcal{O}_{11}
-\frac{4}{3}\,\mathcal{O}_{14}
-\frac{20}{3}\,\mathcal{O}_{15}
-\frac{16}{3}\,\mathcal{O}_{16}
-\frac{2}{3}\,\mathcal{O}_{17}
\,,
&&
\\
\nonumber & 
\hspace{28mm}
\mathcal{O}_{21}=
\frac{1}{16}\,\mathcal{O}_1
+\frac{3}{4}\,\mathcal{O}_2
+\frac{3}{8}\,\mathcal{O}_4
+\frac{5}{8}\,\mathcal{O}_5
+\mathcal{O}_6
-\mathcal{O}_7
-\frac{1}{2}\,\mathcal{O}_9
-\frac{3}{2}\,\mathcal{O}_{10}
&&
\\
\nonumber & \hspace{39mm}
-3\,\mathcal{O}_{11}
-\mathcal{O}_{14}
-2\,\mathcal{O}_{15}
-4\,\mathcal{O}_{16}
-\mathcal{O}_{17}
\,.
&&
\end{flalign}
\end{itemize}
For each of the $O(N)^r$ models, it can be checked explicitly that the matrix $G_{ij}$ is positive definite, implying that the RG flow is indeed monotonic, as required by unitarity. When we send $N$ to $-N$, it can be checked in each case  that $G_{ij}$ either remains positive definite or becomes negative definite, entailing that the RG flows of the corresponding $Sp(N)^r$ theories too are monotonic. When $r$ is odd, however, the fundamental fields of the $Sp(N)^r$ theories obtained by sending $N\rightarrow -N$ are anti-commuting scalars, so that these theories are non-unitary. But in Ref.~\cite{leclair2007semi} it was shown that the Hamiltonian of the ``wrong statistics" $Sp(N)$ theory possesses the property of being pseudo-Hermitian. In light of the monotonic RG flows we observe, we are therefore naturally led to pose the question, does pseudo-unitarity imply monotonicity?

\bibliographystyle{ssg}
\bibliography{tensor}

\end{document}